\documentstyle[amssymbols]{article}
\def\hybrid{\topmargin 0pt      \oddsidemargin 0pt
	\headheight 0pt \headsep 0pt \textheight 9in \textwidth 6.25in
\marginparwidth .875in \parskip 5pt plus 1pt \jot = 1.5ex}

\catcode`\@=11
\def\marginnote#1{}
\def\numberbysection{\@addtoreset{equation}{section}
	\def\theequation{\thesection.\arabic{equation}}}
\catcode`@=12
\relax
\numberbysection
\hybrid
\def\C{{\Bbb C}}

\begin{document}

\noindent
\hfill 
	{PAR--LPTHE 93--26}
\vskip 2cm
\centerline{\Large Integrable mappings and polynomial growth.}
\centerline{\today}
\vskip 2cm
\centerline{S. Boukraa, J-M. Maillard and G. Rollet}
\centerline{Laboratoire de Physique Th\'eorique et des Hautes Energies}
\centerline{Unit\'e associ\'ee au C.N.R.S ( UA 280)}
\centerline{Universit\'e de Paris 6-Paris 7, Tour 16, $1^{\rm er}$
\'etage, bo\^\i te 126}
\centerline{4 Place Jussieu, F--75252 PARIS Cedex 05}

\vskip 2cm
\begin{abstract}
We describe birational representations of discrete groups generated by
involutions,
 having their origin in the theory of exactly solvable vertex-models in
lattice statistical mechanics.
These involutions correspond respectively to two kinds of
transformations
on $q \times q$ matrices:
the inversion of the $q \times q$ matrix and an (involutive)
permutation of
the entries of the matrix.
We concentrate on the case where these permutations are elementary
transpositions of two entries.
In this case the birational transformations fall into six different
classes.
For each class we analyze the factorization properties of the iteration
of
these transformations. These factorization properties enable to define
some
 canonical homogeneous polynomials associated with these factorization
properties.
Some mappings yield a polynomial growth of the complexity of the
iterations.
For three  classes the successive iterates, for $q=4$, actually
lie on elliptic curves.
This analysis also  provides examples of integrable mappings in
arbitrary
dimension, even infinite.
Moreover, for two classes, the homogeneous polynomials are
shown to satisfy non trivial non-linear recurrences.
The relations between factorizations of the iterations,
 the existence of recurrences on one or several variables, as well as
 the
integrability of the mappings are analyzed.
\end{abstract}

\noindent
 {\bf AMS Classification scheme numbers}: 82A68, 82A69, 14E05, 14J50,
16A46, 16A24, 11D41

\vfill
work supported by CNRS


\newpage
\newpage
\section{Introduction}
In previous papers, we have analyzed
birational representations of discrete groups generated by
involutions, having their origin in the theory of exactly solvable
models in lattice statistical mechanics
\cite{bmv1,bmv1b,bmv2,bmv2b,prl,prl2}.

The group
of birational transformations first studied
in~\cite{Ma84,St79,Ba82,Ma86}
 is generated by the so-called {\em inversion
relations}~\cite{Ba81,JaMa82} which amount to combine
two very simple algebraic transformations:
the matrix inversion and permutations
\footnote{In the framework of vertex lattice models. For spin models,
the
groups of birational transformations are also generated by similar
simple
involutions but slightly different~\cite{bmv1,bmv1b,Ma86}.}
 of the entries of a matrix
\cite{bmv1}.  In such a general framework, the dimension
of the lattice and  the lattice itself, only occur through the number
of
inversion relations and the permutations of the entries introduced to
generate the birational transformations \cite{bmv2b,prl,MaRo93}.

This justifies considering the following problem:
to generalize to $q \times q$ matrices and  analyze {\em  birational}
transformations generated by
the matrix inverse and a permutation of the entries of the matrix, and
finally find the permutations of the matrix for which the corresponding
birational transformations yield {\em integrable} mappings.
This analysis
is interesting for itself for the theory of mappings of several
variables and the theory of discrete dynamical systems, disregarding
the
relation with integrable lattice models.
Such an analysis is performed in a series of parallel
publications~\cite{BoMaRo93a,BoMaRo93c,BoMaRo93d}.

In~\cite{BoMaRo93a}, a particular transposition of the entries was
analyzed.
For this very transposition, it has been shown that the iteration of
the
associated birational transformations {\em present some remarkable
factorization  properties}.
Actually {\em the entries of the successive}  $q \times q$
{\em matrices corresponding to the iteration of our transformation, as
well
as the determinants of these matrices, do factorize into homogeneous
polynomials of all the entries of the initial} $q \times q$ {\em
matrix}.
These factorization properties explain~\cite{FaVi93} why the
``complexity''
of these iterations (degree of the successive iterates), instead of
having
the exponential growth one expects at first sight, actually has a {\em
polynomial growth}.
It has also been shown that the homogeneous polynomial factors
occurring in these factorizations {\em do satisfy remarkable
non-linear recurrences} and that these recurrences {\em were actually
integrable recurrences yielding algebraic elliptic
curves}~\cite{BoMaRo93a}.

We will concentrate
here on  simple heuristic examples of permutations: in fact all the
transpositions of {\em two} entries of a $q \times q$ matrix.
In~\cite{BoMaRo93d} it has been shown that the analysis of the
birational
transformations corresponding to {\em all the transpositions of two
entries}, actually reduces to the study of {\em six classes} of such
mappings.

The transposition analyzed in~\cite{BoMaRo93a} corresponds to the first
class with respect to this classification. We will revisit here  the
analysis performed in~\cite{BoMaRo93a} (occurrence of factorizations,
recurrences, ... ) for the five remaining classes.
The  mappings associated with three of these five classes are not
integrable, even for $q=4$~\cite{BoMaRo93d}.
This will clarify the relations between all these structures
and the integrability.
In particular, it  will help to understand {\em to what extend
factorizations yield integrability}.
In fact, it will be shown that the occurrence of factorizations is a
{\em
quite general phenomenon}: it does occur even outside the framework of
integrability.
The existence of factorization of our transformations yields a growth
of
the complexity of the iterations, even when exponential,
smaller than the generic $(q-1)^n$ growth. On the other hand, {\em
integrable mappings only occur with a polynomial growth} of the
complexity
of the iteration.
As far as recurrences are concerned, it is tempting, at first sight, to
see
a close connection between the occurrence of  {\em recurrences} and the
integrability of the birational transformations, since this
integrability
yields  {\em curves}.
The detailed analysis of the five remaining classes rules out such
naive
connections, and will make clear the actual relations between these
various
structures.
{\em As a byproduct it will provide, with the integrable subcase of one
of
these classes} (class IV), {\em an example
of integrable mapping in arbitrary dimension, even infinite}.

\section{Notations}

Let us consider the following $q \times q$ matrix:
\begin{eqnarray}
\label{rq}
R_q=
\pmatrix{
m_{11} & m_{12} & m_{13} & m_{14} & \cdots \cr
m_{21} & m_{22} & m_{23} & m_{24} & \cdots \cr
m_{31} & m_{32} & m_{33} & m_{34} & \cdots \cr
m_{41} & m_{42} & m_{43} & m_{44} & \cdots \cr
\vdots & \vdots & \vdots & \vdots & \ddots \cr }
\end{eqnarray}
Let us introduce the following transformations, the matrix inverse
$\widehat{I}$, the homogeneous matrix inverse $I$:
\begin{eqnarray}
\widehat{I}: R_q \longrightarrow R_q^{-1}
\end{eqnarray}
\begin{eqnarray}
I: R_q \longrightarrow \, \, R_q^{-1} \cdot \det (R_q)
\end{eqnarray}
The homogeneous inverse $I$ is a {\em polynomial} transformation on
each of
the entries $m_{ij}$ which associates to each $m_{ij}$ its
corresponding
cofactor.

In the following, $t$ will denote an arbitrary transposition of two
entries
of the $q \times q$ matrix, and $t_{ij-kl}$ will denote the
transposition
exchanging $m_{ij}$ and $m_{kl}$.

The two transformations $t$ and $\widehat{I}$ are {\em involutions}
whereas
the homogeneous inverse verifies

 $I^2=(\det(R_q))^{q-2}\cdot{\cal I}d$,
where ${\cal I}d$ denotes the identity transformation.

We also introduce the (generically {\em infinite order})
transformations
$K=t \cdot I$ and $\widehat{K}=t \cdot \widehat{I}$.

$K$ is a (homogeneous) {\em polynomial} transformation on the entries
$m_{ij}$, while transformation $\widehat{K}$ is clearly a rational
transformation on the entries  $m_{ij}$.
In fact $\widehat{K}$ is a {\em birational} transformation since its
inverse transformation is $\widehat{I} \cdot t$.

\section{Recalls}

\subsection{Six equivalence classes}
\label{six}

Let us first recall that, as far as the analysis of transformation $K$
is
concerned, {\em all the transpositions can actually be reduced to six
different classes}~\cite{BoMaRo93d} {\em of transpositions}
\footnote{At first sight, one has to study as many mappings as there
are
transpositions $t$, of two elements among the sixteen entries of the
matrix, that is $\left (^2_{16}\right )=120$.}.
One can thus study a single mapping in each class and directly deduce
the
results concerning all the other transformations of the same class.
A first step to prove it amounts to giving an equivalence relation on
these
$120$ transpositions, which does not modify the structure of the
corresponding transformations~\cite{BoMaRo93d}.
This equivalence up to relabelling conjugations, does not modify the
properties of the mappings and yields {\em seven} equivalence classes
(with the notation $[m_{ij}-m_{kl}]$ denoting the transposition
exchanging
 the two entries  $m_{ij}$ and $m_{kl}$ of matrix (\ref{rq})):

- Class ${\cal C}_1 $  corresponds to all the 6 transpositions of the
form
$[m_{ij}-m_{ji}]$

- Class  ${\cal C}_2 $  corresponds to all the 6 transpositions of the
form
$[m_{ii}-m_{jj}]$

- Class  ${\cal C}_3 $  corresponds to all the 12 transpositions of the
form $[m_{ij}-m_{kl}]$

- Class  ${\cal C}_4 $  corresponds to all the 24 transpositions of the
form $[m_{ij}-m_{jk}]$ or $[m_{ji}-m_{kj}]$

- Class  ${\cal C}_5 $  corresponds to all the 24 transpositions of the
form $[m_{ij}-m_{ik}]$ or $[m_{ji}-m_{ki}]$

-Class  ${\cal C}_6 $  corresponds to all the 24 transpositions of the
form
$[m_{ii}-m_{jk}]$

- Class  ${\cal C}_7 $  corresponds to all the 24 transpositions of the
form $[m_{ii}-m_{ij}]$ or $[m_{ii}-m_{ji}]$

where the various indices $i, j, k$ and $l$ are all different.

Moreover, one can actually show~\cite{BoMaRo93d} that classes  ${\cal
C}_1
$ and  ${\cal C}_2 $ {\em yield the same behavior} as far as the
iterations
of their associated birational mappings are concerned:
the transformations $K^2$ respectively associated to classes  ${\cal
C}_1 $
and ${\cal C}_2 $ {\em are conjugated}.
Therefore classes  ${\cal C}_1 $ and  ${\cal C}_2 $ {\em can be brought
together in the same class}, we will denote class I, as far as the
analysis
of the birational transformation is concerned.
The five other classes  $({\cal C}_3, \ldots, {\cal C}_7) $ will be
relabelled classes (II,\ldots, VI) in the same order.

It is important to note that this classification in six classes
 {\em holds for $q \times q$ matrices, for any value of $q \ge 4$}.
For $q=3$ one remarks that class II no longer exists and similarly,
 for $q=2$, classes III, IV, V do not exist anymore.

Let us also remark that any transposition of two entries $m_{i_1 j_1}$
 and $m_{i_2 j_2}$ of a $q \times q$ matrix can be associated with a
 transposition exchanging $m_{\sigma(i_1) \sigma(j_1)}$ and
 $m_{\sigma(i_2) \sigma(j_2)}$, where $\sigma(i_1), \sigma(j_1),
\sigma(i_2)$ and $\sigma(j_2)$ run into $\{1,2,3,4\}$.
One can thus restrict the transposition to one in the $4 \times 4$
block-matrix corresponding to the first four rows and columns.

\subsection{Class I}
\label{rappel}

Let us recall the factorization properties and
the recurrences obtained for transposition $t_{12-21}$~\cite{BoMaRo93a}
which represents one transposition among a set of transpositions which
has
been denoted class I in the exhaustive classification given in the
previous
section.
Let us also recall that this transformation corresponds, for $q=4$, to
{\em
integrable mappings} and yields a foliation of $\C P_{15}$ in {\em
algebraic elliptic curves} given as intersections of
quadrics~\cite{BoMaRo93d}.

Let us first consider the successive matrices obtained by iteration of
the
homogeneous transformation $K$, associated with  $t_{12-21}$, on
generic $q
\times q$ matrix $R_q$ and their determinants:
\begin{eqnarray}
M_0=R_q,\quad M_1=K(M_0)\,, \quad
f_1=\det(M_0)   \nonumber
\end{eqnarray}

The determinant of  matrix $M_1$ {\em remarkably factorizes} enabling
 to introduce a homogeneous polynomial $f_2$:
\begin{eqnarray}
\label{f2q-3}
f_2={{\det(M_1)}\over {f_1^{q-3}}}
\end{eqnarray}

Moreover, $f_1^{q-4}$ also {\em  factorizes in all the entries} of
matrix
$K(M_1)$, leading to introduce a ``reduced'' matrix $M_2$:
\begin{eqnarray}
\label{m2q-4}
M_2= {{K(M_1)}\over {f_1^{q-4}}}
\end{eqnarray}

Again, $\det(M_2)$ factorizes enabling to define a new polynomial
$f_3$:
\begin{eqnarray}
\label{f3m2}
f_3={{\det(M_2)}\over {f_1^3 \cdot f_2^{q-3}}}
\end{eqnarray}

Calculating $K(M_2)$, one can see that $f_1^2 \cdot f_2^{q-4}$
 factorizes in all the entries of this matrix $K(M_2)$,
leading  a new matrix:
\begin{eqnarray}
\label{m3q-4}
M_3={{K(M_2)}\over{f_1^2 \cdot f_2^{q-4}}}
\end{eqnarray}

Again, its determinant factorizes $f_1^{q-1} \cdot f_2^3 \cdot
f_3^{q-3}$,
 yielding the homogeneous polynomial $f_4$:
\begin{eqnarray}
\label{f4m3}
f_4={{\det(M_3)}\over{f_1^{q-1} \cdot f_2^3 \cdot f_3^{q-3}}}
\end{eqnarray}

Calculating $K(M_3)$, one sees that $f_1^{q-2} \cdot f_2^2 \cdot
f_3^{q-4}$
factorizes in all the entries of this matrix $K(M_3)$,
leading to introduce a new matrix:
\begin{eqnarray}
M_4={{K(M_3)}\over{f_1^{q-2} \cdot f_2^2 \cdot f_3^{q-4}}}
\end{eqnarray}

The factorization properties are now stabilized and they reproduce
similarly at any order $n$. Generally,  for $n \ge 1$ and $q \ge 4$,
one
has the factorizations:
\begin{eqnarray}
\label{mkVI}
M_{n+3} \, = \, {{K(M_{n+2})}\over{f_n^{q-2}\; f_{n+1}^2 \;
f_{n+2}^{q-4} }}  \\
\label{detVI}
f_{n+3} \, = \, {{\det(M_{n+2})}\over{f_n^{q-1}\; f_{n+1}^3 \;
f_{n+2}^{q-3} }}
\end{eqnarray}
giving the following relation {\em independent of} $q$:
\begin{eqnarray}
\label{MGMqq}
{{K(M_{n+2})} \over {\det(M_{n+2})}}\, = \, {{M_{n+3}} \over {f_{n}
f_{n+1} f_{n+2} f_{n+3}}}
\end{eqnarray}

Note that $K(M_{n+2})/\det(M_{n+2})$ is nothing but
$\widehat{K}(M_{n+2})$.

This defines the (left) action of the homogeneous transformation
 $K$ on matrices $M_n$ and on the set of polynomials $f_n$.
These polynomials are closely related to determinants of these
matrices, and  are actually the (generically) ``optimal''
 factorizations corresponding to the iterations of the (left) action of
$K$~\cite{BoMaRo93a}.

{\em One can also introduce a right-action of }$K$ on the matrices
$M_n$:
the entries $m_{ij}$ of $M_0$ are replaced by the corresponding entries
of
$K(M_0)$, i.e. $( K(M_0) )_{ij}$ (and similarly  for any algebraic
expression of these entries such as the $f_n$'s for instance).
Amazingly, the {\em right-action} of $K$ on the $f_n$'s and the
matrices
$M_{n}$'s yields a {\em remarkable factorization} of $f_1$ (and only
$f_1$):
\begin{eqnarray}
\label{kfnVI}
(f_n)_K = f_{n+1} \cdot f_1^{\mu_n}
\end{eqnarray}
and:
\begin{eqnarray}
\label{mfnVI}
(M_n)_K = M_{n+1} \cdot f_1^{\nu_n}
\end{eqnarray}

In order to relate the right and left action of $K$, one can also
introduce
the matrices $\widehat{M}_n$ which corresponds to $n$-times the left
(or
right) action of $K$ on $M_0$:
\begin{eqnarray}
\widehat{M}_{n}\, =\, K^n(M_0) \,=\, \Bigr ( M_0 \Bigr )_{K^n}
\end{eqnarray}

One has the following relations:
\begin{eqnarray}
\label{set1}
\widehat{M}_{1} &=& K (M_0), \;\;
\widehat{M}_{2} \, =\,  K (\widehat{M}_{1})
\, = \, f_1^{(q-4)} \cdot M_2, \cdots \nonumber \\
&\vdots&\nonumber \\
\widehat{M}_n &=& K(\widehat{M}_{n-1})  \\
&=& \Bigl ( f_1^{(q-1)^{n-4}} \cdot f_2^{(q-1)^{n-5}}
\cdot f_3^{(q-1)^{n-6}} \cdots f_{n-4}^{(q-1)}
\cdot f_{n-3}\Bigr )^{(q-2)^3} \cdot
f_{n-2}^{(q-2)(q-3)} \cdot f_{n-1}^{q-4} \cdot M_n \nonumber
\end{eqnarray}

Denoting $\alpha_n$ the degree of the determinant of matrix $M_n$ and
$\beta_n$ the degree of the polynomial $f_n$, one immediately gets from
equations (\ref{mkVI}), (\ref{detVI}), (\ref{kfnVI}) and  (\ref{mfnVI})
the
following linear relations (with integer coefficients):
\begin{eqnarray}
\label{aln}
\alpha_{n+2} &=& (q-1)\; \beta_{n} + 3\; \beta_{n+1} + (q-3)\;
\beta_{n+2}
+\beta_{n+3}\; , \nonumber \\
(q-1)\; \alpha_{n+2} &=& \alpha_{n+3} + q(q-2)\;\beta_{n}+2q\;
\beta_{n+1}+q(q-4)\;\beta_{n+2}\; , \nonumber \\
(q-1) \; \beta_n &=& \beta_{n+1} + q \; \mu_n \; , \nonumber \\
(q-1) \; \alpha_n &=& \alpha_{n+1} + q^2 \; \nu_n
\end{eqnarray}

Let us introduce $\alpha(x)$, $\beta(x)$, $\mu(x)$ and $\nu(x)$  the
generating functions of the $\alpha_n$'s, $\beta_n$'s, $\mu_n$'s and
$\nu_n$'s:
\begin{eqnarray}
\label{g}
\alpha(x)= \sum^\infty_{n=0} \alpha_n \cdot x^n  ,\, \,
\beta(x)= \sum^\infty_{n=1} \beta_n \cdot x^n  ,\, \,
\mu(x)= \sum^\infty_{n=1} \mu_n \cdot x^n ,\, \,
\nu(x)= \sum^\infty_{n=1} \nu_n \cdot x^n
\end{eqnarray}
{}From the right-action of $K$ (see factorizations (\ref{kfnVI}) and
(\ref{mfnVI})) one also gets linear relations on the $\alpha_n$'s,
$\beta_n$'s, $\mu_n$'s and $\nu_n$'s:
\begin{eqnarray}
\label{abnmunuI}
(q-1) \; \beta_n \, = \, \beta_{n+1} + q \; \mu_n \, , \quad
(q-1) \; \alpha_n \,= \,
\alpha_{n+1} + q^2 \; \nu_n
\end{eqnarray}
 as well as the corresponding linear relations on the generating
 functions:
\begin{eqnarray}
\label{ggI}
((q-1)\,x-1) \cdot \beta(x) \,= \,q\,x \; \mu(x) - q\,x \,\, ,  \quad
((q-1)\,x-1)
\cdot \alpha(x)\, =\, q^2 \, x \; \nu(x) - q
\end{eqnarray}
The explicit expressions of these generating functions read
respectively:
\begin{eqnarray}
\label{abmnx1}
\alpha(x)&=&{{q\;(1+(q-3)x+3x^2+(q-1)x^3)}\over{(1+x)(1-x)^3}}\; ,
\quad
\beta(x)\,=\, {{q\;x}\over {(1+x)(1-x)^3}} \; , \,\nonumber \\
\mu(x)&=&{{(q-3)+2\,x^2 - x^3)}\over {(1+x)(1-x)^3}}\; ,  \quad
\nu(x)\,=\,{{x\;((q-4)+2x+(q-2)x^2)} \over {(1+x)(1-x)^3}}
\end{eqnarray}
 giving on the $\alpha_n$'s, $\beta_n$'s, $\mu_n$'s and $\nu_n$'s:
\begin{eqnarray}
\label{abmn}
\alpha_n&=&q \; \Bigl ( {{q\,n^2}\over{2}} + {{q}\over{4}} -
{{(-1)^n\,q}\over{4}} + (-1)^n \Bigr )\, ,  \quad
\beta_n ={{q}\over{8}} \; \Bigl ( 2\,n\,(n+2)+1-(-1)^n \Bigr )\; ,
\nonumber \\
\mu_n&=& \beta_n-{{(n+1)\,(n+2)}\over{2}}\, ,  \quad
\nu_n ={{\alpha_n}\over{q}}\, -\, (1+n+n^2)
\end{eqnarray}

On the explicit expressions (\ref{abmn}) of these degrees and
exponents,
 one sees that the iteration of the homogeneous
transformation $K$ yields, as a consequence of
 factorizations (\ref{mkVI}), (\ref{mfnVI}), ...
 a {\em polynomial growth} of
the complexity of the calculations:
the degree of all the homogeneous expressions appearing in the
iterations
(the entries of successive matrices $M_n$, their determinants, ...)
grows
like $n^2$.
On the generating functions $\alpha(x)$, $\beta(x)$, $\mu(x)$ and
$\nu(x)$
(relations (\ref{abmnx1})) this corresponds to the fact that one
{\em only has} $x=\pm 1$ {\em singularities}.

Another important consequence of these factorizations is to introduce
the
(optimal) homogeneous polynomials $f_n$.
Remarkably,  these polynomials {\em do verify, independently  of $q$, a
whole hierarchy of non-linear recurrences}~\cite{BoMaRo93a} such as:
\begin{eqnarray}
\label{recufnqq}
{{f_{n} f_{n+3}^2 - f_{n+4} f_{n+1}^2 }
\over {f_{n-1} f_{n+3} f_{n+4} - f_{n} f_{n+1} f_{n+5}}}\, = \,
{{f_{n-1} f_{n+2}^2 - f_{n+3} f_{n}^2 }
\over {f_{n-2} f_{n+2} f_{n+3} - f_{n-1} f_{n} f_{n+4}}}
\end{eqnarray}
or for instance, among many other recurrences:
\begin{eqnarray}
\label{recufnqq2}
{{f_{n+1} f_{n+4}^2 f_{n+5} - f_{n+2} f_{n+3}^2 f_{n+6} }
\over {f_{n+2}^2 f_{n+3} f_{n+7} - f_{n} f_{n+4} f_{n+5}^2 }}\, = \,
{{f_{n+2} f_{n+5}^2 f_{n+6} - f_{n+3} f_{n+4}^2 f_{n+7} }
\over {f_{n+3}^2 f_{n+4} f_{n+8} - f_{n+1} f_{n+5} f_{n+6}^2 }}
\end{eqnarray}

These recurrences are of course compatible with the linear
recurrences on the $\beta_n$'s (equations (\ref{aln})) and also with
the
right-action of $K$ on the $f_n$'s (see factorization (\ref{kfnVI})).
Moreover these recurrences do have a {\em three parameter symmetry
group}. Introducing the variables $g_n$ which are the product of two
consecutive polynomials $f_n$,  $\,g_n\, = \, f_n \cdot f_{n+1}\,$,
one
can simply verify that all these recurrences are actually invariant
under the
three-parameter symmetry group:
\begin{eqnarray}
\label{symthree}
g_n \, \, \rightarrow \, \, a^{n^2} \cdot b^n \cdot c \cdot \, g_n
\end{eqnarray}
It is therefore tempting to introduce new variables taking these
symmetries
into account (more precisely, variables {\em invariant} under this
three-parameter group):
\begin{eqnarray}
\label{xnfn}
x_n \, = \, {{f_{n-1}^2 \; f_{n+2}} \over {f_{n+1}^2 \;f_{n-2}}}
\end{eqnarray}
In fact, these variables can directly be obtained from the
inhomogeneous
transformation $\widehat{K}$ (see section (\ref{dem})) and read:
\begin{eqnarray}
\label{xndet}
x_n \, =\,  \det\Bigl(\widehat{K}^n(R_{q})\cdot
\widehat{K}^{n+1}(R_{q})\Bigr)
\end{eqnarray}
The equivalence of these two definitions for $x_n$, (\ref{xnfn}) and
(\ref{xndet}), has already been explained in~\cite{BoMaRo93a} and will
not
be recalled here.

With these new variables, recurrence (\ref{recufnqq}) becomes:
\begin{eqnarray}
\label{recuxnqq}
{{{x_{n+2}}-1} \over {{x_{n+1}}\,{x_{n+2}}\,{x_{n+3}}-1}}\, =\,
{{{x_{n+1}}-1} \over {{x_{n}}\,{x_{n+1}}\,{x_{n+2}}-1}} \cdot
{{x_{n}}\,{x_{n+2}}}
\end{eqnarray}

Similarly to the $f_n$'s, one has a {\em whole hierarchy of
recurrences}
on the $x_n$'s.
The analysis of this hierarchy of {\em compatible} non-linear
recurrences
has been performed in~\cite{BoMaRo93a} and will not be detailed here.

It is important to note that these recurrences can be extended to
 {\em any relative integer} $n$. Even more, {\em these
recurrences are invariant under the ``time-reversal'' transformation}:
\begin{eqnarray}
\label{time}
x_n \, \,  \longrightarrow\, \,\, {{1} \over {x_{-n}}}
\end{eqnarray}
 This is a consequence of the fact that the transformations one
 considers, are {\em birational (hence reversible) transformations}.

{\em All these factorizations and recurrences have been proved
in}~\cite{BoMaRo93a}, {\em even for arbitrary $q$}.

Moreover, it has been shown that these recurrences yield algebraic {\em
elliptic curves}~\cite{BoMaRo93d}.
This can be shown relating them to {\em biquadratic} relations,
introducing the (homogeneous) variables $q_n$:
\begin{eqnarray}
q_n \,= \,{{f_n \; f_{n+3}} \over {f_{n+1} \; f_{n+2}}}
\end{eqnarray}
Equation (\ref{recuxnqq}) becomes the {\em biquadratic relation}:
\begin{eqnarray}
\label{step9}
(\rho - q_n - q_{n+1}) ( {q_n \; q_{n+1} + \lambda} )\, = \, \mu
\end{eqnarray}
where $\lambda$, $\rho$ and $\mu$ are constants of integrations
\footnote{Many examples of integrable mappings related to biquadratic
elliptic curves
 have recently been obtained by several
 authors~\cite{QuNi92,QuRoTh88,QuRoTh89,Ra91}.}.

Let us also recall the finite order conditions for recurrence
(\ref{recuxnqq}).
Recalling the biquadratic relation (\ref{step9}) and in particular the
three parameters $\lambda, \mu$ and $\rho$, recurrence (\ref{recuxnqq})
 can be seen to yield finite order orbits, which can be written
 as algebraic conditions bearing alternatively on the $x_n$'s, or the
$q_n$'s,
or even on the three parameters $\lambda, \mu$ and $\rho$.
 These algebraic conditions are given in~\cite{BoMaRo93a} for order
 four,
five, six and seven.
Let us, for instance,  just recall here the conditions of order six
respectively in the variables $q_n, x_n$ or $\lambda, \mu$ and $\rho$
:
\begin{eqnarray}
- x_{n}\, x_{n+1}\, x_{n+2}- x_{n}\, x_{n+1}\, x_{n+2}^{2}+ x_{n
}^{2} x_{n+2}^{2} x_{n+1}-1+ x_{n+2}+ x_{n}\, x_{n+2}\,=\,0
\end{eqnarray}
or
\begin{eqnarray}
\lambda^{3}+\rho^{2}\lambda^{2}-3\,\mu\lambda\rho+2\,\mu^{2}\,=\,0
\end{eqnarray}
or
\begin{eqnarray}
- q_{n}\,q_{n+2}\, q_{n+3}- q_{n}\,q_{n+3}^{2}+ q_{n+1}\,q_{n+3}^{2} -
q_{n}^{2}\, q_{n+2}+ q_{n}^{2}\, q_{n+3} + q_{n}\,q_{n+1}\, q_{n+3}=0
\end{eqnarray}

The relations between these various properties and structures
(factorization properties,
 existence of recurrences, integrability, ...)  have been sketched
 in~\cite{BoMaRo93a}.
The fact that products of a {\em fixed number} of $f_n$'s occur in
relation
(\ref{recufnqq})
 is related to the fact that products of a fixed number of $f_n$'s also
 occur
in the factorizations (\ref{mkVI}), (\ref{detVI}).
 The polynomial growth of these iterations
is, at first sight, in good agreement with a  framework of products of
fixed
number of polynomials
\footnote{However it will be shown in forthcoming publications that
 polynomial growth may occur even with more involved
 factorizations~\cite{BoMaRo93c}}.
To some extend the integrability of the mappings, or more precisely the
occurrence of (algebraic) elliptic curves,
for arbitrary $q$, yield such {\em polynomial growth} of the iterations
(see~\cite{FaVi93,bbb6}).

Transformation $K$ can be seen as a homogeneous transformation bearing
 on $q^2$ entries of the $q \times q$ matrix. For small values of $q$
($q=3$, $q=4$, $q=5$, ...), one can actually
look at the images of the iteration of $K$ and see that these orbits
yield
{\em curves}~\cite{BoMaRo93d}.
For $q=4$ it is possible to show that these  curves are elliptic curves
given as the {\em intersections of fourteen quadrics} in ${\Bbb C}
P_{15}$~\cite{BoMaRo93d}.
These quadrics can be obtained as ``Pl\"ucker-like'' well-suited sums
and
differences
 of $2 \times 2$ minors of the $4 \times 4$ matrix~\cite{BoMaRo93d},
in a very similar way as it occurs in  the sixteen vertex model
\cite{prl2}. These elliptic curves have been seen to be closely related
to biquadratic
relations \cite{BoMaRo93d} which is not surprising recalling
\cite{QuNi92,QuRoTh88,QuRoTh89,Ra91}.
This situation {\em can probably be generalized to} $q \times q$
matrices, the
elliptic curve
in  ${\Bbb C} P_{q^2-1}$ being now the intersection of $q^2-2$
algebraic
expressions of higher degree~\cite{BoMaRo93d}.\footnote{Our
 proof of this statement for arbitrary $q$ is not complete at the
 present
moment.}
The relation between these elliptic curves and the elliptic curves
associated with the
 recurrence on the $f_n$'s or $x_n$'s (see (\ref{recufnqq}) and
 (\ref{recuxnqq}))
 has been analyzed in detail in~\cite{BoMaRo93a}.

Let us finally mention that, for a given initial matrix $M_0$, the
successive iterates of $M_0$ under transformation $K^2$ move in a {\em
five-dimensional affine projective space}:
\begin{eqnarray}
\label{affine}
K^{2\,n}(M_0)\,=\,a^n_0 \cdot M_0\, +\, a^n_1 \cdot P \, +\, \,a^n_2
\cdot M_2\,+\,a^n_3 \cdot M_4\,+\,a^n_4 \cdot M_6\,+\,a^n_5 \cdot M_8
\end{eqnarray}
\begin{eqnarray}
\label{affineodd}
K^{2\,n+1}(M_0)\,=\,b^n_0 \cdot M_1\, +\, b^n_1 \cdot P \, +\, \,b^n_2
\cdot M_3\,+\,b^n_3 \cdot M_5+\,b^n_4 \cdot M_7+\,b^n_5 \cdot M_9
\end{eqnarray}
where matrix $P$ is a fixed matrix, {\em independent of the initial
matrix}
$M_0$, of entries $P_{i,j}= \delta_{i,1} \cdot
\delta_{j,2}-\delta_{i,2}
\cdot \delta_{j,1}$.
Considering the points in  ${\Bbb C} P_{q^2-1}$ associated to the
successive
$q \times q$ matrices corresponding to the iteration of $M_0$ under
transformation $K$ (instead of $K^2$), one thus gets sets of points
which belong to {\em two}
five dimensional affine subspace of  ${\Bbb C} P_{q^2-1}$, {\em which
depend on the initial matrix} $M_0$.
Figure (1) gives one orbit corresponding to the iteration of
transformation
$\widehat{K}$ for a $5 \times 5$ matrices, that is in a 24-dimensional
space. One has
apparently a foliation of this 24-dimensional space in terms of
elliptic curves.

\section{The results for the five other classes}
\label{result}
The analysis performed in section (\ref{rappel}) of the iteration of
transformation $K$ for the transposition $t_{12-21}$ representing class
I
can similarly be performed for the five other classes.

\subsection{Classes II}
\label{classIIfac}

For $q=4$, the analysis of the iteration of the homogeneous
transformation
$K$ for the transpositions of class II {\em yield exactly the same
factorizations} (and therefore the same
generating functions $\alpha(x)$, $\beta(x)$, $\mu(x)$ and $\nu(x)$)
{\em as for class} I.
However, {\em the homogeneous polynomials} $f_n$ (see equations
(\ref{mkVI})
and (\ref{detVI})) {\em do not satisfy any simple recurrence} like
(\ref{recufnqq}).
In fact, it will be shown in section (\ref{demrec}) that there
actually exist recurrences on a finite set of variables which enable,
after
elimination, to get algebraic relations between two variables (namely
$x_n$
and another one). One does not have simple recurrences like
(\ref{recuxnqq})
but still a quite (involved) algebraic relation on these variables.
The orbits of $K$ yield, for $q=4$, {\em algebraic elliptic curves}
which can be seen as intersections of fourteen ``Pl\"ucker-like
quadrics''
in ${\Bbb C} P_{15}$~\cite{BoMaRo93d}.

For class II, the factorizations corresponding to the iterations of
transformation
$K$ detailed in section (\ref{rappel}) (see equations (\ref{f2q-3}),
(\ref{m2q-4}), (\ref{f3m2}), (\ref{m3q-4}), (\ref{f4m3}), ...) for
class I ,
are drastically different, when $q\ge5$, already after two iterations:
\begin{eqnarray}
\begin{array}{cccc}
\hskip -1cm
\left \{ \begin{array}{c}
f_1=\det(M_0)\\ \\ \\M_1=K(M_0)
\end{array}
\right . &
\left \{ \begin{array}{c}
f_2=\det(M_1)/f_1^{q-3}\\ \\ \\M_2= K(M_1)/f_1^{q-4}
\end{array}
\right . &
\left \{ \begin{array}{c}
f_3=\det(M_2)/(f_1^2 \cdot f_2^{q-3})\\ \\ \\M_3=K(M_2)/(f_1 \cdot
f_2^{q-4})
\end{array}
\right . &
\left \{ \begin{array}{c}
f_4=\det(M_3)/(f_1^{q-2} \cdot f_2^2 \cdot f_3^{q-3})\\ \\
\\M_4=K(M_3)/(f_1^{q-3} \cdot f_2 \cdot f_3^{q-4})
\end{array}
\right .
\end{array} \nonumber
\end{eqnarray}
\begin{eqnarray}
\label{facII}
&&f_5={{\det(M_4)}\over{f_1^3 \cdot f_2^{q-2} \cdot f_3^2 \cdot
f_4^{q-3}}},\;\;
M_5={{K(M_4)}\over{f_1^2 \cdot f_2^{q-3} \cdot f_3 \cdot
f_4^{q-4}}},\;\;
f_6={{\det(M_5)}\over{f_1^{q-3} \cdot f_2^3 \cdot f_3^{q-2} \cdot f_4^2
\cdot f_5^{q-3}}},\\
&&M_6={{K(M_5)}\over{f_1^{q-4} \cdot f_2^2 \cdot f_3^{q-3} \cdot f_4
\cdot f_5^{q-4}}},\;\;
f_7={{\det(M_6)}\over{f_1^2 \cdot f_2^{q-3} \cdot f_3^3 \cdot f_4^{q-2}
\cdot f_5^2 \cdot f_6^{q-3}}},\;\;
M_7={{K(M_6)}\over{f_1 \cdot f_2^{q-4} \cdot f_3^2 \cdot f_4^{q-3}
\cdot f_5 \cdot f_6^{q-4}}}\nonumber
\end{eqnarray}
yielding the following factorizations for arbitrary  $n$:
\begin{eqnarray}
\label{detII}
det(M_n)\,=\, f_{n+1} \cdot (f_{n}^{q-3} \cdot f_{n-1}^2 \cdot
f_{n-2}^{q-2} \cdot f_{n-3}^3 )\cdot (f_{n-4}^{q-3} \cdot f_{n-5}^2
\cdot f_{n-6}^{q-2}
\cdot f_{n-7}^3) \cdots f_{1}^{\delta_n}
\end{eqnarray}
where $\delta_n$ depends on the truncation, and:
\begin{eqnarray}
\label{KII}
K(M_n)\, = \,
M_{n+1} \cdot ( f_{n}^{q-4} \cdot f_{n-1} \cdot f_{n-2}^{q-3} \cdot
f_{n-3}^2
) \cdot (f_{n-4}^{q-4} \cdot f_{n-5} \cdot f_{n-6}^{q-3} \cdot
f_{n-7}^2 ) \cdots f_{1}^{\zeta_n}
\end{eqnarray}
where $\zeta_n=q-4\,$  for $n=1$ (mod 4), $\zeta_n=1\, $  for $n=2$
(mod
4),  $\zeta_n=q-3 \,$  for $n=3$ (mod 4) and  $\zeta_n=2 $  for $n=0$
(mod 4).

For factorization (\ref{KII}), one has periodically (with period four)
the sequence
[(q-4)(1)(q-3)(2)] for the exponents of the $f_n$'s of the
``string-like'' factor
 in the right-hand side of (\ref{KII}), while for factorization
 (\ref{detII}),
one has periodically (again with period four) the sequence
[(q-3)(2)(q-2)(3)] for
the exponents of the $f_n$'s of
 the ``string-like'' factor in the right-hand side of relation
 (\ref{detII}).

One notes that the
following factorization {\em independent of $q$} occurs, which is
actually
different from relation (\ref{MGMqq}):
\begin{eqnarray}
\label{facKdetII}
{{K(M_{n})} \over {det(M_{n})}} \,\,=\,\, {{M_{n+1}} \over {f_{1} f_{2}
\; \cdots \; f_{n} f_{n+1}}}
\end{eqnarray}

These factorizations (\ref{detII}) and (\ref{facKdetII}) yield linear
recurrences on the $\alpha_n$'s and $\beta_n$'s:
\begin{eqnarray}
\label{albeII}
\alpha_n &=& \beta_{n+1} + (q-3) \; \beta_{n} +  2\; \beta_{n-1} +
(q-2) \;  \beta_{n-2} + 3 \;  \beta_{n-3}
+ (q-3) \; \beta_{n-4} \nonumber \\
&&+ 2\;  \beta_{n-5} + (q-2) \;  \beta_{n-6} + 3 \;  \beta_{n-7} +
\cdots + \delta_n \; \beta_1
\end{eqnarray}
and:
\begin{eqnarray}
\label{linII}
q\; ( \beta_{1}+\beta_{2}+\cdots+\beta_{n+1}) \,=\, \alpha_{n} +
\alpha_{n+1}
\end{eqnarray}

{}From relation (\ref{albeII}), one gets on the generating functions
$\alpha(x)$ and $\beta(x)$:
\begin{eqnarray}
x\;\alpha(x) \,=\, \beta(x)\cdot \Bigl(  1+ {{(q-3)\;x + 2\;x^2
+(q-2)\;x^3+3\;x^4}\over{1-x^4}} \Bigr)
\end{eqnarray}
and from relation (\ref{linII}):
\begin{eqnarray}
\label{genII}
{{q \; \beta(x)}\over{(1-x)}}\,\, = \,\, (1+x)\cdot \alpha(x)\, - \,q
\end{eqnarray}
The generating functions  $\alpha(x)$ and $\beta(x)$ read:
\begin{eqnarray}
\label{alxII}
\alpha(x) \,=\,{\frac {q\left
(1+2\,x^{4}+x\,q-3\,x+2\,x^{2}+x^{3}\,q-2\,x^{3}\right
)}{(1-x)(1+x)(1-2\,x-2\,x^3)}}, \quad
\beta(x)\,=\,{\frac {q\,x\,\left (1+x^{2}\right )}{1-2\,x-2\,x^3}}
\end{eqnarray}

Again (see equations (\ref{kfnVI}) ...) the {\em right} action of $K$
on the
$f_n$'s and on matrices $M_{n}$'s {\em factorizes} $f_1$ {\em and only}
$f_1$:
\begin{eqnarray}
\label{fnkII}
(f_n)_K \,= \, f_{n+1} \cdot f_1^{\mu_n}
\quad \quad \hbox{and} \quad \quad
(M_n)_K \, =\,  M_{n+1} \cdot f_1^{\nu_n}
\end{eqnarray}

One deduces again, from factorizations (\ref{fnkII}), the linear
recurrences (\ref{abnmunuI}) and relations (\ref{ggI}) on the
generating functions.
The generating functions $\mu(x)$ and $\nu(x)$ read respectively:
\begin{eqnarray}
\label{muxII}
\mu(x)\,=\, {\frac {x\,\left ((q-3) -x +(q-3) \,x^2 \right
)}{1-2\,x-2\,x^{3}}}, \,\,\,\,
\nu(x)\,=\,{\frac {x\,\left (q-4+x+(q-3)\,x^2+2\,x^3\right )}{\left
(1-x\right
)\left (1+x\right )\left (1-2\,x -2 \, x^{3}\right )}}
\end{eqnarray}

{}From equations (\ref{alxII}) and  (\ref{muxII}), it is clear that
 one has an {\em exponential growth} of exponents $\alpha_n$,
 $\beta_n$, $\mu_n$ and $\nu_n$.
They grow like $\lambda^n$ where  $\lambda = 2.359304086 $ ... is the
largest root of $2+2\,z^2-z^3$.

Let us underline that, for a given initial matrix $M_0$, the
successive iterates of $M_0$ under transformation $K^2$ move, in a {\em
three-dimensional affine matrix space}:
\begin{eqnarray}
\label{affineII}
K^{2\,n}(M_0)\,=\,a^n_0 \cdot M_0\, +\, a^n_1 \cdot P \, +\, \,a^n_2
\cdot M_2\,+\,a^n_3 \cdot M_4
\end{eqnarray}
\begin{eqnarray}
\label{affineIIodd}
K^{2\,n+1}(M_0)\,=\,b^n_0 \cdot M_1\, +\, b^n_1 \cdot P \, +\, \,b^n_2
\cdot M_3\,+\,b^n_3 \cdot M_5
\end{eqnarray}

where matrix $P$ is a fixed matrix representing the transposition of
class
II one considers (here $m_{1,2}-m_{3,4}$), {\em independent of the
initial matrix}
$M_0$, of entries $P_{i,j}= \delta_{i,1} \cdot
\delta_{j,2}-\delta_{i,3}
\cdot \delta_{j,4}$.

Figure (2a) and (2b) show (the projection of) two orbits corresponding
to
the iteration of a transformation of class II for $5
\times 5$ matrices.
For class II, though one often gets curves (similar to figure
(1)), one sees, with figure (2b) for instance, that some orbits may lie
on
higher dimensional varieties
\footnote{This gives a strong indication that, when one gets curves,
these
curves are not algebraic.}.

Moreover, a careful look at figure (2b), shows the occurrence of a
``small
island'' in a quite uniform density of points.
This situation is reminiscent of the one encountered with the
Henon-Heiles
mappings or the ``almost'' integrable mappings~\cite{BoMaRo93d,He64}.

The drastically different behavior, one encounters,
for $q=4$ and for $q>4$, shows that the $q$-generalization of
transformation $K$ (we introduce in section (\ref{rappel})) is
certainly non-trivial.

\subsection{Class III}
\label{classIIIfac}

Remarkably, the analysis of the iteration of the homogeneous
transformation
$K$ for the transpositions of class III
{\em yield exactly the same factorizations for arbitrary} $q$ (and
therefore the same
generating functions $\alpha(x)$, $\beta(x)$, $\mu(x)$ and $\nu(x)$)
{\em as for class} I (see section (\ref{rappel})).
However, {\em the homogeneous polynomials} $f_n$ (see equations
(\ref{mkVI})
and (\ref{detVI})) {\em do not satisfy any simple recurrence} like
(\ref{recufnqq}).
In fact, it will be shown in section (\ref{demrec}) that there
actually exist recurrences on a finite set of variables which enable,
after
elimination, to get algebraic relations between two variables (namely
$x_n$
and another one), and finally on a {\em single variable}.
The elimination happens to be less involved than for class II.
Nevertheless one does not have simple recurrences like (\ref{recuxnqq})
but still a quite (involved) algebraic relation on these variables.

Again, the orbits of $K$ yield, for $q=4$, {\em algebraic elliptic
curves}
which can be seen as intersections of fourteen ``Pl\"ucker-like
quadrics''
in ${\Bbb C} P_{15}$~\cite{BoMaRo93d}.

As for class I, the
successive iterates of a given initial matrix $M_0$,  under
transformation $K^2$, move in a {\em
five-dimensional affine projective space}:
\begin{eqnarray}
\label{affineIII}
K^{2\,n}(M_0)\,=\,a^n_0 \cdot M_0\, +\, a^n_1 \cdot P \, +\, \,a^n_2
\cdot M_2\,+\,a^n_3 \cdot M_4+\,a^n_4 \cdot M_6+\,a^n_5 \cdot M_8
\end{eqnarray}
\begin{eqnarray}
\label{affineIIIodd}
K^{2\,n+1}(M_0)\,=\,b^n_0 \cdot M_1\, +\, b^n_1 \cdot P \, +\, \,b^n_2
\cdot M_3\,+\,b^n_3 \cdot M_5+\,b^n_4 \cdot M_7+\,b^n_5 \cdot M_9
\end{eqnarray}

where matrix $P$ is a fixed matrix representing the transposition of
class
III one considers.

Figure (3a) and (3b) show (the projection of) two orbits corresponding
to
the iteration of a transformation of class III for $5
\times 5$ matrices.

Very often the iteration of a transformation of class III for $5
\times 5$ matrices yields curves similar to figure (1).
Figure (3b) {\em however looks like a set of curves lying on a
surface}.
This seems to rule out, for class III, a
foliation of the 24-dimensional parameter space in curves, but it does
not
rule out the fact that these orbits could be algebraic surfaces or
``nice'' higher dimensional algebraic varieties, like abelian varieties
(which is suggested by the polynomial growth).

The occurrence of integrable recurrences, independent of $q$, on the
$f_n$'s associated with
algebraic elliptic curves probably explains the better regularity of
mappings of
class I, compared to mappings of class III.

\subsection{Class IV}
\label{class4}
The factorizations corresponding to the iterations of transformation
$K$ detailed in section (\ref{rappel}) (see equations (\ref{f2q-3}),
(\ref{m2q-4}), (\ref{f3m2}), (\ref{m3q-4}), (\ref{f4m3}), ...) for
class I (and also, for classes II and III),
now read for class IV:
\begin{eqnarray}
\begin{array}{cccc}
\hskip -1cm
\left \{ \begin{array}{c}
f_1=\det(M_0)\\ \\ \\M_1=K(M_0)
\end{array}
\right . &
\left \{ \begin{array}{c}
f_2=\det(M_1)/f_1^{q-2}\\ \\ \\M_2= K(M_1)/f_1^{q-3}
\end{array}
\right . &
\left \{ \begin{array}{c}
f_3=\det(M_2)/(f_1 \cdot f_2^{q-2})\\ \\ \\M_3=K(M_2)/f_2^{q-3}
\end{array}
\right . &
\left \{ \begin{array}{c}
f_4=\det(M_3)/(f_1^{q-1} \cdot f_2 \cdot
f_3^{q-2})\\ \\ \\M_4=K(M_3)/(f_1^{q-2} \cdot f_3^{q-3})
\end{array}
\right .
\end{array} \nonumber
\end{eqnarray}
\begin{eqnarray}
\label{facIV}
&&f_5={{\det(M_4)}\over{f_1^2 \cdot f_2^{q-1} \cdot f_3 \cdot
f_4^{q-2}}},\;\;
M_5={{K(M_4)}\over{f_1 \cdot f_2^{q-2} \cdot f_4^{q-3}}},\;\;
f_6={{\det(M_5)}\over{f_1^{q-2} \cdot f_2^2 \cdot f_3^{q-1} \cdot f_4
\cdot f_5^{q-2}}},\\
&&M_6={{K(M_5)}\over{f_1^{q-3} \cdot f_2 \cdot f_3^{q-2} \cdot
f_5^{q-3}}},\;\;
f_7={{\det(M_6)}\over{f_1 \cdot f_2^{q-2} \cdot f_3^2 \cdot f_4^{q-1}
\cdot f_5 \cdot f_6^{q-2}}},\;\;
M_7={{K(M_6)}\over{f_2^{q-3} \cdot f_3 \cdot f_4^{q-2} \cdot
f_6^{q-3}}}\nonumber
\end{eqnarray}
yielding the following factorizations for arbitrary  $n$:
\begin{eqnarray}
\label{detIV}
det(M_n)\,=\, f_{n+1} \cdot (f_{n}^{q-2} \cdot f_{n-1} \cdot
f_{n-2}^{q-1} \cdot f_{n-3}^2 )\cdot (f_{n-4}^{q-2} \cdot f_{n-5} \cdot
f_{n-6}^{q-1}
\cdot f_{n-7}^2) \cdots f_{1}^{\delta_n}
\end{eqnarray}
where $\delta_n$ depends on the truncation, and:
\begin{eqnarray}
\label{KIV}
K(M_n)\, = \, M_{n+1} \cdot ( f_{n}^{q-3} \cdot f_{n-2}^{q-2} \cdot
f_{n-3} ) \cdot (f_{n-4}^{q-3} \cdot f_{n-6}^{q-2} \cdot f_{n-7} )\cdot
(f_{n-8}^{q-3} \cdot f_{n-10}^{q-2} \cdot f_{n-11} ) \cdots
f_{1}^{\zeta_n}
\end{eqnarray}
where $\zeta_n=q-3\,$  for $n=1$ (mod 4), $\zeta_n=0\, $  for $n=2$
(mod
4),  $\zeta_n=q-2 \,$  for $n=3$ (mod 4) and  $\zeta_n=1 $  for $n=0$
(mod 4).

For factorization (\ref{KIV}), one has periodically (with period four)
the sequence
[(q-3)(0)(q-2)(1)] for the exponents of the $f_n$'s of the
``string-like'' factor
 in the right-hand side of (\ref{KIV}), while for factorization
 (\ref{detIV}),
one has periodically (again with period four) the sequence
[(q-2)(1)(q-1)(2)] for
the exponents of the $f_n$'s of
 the ``string-like'' factor in the right-hand side of (\ref{detIV}).

One notes that the
following factorization {\em independent of $q$} occurs, which is
different from relation (\ref{MGMqq}), but actually identifies with
relation (\ref{facKdetII}):
\begin{eqnarray}
\label{facKdetIV}
{{K(M_{n})} \over {det(M_{n})}} \,\,=\,\, {{M_{n+1}} \over {f_{1} f_{2}
\; \cdots \; f_{n} f_{n+1}}}
\end{eqnarray}

These factorizations (\ref{detIV}) and (\ref{facKdetIV}) yield linear
recurrences on the $\alpha_n$'s and $\beta_n$'s:
\begin{eqnarray}
\label{albeIV}
\alpha_n &=& \beta_{n+1} + (q-2) \; \beta_{n} +  \beta_{n-1} + (q-1)
\;  \beta_{n-2} + 2 \;  \beta_{n-3}
+ (q-2) \; \beta_{n-4} \nonumber \\
&&+  \beta_{n-5} + (q-1) \;  \beta_{n-6} + 2 \;  \beta_{n-7} + \cdots +
\delta_n \; \beta_1
\end{eqnarray}
and:
\begin{eqnarray}
\label{linIV}
q\; ( \beta_{1}+\beta_{2}+\cdots+\beta_{n+1}) \,=\, \alpha_{n} +
\alpha_{n+1}
\end{eqnarray}

{}From relation (\ref{albeIV}), one gets on the generating functions
$\alpha(x)$ and $\beta(x)$:
\begin{eqnarray}
x\;\alpha(x) \,=\, \beta(x)\cdot \Bigl(  1+ {{(q-2)\;x +x^2
+(q-1)\;x^3+2\;x^4}\over{1-x^4}} \Bigr)
\end{eqnarray}
and from relation (\ref{linIV}) one recovers relation (\ref{genII}):
\begin{eqnarray}
\label{genIV}
{{q \; \beta(x)}\over{(1-x)}}\,\, = \,\, (1+x)\cdot \alpha(x) -q
\end{eqnarray}
The generating functions  $\alpha(x)$ and $\beta(x)$ read:
\begin{eqnarray}
\label{alxIV}
\alpha(x) \,=\,{\frac {q\left (1+x^{4}+xq-2\,x+x^{2}+x^{3}q-x^{3}\right
)}{(1-x)(1+x)(1-x-x^3)}}, \quad
\beta(x)\,=\,{\frac {q\,x\,\left (1+x^{2}\right )}{1-x-x^3}}
\end{eqnarray}

Again (see equations (\ref{kfnVI}), (\ref{fnkII}) ...) the {\em right}
action of $K$ on the
$f_n$'s and on matrices $M_{n}$'s {\em factorizes} $f_1$ {\em and only}
$f_1$:
\begin{eqnarray}
\label{fnkIV}
(f_n)_K \,= \, f_{n+1} \cdot f_1^{\mu_n}
\quad \quad  \hbox{and} \quad \quad
(M_n)_K \, =\,  M_{n+1} \cdot f_1^{\nu_n}
\end{eqnarray}
One deduces again, from factorizations (\ref{fnkIV}), the linear
recurrences (\ref{abnmunuI}) and the relations (\ref{ggI}) on the
generating functions.
The generating functions $\mu(x)$ and $\nu(x)$ read respectively:
\begin{eqnarray}
\label{muxIV}
\mu(x)\,=\, {\frac {x\,\Bigl ((q-2)\,(1+x^2)-x\Bigr )}{1-x-x^{3}}},
\,\,\,\,
\nu(x)\,=\,{\frac {x\,\Bigl (q-3+(q-2)\,x^2+x^3\Bigr )}{\left
(1-x\right )\left (1+x\right )\left (1-x-x^{3}\right )}}
\end{eqnarray}

{}From equations (\ref{alxIV}) and (\ref{muxIV}), it is clear that
 one has an {\em exponential growth} of exponents $\alpha_n$,
 $\beta_n$, $\mu_n$ and $\nu_n$.
They grow like $\lambda^n$ where  $\lambda = 1.465571226$ ... is the
largest root of $z^3-z^2-1$.
One remarks that some homogeneous polynomials, similar to the
numerators,
or denominators
 appearing in  recurrences like (\ref{recufnqq}), do {\em satisfy some
 additional factorization properties}:
\begin{eqnarray}
\label{liste}
&& (f_4-f_2 f_3), \qquad
(f_5-f_2 f_4), \qquad
(f_6-f_3 f_5), \qquad
(f_7-f_4 f_6), \qquad
(f_8-f_5 f_7) \nonumber \\
&& (f_1 f_5-f_4 f_3), \qquad
(f_2 f_6-f_5 f_4), \qquad
(f_3 f_7-f_6 f_5 f_1), \qquad
(f_4 f_8-f_7 f_6 f_2),\, \cdots  \nonumber \\
&& (f_2 f_6 f_{10} f_{14} - f_4 f_8 f_{12} f_{13} ), \cdots
\end{eqnarray}

The polynomials $f_n$ for class IV not only satisfy this additional
factorization but {\em actually satisfy, for arbitrary $q$, exact
relations}  where the new polynomials (\ref{liste}) play  a  key role:
\begin{eqnarray}
&&\qquad
(-f_6+f_3\;f_5)\;(-f_3+f_1\;f_2)+(f_1\;f_5-f_4\;f_3)\;(-f_4+f_1\;f_3)\,=\,0
\nonumber \\
&&f_3\;f_7\;(-f_{15}+f_{12}\;f_{14})\;(-f_4\;f_8\;f_{12}+f_2\;f_6\;f_{10}\;f_{11})
\qquad \qquad \nonumber \\
&&\qquad \qquad
+f_1\;f_5\;f_9\;(f_2\;f_6\;f_{10}\;f_{14}-f_4\;f_8\;f_{13}\;f_{12})\;(-f_{13}+f_{10}\;f_{12})\,=\,0
\nonumber \\
&&\vdots \nonumber
\end{eqnarray}

In fact, the $f_n$'s do not satisfy simple recurrences like
(\ref{recufnqq}),
 {\em but ``pseudo-recurrences'', where products from $f_1$ to $f_n$
 occur}.
One of these ``pseudo-recurrences'' can be written as follows :
\begin{eqnarray}
\label{pseudo}
&&{{(f_{n+2} - f_{n-1}f_{n+1})}\over{(f_{n} - f_{n-3}f_{n-1})}}
\;\cdot\; {{f_{n-6} \; f_{n-10} \; f_{n-14} \; \cdots} \over {f_{n-4}
\; f_{n-8} \; f_{n-12} \; \cdots}} \qquad \qquad \nonumber \\
&&\qquad \qquad =\,{{f_{n} \; (f_{n-1} \; f_{n-5} \; f_{n-9} \; \cdots)
- (f_{n+1} \; f_{n-3} \; f_{n-7} \; \cdots)} \over
{f_{n-2} \; (f_{n-3} \; f_{n-7} \; f_{n-11} \; \cdots) - (f_{n-1} \;
f_{n-5} \; f_{n-9} \; \cdots)}}
\end{eqnarray}

The polynomials occurring in the numerator and the denominator of
 the ``pseudo-recurrence'' (\ref{pseudo}) suggests the following
 recurrence on the $\beta_n$'s:
\begin{eqnarray}
\beta_{n+3} - \beta_{n} - \beta_{n+2}\, =\, 0
\end{eqnarray}
This recurrence would have suggested, since the beginning, a
$\,\,1-x-x^3=0 \,$
singularity (see relations (\ref{alxIV}) and (\ref{muxIV})).

Though, one does not have recurrences on the $f_n$'s but
pseudo-recurrences
such as (\ref{pseudo}), the previous variables $x_n$ (see
(\ref{xndet})),
which can always be defined,
 {\em remarkably satisfy very simple recurrences}
(see the demonstration in section (\ref{demrec})).
As for class I, the recurrences on the $x_n$'s are {\em independent of}
$q$:
this independence will be understood in section (\ref{demrec}).
One of these recurrences  reads:
\begin{eqnarray}
\label{xnIV}
{{x_{n+3} -1}\over{x_{n+2}\; x_{n+4}-1}}\,=\,{{x_{n+1} -1}\over{x_n \;
x_{n+2}-1}} \cdot x_n \; x_{n+3}
\end{eqnarray}

Studying the iteration of $\widehat{K}$ in the $q^2 -1$-dimensional
space ${\Bbb C} P_{q^2-1}$,
one can show that these orbits actually belong to remarkable two
dimensional subvarieties (given by intersection of quadrics in
  ${\Bbb C} P_{15}$~\cite{BoMaRo93d}), namely {\em planes}
(see section (\ref{demrecIV}) in the
following and  see also~\cite{BoMaRo93d}). Inside these
planes, {\em which depend on the initial point} in the $q^2
-1$-dimensional space
(that is the initial matrix), the orbits look like
curves for many of the trajectories  (see~\cite{BoMaRo93d}).

\subsection{Class V}

The factorizations corresponding to the iterations of transformation
$K$ read for class V:
\begin{eqnarray}
\begin{array}{cccc}
\hskip -1cm
\left \{ \begin{array}{c}
f_1=\det(M_0)\\ \\ \\M_1=K(M_0)
\end{array}
\right . &
\left \{ \begin{array}{c}
f_2=\det(M_1)/f_1^{q-3}\\ \\ \\M_2= K(M_1)/f_1^{q-4}
\end{array}
\right . &
\left \{ \begin{array}{c}
f_3=\det(M_2)/(f_1 \cdot f_2^{q-3})\\ \\
\\M_3=K(M_2)/(f_2^{q-4})
\end{array}
\right . &
\left \{ \begin{array}{c}
f_4=\det(M_3)/(f_1^{q-1} \cdot f_2 \cdot f_3^{q-3})\\ \\
\\M_4=K(M_3)/(f_1^{q-2} \; f_3^{q-4})
\end{array}
\right .
\end{array} \nonumber
\end{eqnarray}
The factorizations are now stabilized, yielding for arbitrary  $n$:
\begin{eqnarray}
\label{detIII}
det(M_{n+2}) = f_n^{q-1} \cdot f_{n+1} \cdot f_{n+2}^{q-3} \cdot
f_{n+3}
\end{eqnarray}
\begin{eqnarray}
\label{KIII}
K(M_{n+2})\, = \, f_{n}^{q-2} \cdot f_{n+2}^{q-4} \cdot M_{n+3}
\end{eqnarray}
One notes that again, {\em as well as for classes} I and III (see
equation (\ref{MGMqq})),
the following factorizations, {\em independent of $q$}, occur:
\begin{eqnarray}
\label{kdetVV}
{{K(M_{n+2})} \over {det(M_{n+2})}}\, =\, {{M_{n+3}} \over {f_{n}
f_{n+1} f_{n+2} f_{n+3}}}
\end{eqnarray}

Factorizations (\ref{detIII}) and (\ref{KIII}) yield linear recurrences
on the $\alpha_n$'s and $\beta_n$'s:
\begin{eqnarray}
\label{alIII}
\alpha_{n+2} \,=\, (q-1)\; \beta_{n} + \beta_{n+1} + (q-3)\;
\beta_{n+2} +\beta_{n+3}
\end{eqnarray}
\begin{eqnarray}
\label{beIII}
(q-1)\; \alpha_{n+2}\, =\, \alpha_{n+3} +
q\,(q-2)\;\beta_{n}+q\,(q-4)\;\beta_{n+2}
\end{eqnarray}
The two generating functions $\alpha(x)$ and $\beta(x)$ read:
\begin{eqnarray}
\label{alxIII}
\alpha(x)\,=\,{{q\;(1+(q-3)\,x+x^2+(q-1)\,x^3)}\over{(1+x)(1-3x+x^2-x^3)}},
\quad
\beta(x)\,=\,{{q\;x}\over{(1+x)(1-3x+x^2-x^3)}}
\end{eqnarray}

Remarkably, similarly to what happened for classes I, II, III and IV
, the right-action of $K$ on the
$f_n$'s, or the $M_n$'s,
 factorizes $f_1$ and only $f_1$:
the factorizations (\ref{kfnVI}), (\ref{mfnVI}), the linear relations
(\ref{abnmunuI})
on the exponents $\alpha_n$, $\beta_n$, $\mu_n$ and $\nu_n$,
 as well as the linear relations (\ref{ggI}) on the generating
 functions, {\em are still valid
for} class V.
The two generating functions $\mu(x)$ and $\nu(x)$ read :
\begin{eqnarray}
\label{muxIII}
\mu(x)\,=\,{{x\;((q-3)-2\,x-x^3)}\over{(1+x)(1-3\,x+x^2-x^3)}}, \quad
\nu(x)\,=\,{{x\;(q-4+(q-2)x^2)}\over{(1+x)(1-3\,x+x^2-x^3)}}
\end{eqnarray}
One notes that the {\em roots of the denominator} of
 $\alpha(x)$, $\beta(x)$, $\mu(x)$ and $\nu(x)$ are {\em not on the
 unit circle}.
Thus one has an {\em exponential growth of the complexity} of the
calculation since the degree of all the polynomials one deals with
 (that is the exponents $\alpha_n$, $\beta_n$, $\mu_n$ and $\nu_n$)
 {\em grow
exponentially} with $n$, like $\lambda^n$ with  $\lambda =2.769292354 $
...
\footnote{This value of $\lambda$ is the largest root of $P(z)= -1+\,z
-3\,
z^2+z^3$ (let us note the change $x \rightarrow 1/z$).}.
For instance, expanding $\beta(x)$, one gets:
\begin{eqnarray}
\beta(x)&=&
qx+2\,qx^{2}+6\,qx^{3}+16\,qx^{4}+45\,qx^{5}+124\,qx^{6}+344\,qx^{7}+952\,qx^{8}+2637\,qx^{9}+\cdots
\nonumber
\end{eqnarray}
On this example, one sees that it is possible to {\em have
factorizations
involving products of a fixed  number of polynomials} $f_n$ and, in the
same time, an {\em exponential growth of the calculations} of the
iterations.

Again, one can study the iteration of $\widehat{K}$ seen as a
birational
transformation in ${\Bbb C} P_{q^2-1}$.
These orbits look like {\em curves} in some domain of ${\Bbb C}
P_{q^2-1}$~\cite{BoMaRo93d}.
For $q=4$, these orbits can be seen to lie on a subvariety which is the
intersection of {\em twelve} Pl\"ucker-like quadrics in ${\Bbb C}
P_{15}$
 and, more generally, at most,  $q^2-4$ algebraic expressions for $q
 \times q$ matrices~\cite{BoMaRo93d}.

\subsection{Class VI}

For class VI, the factorizations corresponding to the iterations of the
homogeneous
transformation $K$ read as follows:
\begin{eqnarray}
\begin{array}{cccc}
\hskip -1cm
\left \{ \begin{array}{c}
f_1=\det(M_0)\\ \\ \\M_1=K(M_0)
\end{array}
\right . &
\left \{ \begin{array}{c}
f_2=\det(M_1)/f_1^{q-2}\\ \\ \\M_2= K(M_1)/f_1^{q-3}
\end{array}
\right . &
\left \{ \begin{array}{c}
f_3=\det(M_2)/(f_1 \cdot f_2^{q-2})\\ \\
\\M_3=K(M_2)/(f_2^{q-3})
\end{array}
\right . &
\left \{ \begin{array}{c}
f_4=\det(M_3)/(f_1^{q-2} \cdot f_2 \cdot f_3^{q-2})\\ \\
\\M_4=K(M_3)/(f_1 \; f_3)^{q-3}
\end{array}
\right .
\end{array} \nonumber
\end{eqnarray}
\begin{eqnarray}
\hskip -.5cm
f_5={{\det(M_4)}\over{f_1 \cdot f_2^{q-2} \cdot f_3 \cdot
f_4^{q-2}}},\;\; M_5={{K(M_4)}\over{(f_2 \;
f_4)^{q-3}}},\;\;f_6={{\det(M_5)}\over{ f_1^{q-2} \cdot f_2 \cdot
f_3^{q-2} \cdot f_4
\cdot f_5^{q-2}}},\;\; M_6={{K(M_5)}\over{(f_1 \; f_3
\;f_5)^{q-3}}},\,\, \ldots \nonumber
\end{eqnarray}
yielding the following {\em ``string-like'' factorizations for
arbitrary $n$}:
\begin{eqnarray}
\label{facKI}
K(M_n) \,=\, M_{n+1} \cdot ( f_{n} \cdot f_{n-2} \cdot f_{n-4} \cdot
f_{n-6} \cdots f_{\xi_n} )^{q-3}
\end{eqnarray}
where $\xi_n=1$ for $n$ odd and $\xi_n=2$ for $n$ even.
\begin{eqnarray}
\label{facdetI}
det(M_n) \,= \, f_{n+1} \cdot f_{n}^{q-2} \cdot f_{n-1} \cdot
f_{n-2}^{q-2}
\cdot f_{n-3} \cdot f_{n-4}^{q-2} \cdots f_{1}^{\zeta_n}
\end{eqnarray}
where $\zeta_n=1\,$ for $n$ even and $\,\zeta_n=q-2\,$ for $n$ odd
\footnote{
Note that, one has no factorization for $q=3$ for $K(M_n)$.}.

Equations (\ref{facKI}) and (\ref{facdetI}) yield the following simple
``string-like'' relation {\em independent of $q$}, which amazingly
happens to be {\em the same relation as for class} IV (see equation
(\ref{facKdetIV})) and the same relation as for class II for $q \ge 5$
(see
factorization (\ref{facKdetII})):
\begin{eqnarray}
\label{KdetI}
{{K(M_n)} \over {det(M_n)}}\,=\, {{M_{n+1}} \over {f_1 \cdot f_2 \;
\cdots
\; f_n \cdot f_{n+1}}}
\end{eqnarray}
In fact, one notices the occurrence of ``string-like'' factorization
relations (like (\ref{facdetI})
or (\ref{KdetI})), instead of factorizations {\em with a fixed number
of polynomials} ( see
relations (\ref{kfnVI}), (\ref{mfnVI}), or (\ref{detVI})), for the
two classes IV and VI, for which the transposition {\em permutes
entries
belonging to the same column or to the same row}, but also for class II
(for $q \ge 5$), the transposition of which involving two rows and two
columns.

Equations (\ref{facdetI}) and (\ref{KdetI}) also yield the following
linear recurrences on the $\alpha_n$'s and $\beta_n$'s:
\begin{eqnarray}
\label{albeI}
\alpha_n = \beta_{n+1} + (q-2)\, \beta_{n} +  \beta_{n-1} + (q-2)\,
\beta_{n-2} +  \beta_{n-3} + (q-2)\, \beta_{n-4} +
\cdots  + \zeta_n \; \, \beta_1
\end{eqnarray}
and:
\begin{eqnarray}
\label{lina1b1}
q\; ( \beta_{1}+\beta_{2}+\cdots+\beta_{n+1}) \,=\, \alpha_{n} +
\alpha_{n+1}
\end{eqnarray}

Introducing the ``odd'' and ``even'' generating functions
$\alpha_{odd}(x)$, $\alpha_{even}(x)$ and $\beta_{odd}(x)$,
$\beta_{even}(x)$:
\begin{eqnarray}
\alpha_{odd}(x)\,=\, \alpha_1 \; x +  \alpha_3 \; x^3 +  \alpha_5 \;
x^5 +
\cdots, \quad
\alpha_{even}(x)\,=\, \alpha_0 +  \alpha_2 \; x^2 +  \alpha_4 \; x^4 +
\cdots \nonumber
\end{eqnarray}
and similarly:
\begin{eqnarray}
\beta_{odd}(x) \,=\,\beta_1 \; x +  \beta_3 \; x^3 +  \beta_5 \; x^5 +
\cdots, \quad
\beta_{even}(x) \,=\,\beta_0 +  \beta_2 \; x^2 +  \beta_4 \; x^4 +
\cdots \nonumber
\end{eqnarray}
One deduces from (\ref{albeI}) the following relations on the partial
generating functions $\alpha_{even}(x)$ and $\alpha_{odd}(x)$,
$\beta_{even}(x)$ and $\beta_{odd}(x)$:
\begin{eqnarray}
\alpha_{even}(x)\,= \,{{1}\over{1-x^2}} \,\cdot \Bigl(  (q-2) \;
\;\;\beta_{even}(x) + {{\beta_{odd}(x)}\over{x}})  \Bigr) \\
\alpha_{odd}(x)\,=\,{{1}\over{1-x^2}} \,\cdot \Bigl(  (q-2) \;
\;\;\beta_{odd}(x) + {{\beta_{even}(x)}\over{x}})  \Bigr)
\end{eqnarray}
yielding on the generating functions $\,
\alpha(x)=\alpha_{even}(x)+\alpha_{odd}(x)\,$
 and $\,\beta(x)=\beta_{even}(x)+\beta_{odd}(x)$:
\begin{eqnarray}
\label{genea1b1}
\alpha (x)\,=\, {{1}\over{1-x^2}} \cdot \Bigl(  (q-2) \;\;\; \beta (x)
+ {{\beta (x)}\over{x}})  \Bigr)
\end{eqnarray}

One also recovers (\ref{genII}) or (\ref{genIV}) from relation
(\ref{lina1b1}):
\begin{eqnarray}
\label{gena1b1}
{{q \; \beta(x)}\over{(1-x)}}\,\,=\,\, (1+x)\cdot \alpha(x) -q
\end{eqnarray}
Then, the generating functions $\alpha(x)$ and $\beta(x)$ read:
\begin{eqnarray}
\label{alxI}
\alpha(x)\,=\,{\frac {q\left( 1+(q-2)\,x \right )}{\left (1+x\right
)\left
(1-2\,x\right )}}, \quad
\beta(x)\,=\,{\frac {q\,x\,\left (1-x\right )}{1-2\,x}}
\end{eqnarray}

Similarly to what happened for all the other classes, the {\em
right-action of} $K$ on the
$f_n$'s, or the $M_n$'s,
 factorizes $f_1$ {\em and only} $f_1$:
the factorizations (\ref{kfnVI}), (\ref{mfnVI}), the linear relations
on the exponents $\alpha_n$, $\beta_n$, $\mu_n$ and $\nu_n$
(\ref{abnmunuI}) as well as the linear relations on the generating
functions
 (\ref{ggI}) {\em are still valid
for} class VI.
In particular one still has the two functional relations:
\begin{eqnarray}
\label{gg}
((q-1)\,x-1) \cdot \beta(x) \,= \,q\,x \; \mu(x) - q\,x, \quad
((q-1)\,x-1) \cdot \alpha(x)\, =\, q^2 \, x \; \nu(x) - q
\end{eqnarray}
yielding the following expressions for $\mu(x)$ and $\nu(x)$:
\begin{eqnarray}
\mu(x)\,=\,{\frac {\left (q-2-(q-1)\,x\right )\,x}{1 - 2\,x}}\, , \quad
\nu(x)\,=\,{\frac {\left (q-3\right )\,x}{\left (1+x\right )\left
(1-2\,x\right )}}
\end{eqnarray}

Since $z\,=\, 1/x\,=\,2$ is the only root of all these generating
functions which is not on
the unit circle $\alpha_n$, $\beta_n$, $\mu_n$ and $\nu_n$ clearly {\em
grow
exponentially} like $2^n$.
For instance $\beta (x)$ reads:
\begin{eqnarray}
\beta (x)=q\,x\,\left ( 1+ \sum_{n=0}^{\infty}\;2^n\,x^{n+1} \right )
\end{eqnarray}
Let us also note, for example, that $\mu_{n+1}=2\;\mu_n$ (for $n \ge
2$).
The fact that ``string-like''
factorizations occur is, at first sight, not compatible with the
existence of
simple recurrences on the $f_n$'s like (\ref{recufnqq}) where products
of a fixed number of $f_n$'s occur.  Actually, we have not been able to
find any simple
recurrences on the $f_n$'s\footnote {One may also think, at first
sight,
that such unpleasant ``string-like'' factorizations rule out any
possible polynomial growth and automatically yield exponential growth:
in fact {\em this
is not true}~\cite{BoMaRo93c}}.
One should however note the following point: expressions, similar
to the numerators or denominators appearing in recurrences like
(\ref{recufnqq}), do {\em satisfy some nice additional factorization
properties},
which enable to introduce new polynomials $f^{(1)}_n$, $f^{(2)}_n$,
$f^{(3)}_n$, $f^{(4)}_n$:
\begin{eqnarray}
\label{adfac}
&& f^{(1)}_4= (f_4-f_3^2), \qquad f^{(1)}_5= (f_5-f_4^2)/f_1, \qquad
f^{(1)}_6= (f_6-f_5^2)/(f_1\,f_2), \nonumber \\ && f^{(1)}_7=
(f_7-f_6^2)/(f_1^2\,f_2\,f_3), \qquad f^{(1)}_8=
(f_8-f_7^2)/(f_1^2\,f_2^2\,f_3\,f_4), \nonumber \\ && f^{(2)}_5=
(f_5-f_4 f_3^2), \qquad f^{(2)}_6= (f_6-f_5 f_4^2)/f_1, \qquad
f^{(2)}_7= (f_7-f_6 f_5^2)/(f_1\,f_2), \nonumber \\ && f^{(2)}_8=
(f_8-f_7 f_6^2)/(f_1^2\,f_2\,f_3), \nonumber \\ && f^{(3)}_6=
(f_6-f_4^4)/f_1, \qquad f^{(3)}_7= (f_7-f_5^4)/(f_1\,f_2), \qquad
f^{(3)}_8= (f_8-f_6^4)/(f_1^2\,f_2\,f_3), \nonumber \\ &&f^{(4)}_7=
(f_7-f_4^8)/f_1,\,\,\, \cdots
\end{eqnarray}
Moreover, there does exist other additional factorizations.  For
example the following polynomials {\em do factorize} but their factors
are
not the polynomials $f_n$, and not even the new
polynomials (\ref{adfac}):
\begin{eqnarray}
\label{adfac2}
&& f^{(5)}_4= (f_2 f_4-f_3^2 f_1),\;\; f^{(5)}_5= (f_3 f_5-f_4^2 f_1
f_2),\;\; f^{(5)}_6= (f_4 f_6-f_5^2 f_1 f_2 f_3),\nonumber \\ &&
f^{(6)}_4= (f_4-f_1 f_2 f_3),\;\; f^{(6)}_5= (f_5-f_1 f_2 f_3
f_4),\;\; f^{(6)}_6= (f_6-f_1 f_2 f_3 f_4 f_5),\,\,\, \cdots
\end{eqnarray}
However, though the situation seems very similar to the one encountered
for
class IV  (see
equations (\ref{liste})), we have not been able to find pseudo
recurrences
like (\ref{pseudo}), neither recurrences on the $x_n$'s.
These factorizations (\ref{adfac}) and (\ref{adfac2}) suggest
 the following linear recurrences on the $\beta_n$'s {\em  only valid
 for} $n \ge 2$:
\begin{eqnarray}
\label{eq1}
\beta_{n+1} -2 \; \beta_{n} \,=\,0, \qquad
\beta_{n+2} - \beta_{n+1}  - 2\;\beta_{n} \,=\,0 \\
\label{eq2}
\beta_{n+1} \,=\, \beta_{1} + \beta_{2} + \cdots + \beta_{n}
\end{eqnarray}
For $n\,=\,1$ one has $\beta_1 \, = \beta_2 \, = q $ which is not in
agreement with (\ref{eq1}).
{}From relation (\ref{eq2}), one gets:
\begin{eqnarray}
\beta(x) \,\,= \,\,{{x \; \beta(x) }\over{1-x}} + q\,x
\end{eqnarray}
which is satisfied by the exact expression of $\beta(x)$, namely
equation (\ref{alxI}).

Many more compatibilities between linear recurrences on the exponents
and factorizations (\ref{facKI}), (\ref{facdetI}), (\ref{KdetI}) or
``additional'' factorizations (\ref{adfac}) and (\ref{adfac2}) can be
verified.
In particular, despite the fact
that the iteration corresponding to class VI seems to be involved, it
is
nevertheless possible to associate to these iterations recurrences
bearing on a {\em fixed number} of variables including the variable
$x_n$ (see section (\ref{demrec}) in the following).

Again, one can study the iteration of $\widehat{K}$ seen as a
birational
transformation in ${\Bbb C} P_{q^2-1}$.
For $q=4$
these orbits look like {\em curves} in some domain of ${\Bbb C}
P_{15}$~\cite{BoMaRo93d}.
In fact, these orbits can be seen to belong to a three-dimensional
subvariety which is
the {\em intersection of only twelve Pl\"ucker-like quadrics} in
${\Bbb C} P_{15}$ and more generally the intersection of, at most,
$q^2-4$
algebraic expressions in ${\Bbb C} P_{q^2-1}$ for
$q \times q$ matrices~\cite{BoMaRo93d}.

\section{Demonstration}
\label{dem}

Let us prove here all the results given previously, in particular the
{\em factorization results} and the existence of {\em recurrences on a
fixed number
of variables} and sometimes, on a single variable.

Let $t$ denote the transposition  exchanging $m_{i_2 j_2}$ and $m_{i_1
j_1}$.
Let $P$ be a fixed matrix associated to $t$, for which all entries are
equal to zero, except the two entries which are permuted by $t$:
\begin{eqnarray}
\label{Pij}
P_{i_1 j_1}=1,\quad P_{i_2 j_2}=-1
\end{eqnarray}
$\Delta_0$ will denote the difference between the two entries $m_{i_2
j_2}$ and $m_{i_1 j_1}$:
\begin{eqnarray}
\label{delta0ij}
\Delta_0 \,=\, \, m_{i_2 j_2}-m_{i_1 j_1}
\end{eqnarray}
$\Delta_1$ denotes the difference between the two entries
$\widehat{K}(R_q)_{i_2 j_2}$ and $\widehat{K}(R_q)_{i_1 j_1}$, and
generally $\Delta_n$ denotes the difference between the two entries
$\widehat{K}^n(R_q)_{i_2 j_2}$ and $\widehat{K}^n(R_q)_{i_1 j_1}$.
With these notations transposition $t$ reads on a generic matrix
$R_q$:
\begin{eqnarray}
\label{tdp}
t(R_q) \,=\, R_q+\Delta_0 \cdot P
\end{eqnarray}
Replacing in (\ref{tdp}) matrix $R_q$ by matrix
$\widehat{I}(R_q)$, the inhomogeneous transformation $\widehat{K}$,
can also be seen as a ``deformation'' of the matricial inverse
$\widehat {I}$ :
\begin{eqnarray}
\widehat{K}(R_{q}) \,=\,\, \widehat{I}(R_{q})-\Delta_1 \cdot P
\end{eqnarray}
Noticing that:
\begin{eqnarray}
\Delta_0(\widehat{I}(R_{q})) \,=\,
-\Delta_0(\widehat{K}(R_{q}))=\,-\Delta_1
\end{eqnarray}
Let us introduce matrix $U=R_{q} \cdot \widehat{K}(R_{q})$, which is,
by
construction, close from the identity matrix. We will first assume
that $j_1 \ne j_2$ (and of course $i_1 \ne i_2$ \footnote{If $i_1=i_2$
one
can choose another element of the same class, which satisfies
$j_1=j_2$.}):
\pagebreak
\begin{eqnarray}
\label{Uij}
U \,=\, {\cal I}d_{q}-\Delta_1 \; R_{q} \cdot P \, =
\end{eqnarray}
\begin{eqnarray}
\left [\begin {array}{cccccccccccccc}
1&0&\cdots &0&-\Delta_1\,m_{1 i_1}&0&\cdots &\cdots &0&\Delta_1\,m_{1
i_2}&0&\cdots &\cdots &0\\
0&1&\ddots &0&-\Delta_1\,m_{2 i_1}&0&\cdots &\cdots &0&\Delta_1\,m_{2
i_2}&0&\cdots &\cdots &0\\
\vdots&\vdots&\ddots&\vdots&\vdots&\vdots&\cdots &\cdots
&\vdots&\vdots&\vdots&\cdots &\cdots &\vdots\\
0&\cdots&0&1&-\Delta_1\,m_{(j_1-1) i_1}&0&\cdots &\cdots
&0&\Delta_1\,m_{(j_1-1) i_2}&0&\cdots &\cdots &0\\
0&\cdots&\cdots &0&1-\Delta_1\,m_{j_1 i_1}&0&\cdots &\cdots
&0&\Delta_1\,m_{j_1 i_2}&0&\cdots &\cdots &0\\
0&\cdots&\cdots &0&-\Delta_1\,m_{(j_1+1)
i_1}&1&0&\cdots&0&\Delta_1\,m_{(j_1+1) i_2}&0&\cdots &\cdots &0\\
0&\cdots&\cdots &0&-\Delta_1\,m_{(j_1+2)
i_1}&0&1&\ddots&0&\Delta_1\,m_{(j_1+2) i_2}&0&\cdots &\cdots &0\\
\vdots&\cdots&\cdots
&\vdots&\vdots&\vdots&\ddots&\ddots&\vdots&\vdots&\vdots&\cdots &\cdots
&\vdots\\
0&\cdots&\cdots &0&-\Delta_1\,m_{(j_2-1)
i_1}&0&0&\cdots&1&\Delta_1\,m_{(j_2-1) i_2}&0&\cdots &\cdots &0\\
0&\cdots&\cdots &0&-\Delta_1\,m_{j_2
i_1}&0&\cdots&\cdots&0&1+\Delta_1\,m_{j_2 i_2}&0&\cdots &\cdots &0\\
0&\cdots&\cdots &0&-\Delta_1\,m_{(j_2+1)
i_1}&0&\cdots&\cdots&0&\Delta_1\,m_{(j_2+1) i_2}&1&0&\cdots &0\\
0&\cdots&\cdots &0&-\Delta_1\,m_{(j_2+2)
i_1}&0&\cdots&\cdots&0&\Delta_1\,m_{(j_2+2) i_2}&0&1&\ddots &0\\
\vdots&\cdots&\cdots
&\vdots&\vdots&\vdots&\cdots&\cdots&\vdots&\vdots&\vdots&\ddots&\ddots
&\vdots\\
0&\cdots&\cdots &0&-\Delta_1\,m_{q
i_1}&0&\cdots&\cdots&0&\Delta_1\,m_{q i_2}&0&0&\ddots &1\\
\end {array}\right ] \nonumber
\end{eqnarray}
This expression of $U$ gives, at once, the determinant :
\begin{eqnarray}
\label{x0ij}
det(U)=\,x_0
&=& \left (1-\Delta_{1}\,m_{j_1 i_1}\right )\left (1+\Delta_{1}\,m_{j_2
i_2}\right )+\Delta_{1}^{2}m_{j_1 i_2}\,m_{j_2 i_1} \nonumber \\
&=& 1 +\left (m_{j_2 i_2}-m_{j_1 i_1}\right )\Delta_{1} + \left (m_{j_1
i_2}\,m_{j_2 i_1}-m_{j_1 i_1}\,m_{j_2 i_2}\right )\Delta_{1}^{2}
\nonumber \\
&=&1+T_{0}\,\Delta_{1}+N_0 \,\Delta_{1}^{2}
\end{eqnarray}
where $N_0=\left (m_{j_1 i_2}\,m_{j_2 i_1}-m_{j_1 i_1}\,m_{j_2
i_2}\right
)$ (that is the $2 \times 2$
 minor corresponding to rows $j_1$ and $j_2$ and columns $i_1$ and
 $i_2$ of
matrix $R_{q}$)
 and $T_0=m_{j_2 i_2}-m_{j_1 i_1}$ ($T_0$ corresponds to the difference
 of
the two entries exchanged by $t$
for the transposed matrix).
One is now able to easily calculate the second step of the iteration :
\begin{eqnarray}
\label{K2ij}
\widehat{K}^{2}(R_{q})\,=\,t(\widehat{I}(\widehat{K}(R_{q})))\,=\,t(\widehat{I}(U).R_{q}
)
\end{eqnarray}
where $\widehat{I}(U)$  also differs from the identity matrix by the
two columns $j_1$ and $j_2$.
Each entry of these columns is the ratio by $x_0$ of a polynomial
quadratic in $\Delta_1$.
Let us calculate explicitly $\widehat{I}(U)$  as a polynomial in
$\Delta_1$.
{}From relation (\ref{Uij}) one directly gets:
\begin{eqnarray}
\widehat{I}(U) \,=\, \sum_{n=0}^{\infty}\;\Delta_1^n\,(R_{q} \cdot P)^n
\end{eqnarray}
Matrix $(R_{q} \cdot P)$ being of a quite simple form, it is easy to
calculate its minimal
polynomial which reads:
\begin{eqnarray}
x\,\cdot (x^2+T_0\,x+N_0)
\end{eqnarray}
One can thus obtain the expression of the matrices $(R_{q} \cdot P)^n$
in terms of
$(R_{q} \cdot P)^2$, of $(R_{q} \cdot P)$ and of the identity matrix.
After straightforward calculations one gets:
\begin{eqnarray}
\label{IUij}
\widehat{I}(U)\,=\,\,{\cal
I}d\,+\,{{\Delta_1\,(1+T_0\,\Delta_1)}\over{x_0}} \cdot (R_{q} \cdot
P)+{{\Delta_1^2}\over{x_0}}\;(R_{q} \cdot P)^2
\end{eqnarray}

Let us now revisit these equations when $j_1=j_2=j$ (or equivalently
$i_1=i_2$), that is for classes IV and VI.
 In this $j_1=j_2=j$ case, $U$ reads:
\begin{eqnarray}
\label{Uij1}
U=\left [\begin {array}{ccccccccc}
1&0&\cdots &0&\Delta_1\,(m_{1 i_2}-m_{1 i_1})&0&\cdots &\cdots &0\\
0&1&\ddots &0&\Delta_1\,(m_{2 i_2}-m_{2 i_1})&0&\cdots &\cdots &0\\
\vdots&\vdots&\ddots&\vdots&\vdots&\vdots&\cdots &\cdots &\vdots\\
0&\cdots&0&1&\Delta_1\,(m_{(j-1) i_2}-m_{(j-1) i_1})&0&\cdots &\cdots
&0\\
0&\cdots&\cdots &0&1+\Delta_1\,(m_{j i_2}-m_{j i_1})&0&\cdots &\cdots
&0\\
0&\cdots&\cdots &0&\Delta_1\,(m_{(j+1) i_2}-m_{(j+1)
i_1})&1&0&\cdots&0\\
0&\cdots&\cdots &0&\Delta_1\,(m_{(j+2) i_2}-m_{(j+2)
i_1})&0&1&\ddots&0\\
\vdots&\cdots&\cdots &\vdots&\vdots&\vdots&\ddots&\ddots&\vdots\\
0&\cdots&\cdots &0&\Delta_1\,(m_{q i_2}-m_{q i_1})&0&0&\cdots&1\\
\end {array}\right ] \nonumber
\end{eqnarray}
This expression of $U$ gives at once the determinant :
\begin{eqnarray}
\label{x0ij1}
det(U)\,=\,\,x_0
&=& 1+\Delta_1\,(m_{j i_2}-m_{j i_1})\nonumber \\
&=&1+T_{0}\,\Delta_{1}
\end{eqnarray}
where $T_0=m_{j i_2}-m_{j i_1}$ ($T_0$ still correspond to the
difference of
the two entries exchanged by $t$
for the transposed matrix).
Equation (\ref{IUij}) becomes:
\begin{eqnarray}
\label{IUij1}
\widehat{I}(U)\,=\,{\cal I}d\,+\,{{\Delta_1}\over{x_0}} \cdot (R_{q}
\cdot P)
\end{eqnarray}
Relation (\ref{K2ij}) is still valid and enables to calculate the
second step of the
iteration.

Let us now give a proof of the factorization properties for the various
classes defined in sections
(\ref{rappel}) and (\ref{result}).

\subsection{Demonstration of the factorizations}
\label{demfac}

The demonstration of the factorizations has been done for class I
in~\cite{BoMaRo93a}: we will here just recall (and generalize to all
the other classes) the main steps.

Factorization properties are obviously associated with the homogeneous
 matrices $K^n(R_q)$ (instead of matrices $\widehat{K}^n(R_{q})$, which
 do
not have polynomial entries):
\begin{eqnarray}
K(R_{q})\,=\,\det(R_{q}) \cdot \widehat{K}(R_{q}) \nonumber
\end{eqnarray}
$K$ being a homogeneous transformation of degree $q-1$, one obtains :
\begin{eqnarray}
\label{K^2}
K^2(R_{q})\,=\, x_0\,\det(R_{q})^{q-2} \cdot \widehat{K}^2(R_{q})
\end{eqnarray}

Let us first study classes I, II, III and V for which $j_1 \ne j_2$ and
$i_1 \ne i_2$.
Equations (\ref{K2ij}) and (\ref{IUij}) give the form for
$x_0\,\widehat{K}^2(R_{q})$.
One remarks that its entries are polynomials in the entries of the
matrix
$R_{q}$ and {\em quadratic} in $\Delta_1$.
The definition of $\Delta_1$ straightforwardly shows that its
denominator is $det(R_{q})$.
Thus matrix $\widehat{K}^2(R_{q})$ reads:
\begin{eqnarray}
\widehat{K}^2(R_{q})\,=\,{{M_2}\over{x_0\, \cdot det(R_{q})^2}}
\end{eqnarray}
where $M_2$ is a matrix with polynomial entries. Equation (\ref{K^2})
thus proves the first step of the factorization :
\begin{eqnarray}
\label{facM1}
K^2(R_{q})\,=\,det(R_{q})^{q-4} \cdot M_2
\end{eqnarray}
A similar demonstration can be performed on $\det(K(R_q))$ and yields
:
\begin{eqnarray}
\label{detK}
det(K(R_q))\,=\,x_0 \cdot det(R_{q})^{q-1}
\end{eqnarray}
The expression of $x_0$, namely (\ref{x0ij}), is also quadratic in
$\Delta_1$. One thus has the following factorization :
\begin{eqnarray}
\label{facd1}
det(K(R_q))\,= \, det(R_{q})^{q-3}\, \cdot f_2
\end{eqnarray}
As far as classes IV and VI are concerned, {\em for which} $j_1=j_2$
{\em or} $i_1=i_2$, $x_0$, as well as the entries of matrix $x_0\,
\widehat{K}^2(R_{q})$, given by
equations (\ref{K2ij}) and (\ref{IUij1}), are polynomials in the
entries of the matrix
$R_{q}$ and {\em linear} in $\Delta_1$.
Matrix $\widehat{K}^2(R_{q})$ reads:
\begin{eqnarray}
\widehat{K}^2(R_{q})\,=\,{{M_2}\over{x_0\, \cdot det(R_{q})}}
\end{eqnarray}
where $M_2$ has polynomial entries. Equation (\ref{K^2}) thus proves
the first step of the factorization :
\begin{eqnarray}
\label{facM11}
K^2(R_{q})\,=\,det(R_{q})^{q-3}\, \cdot M_2
\end{eqnarray}
{}From relation (\ref{detK}) one also gets:
\begin{eqnarray}
\label{facd11}
det(K(R_q))\,=\,det(R_{q})^{q-2}\, \cdot  f_2
\end{eqnarray}
Notice that factorization (\ref{facM1}) is only valid for $q>3$, and
(\ref{facd1}) for $q>2$,
while (\ref{facM11}) is valid for $q>2$, and (\ref{facd11}) for $q>1$.

Considering successively the explicit expressions of $K^n(R_{q})$ and
of
their determinants, one notices that there are
further factorizations (see for instance equation (\ref{detVI})), that
could be obtained the same way.
However these further factorizations depend on the class one considers.
We
will thus just assume these factorizations (however the first steps of
the factorizations have been strictly
obtained by formal computer calculations and their general form has
been
proved recursively in~\cite{BoMaRo93a}).
Their generic form reads:
\begin{eqnarray}
\label{fr}
f_n(K)&=&f_1^{\mu_n}\, \cdot f_{n+1} \\
\label{dl}
\det (M_n)&=&  f_1 ^{v_n}\cdot f_2 ^{v_{n-1}}\cdot f_3 ^{v_{n-2}}\cdots
f_{n-1}^{v_2} \cdot f_{n}^{v_1}\cdot f_{n+1} \\
\label{mr}
\Bigl(M_n\Bigr)_{K}&=&f_1^{\nu_n}\, \cdot  M_{n+1} \\
\label{ml}
K (M_n)&=& M_{n+1}\cdot f_1 ^{u_n}\cdot f_2 ^{u_{n-1}}\cdot f_3
^{u_{n-2}}\cdots f_{n-1}^{u_2} \cdot f_{n}^{u_1}
\end{eqnarray}
with the following relations between the different exponents:
\begin{eqnarray}
\label{recnun}
 \nu_{n+1}&=&(q-1)\,\nu_n + u_{n+1}-(u_1\,\mu_n +u_2\, \mu_{n-1}+
\cdots+u_n\,\mu_1) \\
\label{recmun}
\mu_{n+1}&=&v_{n+1}+q\,\nu_n -(v_n\,\mu_1+v_{n-1}\,\mu_2+\cdots+
\mu_n\,v_1)
\end{eqnarray}
Moreover, it can be shown that (\ref{ml}), the factorization relation
on $K(M_n)$,
necessarily  yields relation (\ref{dl}), the factorization of the
determinant
(and also the inequalities $v_n \ge
1+u_{n}$ when $u_n \ne 0$).
The left factorizations, (\ref{dl}) and (\ref{ml}), and the right
factorizations, (\ref{fr}) and (\ref{mr}), are equivalent when assuming
(\ref{recnun}) and (\ref{recmun}). The proof is given
in~\cite{BoMaRo93a}.

\subsection{Demonstration of the recurrences}
\label{demrec}

Let us briefly sketch the demonstration of the existence of recurrences
{\em independent of $q$} (like recurrences (\ref{recuxnqq}) or
(\ref{xnIV})), on a
{\em finite} set of variables including variable $x_n$ (equation
(\ref{recuxnqq})).

Such a demonstration has already been performed for class I
in~\cite{BoMaRo93a}.
Therefore one will not recall this demonstration but
 one will only sketch  the demonstration for the other classes.

\subsubsection{Demonstration of the recurrences for class II}

In order to represent class II, let us take the transposition $t$
exchanging $m_{12}$ and $m_{34}$.
Definition (\ref{delta0ij}) now reads:

$\Delta_0=[R_{q}]_{3\,4}-[R_{q}]_{1\,2}=\,m_{34}-m_{12}$, relation
(\ref{Pij}) becomes:
\begin{eqnarray}
P=\left [\begin {array}{ccccc}
{ 0}&{ 1}&{ 0}&{ 0}&\ldots\\
{ 0}&{ 0}&{ 0}&{ 0}&\ldots\\
0&{ 0}&{ 0}&-1&\ldots\\
{ 0}&{ 0}&{ 0}&{ 0}&\ldots\\
\vdots & \vdots & \vdots & \vdots&\ddots
\end {array}\right ]
\end{eqnarray}
and  $\Delta_1$ reads:
\begin{eqnarray}
\label{Delta1II}
\Delta_1=\,[ \widehat{K}(R_{q}) ]_{3\,4} - [ \widehat{K}(R_{q})
]_{1\,2}
=\,[ \widehat{I}(R_{q}) ]_{1\,2} - [ \widehat{I}(R_{q}) ]_{3\,4}
\end{eqnarray}
 Here matrix $U=R_{q} \cdot \widehat{K}(R_{q})$ reads:
\begin{eqnarray}
\label{UII}
U=
\left [\begin {array}{ccccccc}
1&- \Delta_{1}\, m_{11}&0& \Delta_{1}\, m_{13}&0&0&\ldots\\
0&1- \Delta_{1}\, m_{21}&0& \Delta_{1}\, m_{23}&0&0&\ldots\\
0&- \Delta_{1}\,m_{31}&1& \Delta_{1}\, m_{33}&0&0&\ldots\\
0&- \Delta_{1}\, m_{41}&0&1+\Delta_{1}\, m_{43}&0&0&\ldots\\
\vdots&- \Delta_{1}\, m_{51}&\vdots&\Delta_{1}\, m_{53}&1&0&\ldots\\
\vdots&\vdots&\vdots&\vdots&0&1&\ddots\\
\vdots&\vdots&\vdots&\vdots&\vdots&\ddots&\ddots
\end {array}\right ]
\nonumber
\end{eqnarray}

This transposition involving two columns (and two rows), the
determinant
$x_0$ of matrix $U$ is given by relation (\ref{x0ij}), as a {\em
quadratic}
expression of $\Delta_1$:
\begin{eqnarray}
\label{x0II}
det(U)\,=\,x_0\,=\,1\,+\,T_{0}\,\Delta_{1}\,+\,N_0 \,\Delta_{1}^{2}
\end{eqnarray}
where $\, N_0\,=\,\left (m_{41}\,m_{23}-m_{43}\,m_{21}\right )\,$ and
$\,T_0\,=\,m_{43}-m_{21}$.

Relation (\ref{IUij}) yields:
\begin{eqnarray}
x_0&\cdot&\widehat{I}(U)=\\
&&\left [\begin {array}{ccccccc}
x_0& \Delta_{1}\,\left ( \Delta_{1}\, m_{11}\, m_{43}+ m_{11}-
\Delta_{1}\, m_{41}\, m_{13}\right )&0&- \Delta_{1}\,\left (
\Delta_{1}\, m_{11}\, m_{23}- \Delta_{1}\, m_{13}
\, m_{21}+ m_{13}\right )&0&0&\ldots\\
0&1+\Delta_{1}\, m_{43}& 0&- \Delta_{1}\,m_{23}&0&0&\ldots\\
0& \Delta_{1}\,\left ( \Delta_{1}\, m_{31}\, m_{43}+m_{31}-
\Delta_{1}\, m_{33}\, m_{41}\right )& x_0& \Delta_{1}\,\left (
\Delta_{1}\, m_{33}\, m_{21}- m_{33}- \Delta_{1}\, m_{23}\,m_{31}\right
)&0&0&\ldots\\
0& \Delta_{1}\, m_{41}&0&1- \Delta_{1}\, m_{21}&0&0&\ldots\\
\vdots&\Delta_{1}\, m_{51}&\vdots&-\Delta_{1}\, m_{53}&x_0&0&\ldots\\
\vdots&\vdots&\vdots&\vdots&0&x_0&\ddots\\
\vdots&\vdots&\vdots&\vdots&\vdots&\ddots&\ddots
\end {array}\right ]\nonumber
\end{eqnarray}

Matrix $\widehat{K}^2(R_q)$ is obtained from relation (\ref{K2ij}).
Its explicit form is quite involved, and will not be given here.

We will just concentrate on a {\em fixed finite} number of variables,
enabling to understand the evolution of $T_0, \, N_0$ and $\Delta_0$,
the
action of $\widehat{K}^2$ preserving this
set of variables.
Namely $\widehat{K}^2$ transforms the following variables as follows:
\begin{eqnarray}
\label{K2VIIr}
m_{21}\, \rightarrow \,\, {{m_{21}-\Delta_1\,N_0}\over{x_0}} \; ,
\;\nonumber \\
m_{23}\, \rightarrow \, \,{{m_{23}}\over{x_0}} \; , \; \nonumber \\
m_{41}\, \rightarrow \,\, {{m_{41}}\over{x_0}}  \; , \;\nonumber \\
m_{43}\, \rightarrow \,\, {{m_{43}+\Delta_1\,N_0}\over{x_0}}
\end{eqnarray}
{}From (\ref{K2VIIr}) one gets the equations:
\begin{eqnarray}
\label{t2n2II}
T_2&=&{{T_0+2\,\Delta_1\,N_0}\over{x_0}} \; , \; \nonumber \\
N_2&=&{{N_0-\Delta_1\,N_0\,(m_{21}-m_{43})+\Delta_1^2\,N_0^2}\over{x_0^2}}
\,\,=\,\,{{N_0\,(1+\Delta_1\,T_0+\Delta_1^2\,N_0)}\over{x_0^2}}\,\,=\,\,{{N_0}\over{x_0}}
\end{eqnarray}
Let us introduce the two variables:
\begin{eqnarray}
\label{var72}
F_0\,=\,m_{11}\,\,m^{<2>}_{22}+m_{33}\,\,m^{<2>}_{44}-m_{13}\,\,m^{<2>}_{42}-m_{31}\,\,m^{<2>}_{24}
\end{eqnarray}
\begin{eqnarray}
\label{var73}
G_0\,=\,m_{22}\,\,m^{<2>}_{11}+m_{44}\,\,m^{<2>}_{33}-m_{42}\,\,m^{<2>}_{13}-m_{24}\,\,m^{<2>}_{31}
\end{eqnarray}
where $m^{<2>}_{ij}$ denotes the entries of matrix
$\widehat{K}^2(R_q)$.
These two variables happen to be equal.
As a consequence of this remarkable equality, $\Delta_2$ satisfies the
following relation:
\begin{eqnarray}
\label{delta2II}
{{\Delta_2+\Delta_0}\over{\Delta_1}}\,=\,F_0\,=\,G_0\,=\,{{F_0+G_0}\over{2}}
\end{eqnarray}

One thus has to calculate the action of  $\widehat{K}^2$ on many new
entries $m_{\alpha, \beta}$ occurring in the right-hand side of
(\ref{var72}) and (\ref{var73}).
Remarkably {\em this action is the same on all these} $m_{\alpha,
\beta}$'s, namely:
\begin{eqnarray}
m_{\alpha, \beta} \, \,\rightarrow \,\,\,\,  m^{<2>}_{\alpha, \beta}
\,\, = \,\, {{m_{\alpha, \beta}+\Delta_1\,P_0^{(\alpha,
\beta)}}\over{x_0}}
\end{eqnarray}
where the $P_0^{(\alpha, \beta)}$'s  are $2\times2$ minors, which,
remarkably,
also transform similarly under $\widehat{K}^2$:
\begin{eqnarray}
P_0^{(\alpha, \beta)} \, \rightarrow \,\,\,\,  P_2^{(\alpha, \beta)}
\,=\, {{P_0^{(\alpha, \beta)}}\over{x_0}}
\end{eqnarray}
yielding:
\begin{eqnarray}
\label{m4}
x_2 \cdot m^{<4>}_{\alpha, \beta}\,\, = \,\, m^{<2>}_{\alpha,
\beta}\,\,{{\Delta_1+\Delta_3}\over{\Delta_1}}\,-\,m_{\alpha,
\beta}\,\,{{\Delta_3}\over{x_0\, \Delta_1}}
\end{eqnarray}
{}From relation (\ref{m4}) one can get the action of $\widehat{K}^2$ on
$(F_0+G_0)/2$:
\begin{eqnarray}
{{F_0+G_0}\over{2}}\, \, \rightarrow \, \,
{{F_2+G_2}\over{2}}\,\,=\,\,{{\Delta_1+\Delta_3}\over{x_2\,\Delta_1}}
\, E_2\,
-\,{{\Delta_3}\over{x_0\, x_2\,\Delta_1}} \,{{F_0+G_0}\over{2}}
\end{eqnarray}
where
$E_0\,=\,m_{11}\,m_{22}+m_{33}\,m_{44}-m_{13}\,m_{42}-m_{31}\,m_{24}$,
the other $E_n$'s being deduced from $E_0$ by the successive
right-action
of $\widehat{K}^2$.

Recalling equations (\ref{delta2II}) one gets:
\begin{eqnarray}
\label{ts10VII}
 {{(\Delta_{n+2}+\Delta_{n+4})}\over { \Delta_{n+3}}}\,
+\, {{\Delta_{n+3}\,(\Delta_n+\Delta_{n+2})}\over
{\Delta_{n+1}^2\;x_n\;x_{n+2}}}
\,-\, {{(E_{n+2}\,(\Delta_{n+1}+\Delta_{n+3}))}\over {
x_{n+2}\,\Delta_{n+1}}}
\,=\,0
\end{eqnarray}
One also needs the  right-action of  $\widehat{K}^2$ on $E_0$. It can
be
deduced from relation (\ref{m4}):
\begin{eqnarray}
\label{ts11VII}
&&x_{n+2}\,E_{n+4}-
{{(\Delta_{n+2}+\Delta_{n+4})\,(\Delta_{n+1}+\Delta_{n+3})}
\over
{ \Delta_{n+1}\,\Delta_{n+3}}}\nonumber \\
&&\quad -{{\Delta_{n+3}}\over { \Delta_{n+1}\;x_n\;x_{n+2}}}
\,\left (
{{\Delta_{n+3}\,E_n}\over {\Delta_{n+1}\,x_n }}
-{{(\Delta_{n+1}+\Delta_{n+3})(\Delta_{n}+\Delta_{n+2})} \over
{\Delta_{n+1}^2}}
\right )\,=\,0
\end{eqnarray}
Equations (\ref{ts10VII}) and  (\ref{ts11VII}) enable to eliminate the
$E_n$'s.
Introducing the well-suited variables $\delta_n=\Delta_{n+2}/\Delta_n$
one gets:
\begin{eqnarray}
\label{camel}
&&{{ x_{n+4}\, x_{n+6}\,\left (1+ \delta_{n+6}\right )}\over {1+
\delta_{n+5}}}
+{{ \delta_{n+5}^{2}\left ( \delta_{n+4}+1\right )}\over {
\delta_{n+4}\,\left (1+
\delta_{n+5}\right )}}
-{{ \delta_{n+5}\,\left (1+ \delta_{n+3}\right )\left (
\delta_{n+4}+1\right )}\over { \delta_{n+4}}}
-{{ \delta_{n+3}^{3} \delta_{n+5}\,\left (1+
\delta_{n+2}\right )}\over{ \delta_{n+2}\, \delta_{n+4}\, x_{n+2}\,
x_{n+4}\,\left (1+ \delta_{n+1}
\right )}}  \nonumber \\
&&-{{ \delta_{n+3}^{3} \delta_{n+5}\, \delta_{n+1}^{2}\left (
\delta_{n}+1
\right )}\over { \delta_{n}\, \delta_{n+2}\, \delta_{n+4}\, x_{n}\,
x_{n+2}^{2} x_{n+4}
\,\left (1+ \delta_{n+1}\right )}}
+{{ \delta_{n+3}^{2} \delta_{n+5}\,\left (1+ \delta_{n+3}\right )\left
(1+ \delta_{n+2}\right
)}\over { \delta_{n+2}\, \delta_{n+4}\, x_{n+2}
\, x_{n+4}}}
\,=\,0
\end{eqnarray}
Finally, coming back to equations (\ref{x0II}) and (\ref{t2n2II}), one
can
eliminate the $T_n$'s and $N_n$'s, and get, with the same variables
$\delta_n$'s,
another relation between the $x_n$'s and the $\delta_n$'s:
\begin{eqnarray}
\label{ts12VII}
\left ({\frac {(x_{n+4}-1)}{\delta_{n+3}}}-{\frac
{(x_{n+2}-1)}{x_{n+2}}}\right
)\, \, =\,\,{\frac
{(1+\delta_{n+1})\,(1+\delta_{n+3})}{x_{n+2}}}\,\left ({\frac
{(x_{n+2}-1)}{\delta_{n+1}}}-{\frac {(x_{n}-1)}{x_{n}}}\right )
\end{eqnarray}

One can in principle eliminate $x_n$ between (\ref{camel}) and
(\ref{ts12VII}): it yields ``huge'' calculations. In contrast the
elimination of $\delta_n$ seems out of range.

Let us just note that, though such a system of recurrences is quite
involved,
one can however get some {\em  finite order conditions} for these
recurrences of class II, namely the orbits of order three and four :

- order three :
\begin{eqnarray}
\label{finiVII}
 x_{n}\, x_{n+1}\, x_{n+2}-1\,=\,0\,\, ,\,\,\,\,\,\hbox{or}\,\,\,\, \,
 \, \, 1+ \delta_{n+1}+ \delta_{n}\, \delta_{n+1}\,=\,0
\end{eqnarray}

- order four :
\begin{eqnarray}
\label{finiVII2}
 x_{n}\, x_{n+2}-1\,=\,0\,\,,\,\,\,\,\hbox{or}\,\,\,\, \, \, \,\,
 \delta_{n}\, \delta_{n+2}+1\,=\,0
\end{eqnarray}

\subsubsection{Demonstration of the recurrences for class III}

One will just sketch here briefly the demonstration of the recurrences
for class III.
Let us represent class III, with transposition $t$ exchanging $m_{12}$
and $m_{31}$.
Then (\ref{delta0ij}) reads:
$\Delta_0=[R_{q}]_{3\,1}-[R_{q}]_{1\,2}=m_{31}-m_{12}$, and matrix $P$
defined by (\ref{Pij}) becomes:
\begin{eqnarray}
P=\left [\begin {array}{ccccc}
{ 0}&{ 1}&{ 0}&{ 0}&\ldots\\
{ 0}&{ 0}&{ 0}&{ 0}&\ldots\\
{ -1}&{ 0}&{ 0}&{ 0}&\ldots\\
{ 0}&{ 0}&{ 0}&{ 0}&\ldots\\
\vdots & \vdots & \vdots & \vdots&\ddots
\end {array}\right ]
\end{eqnarray}
$\Delta_1$ reads:
\begin{eqnarray}
\label{Delta1III}
\Delta_1\,=\,[ \widehat{K}(R_{q}) ]_{3\,1} - [ \widehat{K}(R_{q})
]_{1\,2}
\,=\,[ \widehat{I}(R_{q}) ]_{1\,2} - [ \widehat{I}(R_{q}) ]_{3\,1}
\end{eqnarray}
Here matrix $U\,=\,R_{q} \cdot \widehat{K}(R_{q})$ has a simple form
similar to (\ref{UII}). The determinant
$x_0$ of matrix $U$ is again given by equation (\ref{x0ij}), namely:
\begin{eqnarray}
\label{x0III}
det(U)\,=\,x_0=\,1+T_{0}\,\Delta_{1}+N_0 \,\Delta_{1}^{2}
\end{eqnarray}
with $N_0=\left (m_{11}\,m_{23}-m_{13}\,m_{21}\right )$  and
$\,T_0=m_{13}-m_{21}$.

Thus one can calculate the explicit form of
$\widehat{I}(\widehat{K}(R_{q}))$:
\begin{eqnarray}
\label{IKIII}
x_0&\cdot&\widehat{I}(\widehat{K}(R_{q}))=\\
&&\left [\begin {array}{cccc}
m_{11}&m_{21}+\Delta_1\,(m_{11}\,m_{22}-m_{12}\,m_{21})&m_{13}+\Delta_1\,N_0&\ldots\\
m_{21}-\Delta_1\,N_0&m_{22}+\Delta_1\,(m_{13}\,m_{22}-m_{12}\,m_{23})&m_{23}&\ldots\\
m_{31}+\Delta_1\,(m_{31}\,m_{13}-m_{33}\,m_{11})&w&m_{33}+\Delta_1\,(m_{31}\,m_{23}-m_{33}\,m_{21})&\ldots\\
\vdots & \vdots & \vdots &\ddots
\end {array}\right ] \nonumber
\end{eqnarray}
where $w=m_{32}\,x_0+\Delta_1\,\left [
m_{22}\,(m_{31}+\Delta_1\,(m_{31}\,m_{13}-m_{33}\,m_{11}))-m_{12}\,(m_{33}+\Delta_1\,(m_{31}\,m_{23}-m_{33}\,m_{21}))\right
]$.

Matrix $\widehat{K}^2(R_q)$ is obtained  permuting entries $m_{12}$ and
$m_{31}$ in relation (\ref{IKIII}).

This yields the following expression for $\Delta_2$:
\begin{eqnarray}
\Delta_2 \,\,= \,\,
{{-\Delta_0+\Delta_1\,(m_{11}\,m_{22}-m_{12}\,m_{21}+m_{33}\,m_{11}-m_{31}\,m_{13})}\over
{x_0}}
\end{eqnarray}
Let us denote $P_0$ the expression appearing in the right-hand side of
$\Delta_2$:
\begin{eqnarray}
P_0\,=\,m_{11}\,m_{22}-m_{12}\,m_{21}+m_{33}\,m_{11}-m_{31}\,m_{13}
\end{eqnarray}

Similarly $N_0$, $T_0$ and $P_0$ transform, under $K^{2}$, as follows
:
\begin{eqnarray}
&&T_0\,\,\rightarrow \, \, \,T_2\,=\,{{T_0+2\,\Delta_1\,N_0}\over
{x_0}} \, ,\,\nonumber \\
&&N_0 \, \, \rightarrow \,\, \,N_2\,=\,{{N_0}\over {x_0}} \,
,\,\nonumber \\
&&P_0 \,\, \rightarrow  \, \, \,P_2\, =\, {{P_0}\over {x_0}}-\Delta_2
\, T_2
\end{eqnarray}

Since these results have been calculated on a generic matrix $R_{q}$,
they
can be applied successively on each matrix $\widehat{K}^{n}(R_{q})$ and
 thus all the equations given above are {\em actually recurrence
 relations}.
The expressions of $\widehat{K}(R_{q})$ and $\widehat{K}^2(R_{q})$
again allow
us to prove the recurrence on $x_n$'s and $\Delta_n$'s.

Introducing the well-suited variable $\delta_n$ :
\begin{eqnarray}
\delta_n\,=\,{{\Delta_{n+2}}\over{\Delta_n}}
\end{eqnarray}
and eliminating $N_n$'s, $T_n$'s and $P_n$'s, one gets the following
relations on the $x_n$'s and $\delta_n$'s :
\begin{eqnarray}
\label{rel1uq}
\begin{array}{c}
x_{n}\, x_{n+2}\, x_{n+4}+ x_{n}\, x_{n+2}\, x_{n+4}\,
\delta_{n+1}-x_{n+2}\, x_{n}- x_{n}\, x_{n+2}\, \delta_{n+1}- x_{n}\,
\delta_{n+3}\, x_{n+2}\cr \cr
-2\, x_{n}\, \delta_{n+3}\, x_{n+2}\, \delta_{n+1}+ x_{n}\,
\delta_{n+3}+2\, x_{n}\, \delta_{n+3}\, \delta_{n+1}- \delta_{n+3}^{2}
\delta_{n+1}\, x_{n}\, x_{n+2}+ \delta_{n+3}^{2} \delta_{n+1}\,
x_{n}\cr \cr
+ \delta_{n+3}\, \delta_{n+1}^{2} x_{n}+ \delta_{n+3}^{2}
\delta_{n+1}^{2} x_{n}- \delta_{n+3}\, \delta_{n+1}^{2}-
\delta_{n+3}^{2} \delta_{n+1}^{2}\,=\,0
\end{array}
\end{eqnarray}
and :
\begin{eqnarray}
\label{un+2}
x_{n}\,x_{n+2}\,=\,{\frac
{\delta_{n+1}\,(\delta_{n+1}\,\delta_{n}+\delta_{n+1}+1)}{ \delta_{n}\,
(\delta_{n+2}\,\delta_{n+1}+\delta_{n+2}+1)}}
\end{eqnarray}
In order to write down this last equation in a more handable way, let
us
introduce a new variable $R_n$ :
\begin{eqnarray}
\label{Rn}
R_{n}={{1+ \delta_{n+1}+  \delta_{n}\,\delta_{n+1}}\over {\delta_n}}
\end{eqnarray}
With these new variables $R_{n}$, equation (\ref{un+2}) read:
\begin{eqnarray}
\label{xxrr}
x_n\,x_{n+2}\,=\,{{R_n}\over{R_{n+1}}}
\end{eqnarray}
{}From this last equation one can get $x_{n+2}$ in terms of $x_n$, as
well as
$x_{n+4}$ in terms of $x_{n+2}$ (and therefore of $x_n$).
Let us use, in equation (\ref{rel1uq}), these expressions of $x_{n+2}$
and $x_{n+4}$.
Remarkably one obtains $x_n$ as a function of the $\delta_n$'s.
This expression of $x_n$ takes a very simple form when introducing some
new variables $Q_n$ :
\begin{eqnarray}
\label{Qn}
Q_{n}={{1+ \delta_{n+1}+ \delta_{n+1}\, \delta_{n+3}+
\delta_{n}\,\delta_{n+1}+
\delta_{n}\,\delta_{n+1}\,\delta_{n+3}+ \delta_{n}\,\delta_{n+1}\,
\delta_{n+2}\, \delta_{n+3}}\over {\delta_n\,\delta_{n+1}}}
\end{eqnarray}
$x_n$ reads :
\begin{eqnarray}
\label{last0}
x_{n}\,=\,{{R_{n+3}\,Q_{n}}\over { R_{n+1}\,Q_{n+1}}}
\end{eqnarray}
One straightforwardly obtains $x_{n+2}$ shifting $n$ by two in equation
(\ref{last0}):
\begin{eqnarray}
\label{last1}
x_{n+2}\,=\,{{R_{n+5}\,Q_{n+2}}\over {R_{n+3}\,Q_{n+3}}}
\end{eqnarray}
One can now get the product $\,x_{n}\,x_{n+2}\,$ from (\ref{last0}) and
(\ref{last1}):
\begin{eqnarray}
\label{last2}
x_n\,x_{n+2}\,=\,{{R_{n+5}\,Q_{n}\,Q_{n+2}}\over
{R_{n+1}\,Q_{n+1}\,Q_{n+3}}}
\end{eqnarray}
The two equations (\ref{xxrr}) and (\ref{last2}) yield an {\em
algebraic
relation} between the $\delta_n$'s, namely between $\delta_n,
\,\delta_{n+1}, \, \ldots,\delta_{n+6}$ :
\begin{eqnarray}
\label{last}
Q_{n+1}\,Q_{n+3}\,R_n-Q_n\,Q_{n+2}\,R_{n+5}\,=\,0
\end{eqnarray}
This relation can be written in terms of an invariant expression,
yielding
a first integration:
\begin{eqnarray}
\label{lastplus}
{{Q_{n+1}\,Q_{n+3}}\over
{R_{n+1}\,R_{n+2}\,R_{n+3}\,R_{n+4}\,R_{n+5}}}\,=\,
{{Q_n\,Q_{n+2}}\over
{R_{n}\,R_{n+1}\,R_{n+2}\,R_{n+3}\,R_{n+4}}}\,=\,\lambda
\end{eqnarray}

Similarly to what has been recalled in section (\ref{rappel}),
 the first {\em finite order conditions} for the recurrences of class
 III happen
 to be simple relations when written in terms of these new variables
 $R_n$ and $Q_n$.
These finite order conditions read respectively for order three, four,
five, six and eight :
\begin{eqnarray}
R_{n}\,=\,0,  \,\,R_{n}\, =\, 1, \,\,Q_{n}\,=\,1,
\,\,Q_{n}\,=\,R_{n+1}\,\,\hbox{and}\,\, \,Q_{n}=0
\end{eqnarray}

\subsubsection{Demonstration of the recurrences for class IV}
\label{demrecIV}
Transposition $t$ for class IV permutes two entries of the same column
(or
of the same row). Let us, for instance, represent class IV by
transposition
$t$ which exchanges $m_{12}$ and $m_{32}$.

One has $\Delta_0=[R_{q}]_{3\,2}-[R_{q}]_{1\,2}=m_{32}-m_{12}$, and
matrix $P$ reads:
\begin{eqnarray}
\label{pIV}
P=\left [\begin {array}{ccccc}
{ 0}&{ 1}&{ 0}&{ 0}&\ldots\\
{ 0}&{ 0}&{ 0}&{ 0}&\ldots\\
{ 0}&{ -1}&{ 0}&{ 0}&\ldots\\
{ 0}&{ 0}&{ 0}&{ 0}&\ldots\\
\vdots & \vdots & \vdots & \vdots&\ddots
\end {array}\right ]
\end{eqnarray}
$\Delta_1(R_{q})$ still denotes $\Delta_0(\widehat{K}(R_{q}))$:
\begin{eqnarray}
\label{Delta1IV}
\Delta_1\,=\,[ \widehat{K}(R_{q}) ]_{3\,2} - [ \widehat{K}(R_{q})
]_{1\,2}
\,=\,[ \widehat{I}(R_{q}) ]_{1\,2} - [ \widehat{I}(R_{q}) ]_{3\,2}
\end{eqnarray}
Recalling relation (\ref{Uij1}), the determinant of matrix $U$ reads:
\begin{eqnarray}
\label{x0IV}
det(U)\,=\,x_0\,=\,1+\Delta_1\,T_0
\end{eqnarray}
where  $T_0=m_{23}-m_{21}$.
One has the following explicit form for $\widehat{K}^2(R_{q})$ yielding
$\Delta_2$
and $T_2$:
\begin{eqnarray}
\label{K2IV}
&&\widehat{K}^2(R_{q}) \, = \\
&&{{1}\over{x_0}}\quad\left [\begin {array}{cc}
 m_{11}-\Delta_{1}\,( m_{11}\, m_{23}-m_{21}\, m_{13})&
x_0\,m_{32}- \Delta_{1}\,m_{22}\,( m_{31}-m_{33})
\\
 m_{21}&
 m_{22}
\\
m_{31}-\Delta_{1}\,(m_{31}\, m_{23}-m_{21}\, m_{33})&
x_0\,m_{12}-\Delta_{1}\,m_{22}\,( m_{11}-m_{13})
\\
m_{41}-\Delta_{1}\,( m_{41}\, m_{23}-m_{21}\, m_{43})&
x_0\,m_{42}-\Delta_{1}\, m_{22}\,( m_{41}-m_{43})
\\
\vdots & \vdots
\end {array}\right . \nonumber \\
\noalign {\vskip .5cm}
&&\hskip 2.5cm\left .\begin {array}{ccc}
m_{13}-\Delta_{1}\,(m_{11}\, m_{23}- m_{21}\, m_{13})&
x_0\,m_{14}-\Delta_{1}\, m_{24}\,( m_{11}- m_{13})&\ldots
\\
m_{23}&
m_{24}&\ldots
\\
m_{33}-\Delta_{1}\,( m_{31}\, m_{23}-m_{21}\, m_{33})&
x_0\,m_{34}-\Delta_{1}\, m_{24}\,( m_{31}-m_{33})&\ldots
\\
m_{43}-\Delta_{1}\,( m_{41}\, m_{23}-m_{21}\, m_{43})&
x_0\,m_{44}-\Delta_{1}\, m_{24}\,( m_{41}-m_{43})&\ldots
\\
\vdots & \vdots&\ddots

\end {array}\right ] \nonumber
\end{eqnarray}
\begin{eqnarray}
\Delta_2&=& {
{{\Delta_1}\,{ m_{22}}\,\left ({ m_{33}}-{ m_{31}}+{ m_{11}}-{ m_{13}}
\right )} \over
{{ x_0}}}
\,+\,{ m_{32}}\,-\,{ m_{12}} \nonumber \\
&=&
{{{ \Delta_1}\,{ m_{22}}\,\left ({ m_{33}}-{ m_{31}}+{ m_{11}}-{
m_{13}} \right )} \over
{{ x_0}}}
\,-\,{ \Delta_0}\, ,\,  \nonumber \\
T_2&=& {{T_0}\over{x_0}}
\end{eqnarray}

Similarly, one can write how various quantities such as
  $\left ({ m_{21}} - { m_{23}}\right ), \,{ m_{22}}, \, \left ({
  m_{11}}
-{ m_{13}}\right)$ and $\, \left ( { m_{33}}-{m_{31}}\right )$
  transform under  $\widehat{K}^{2}$:
\begin{eqnarray}
\label{recfinIV}
{ m_{22}}& \rightarrow &{{ m_{22}}\over x_0}\, , \, \nonumber \\
{ m_{33}}-{ m_{31}}& \rightarrow &{{ m_{33} -  m_{31}} \over {x_0}}\, ,
\, \nonumber \\
{ m_{13}}-{ m_{11}}& \rightarrow &{{ m_{13}- m_{11}}\over{ x_0}}
\end{eqnarray}
These various quantities can easily be eliminated, yielding:
\begin{eqnarray}
\label{del4}
{{(\Delta_4+\Delta_2)\,\,x_2}\over
{\Delta_3}}&=&{{(\Delta_2+\Delta_0)}\over
{x_0\,\Delta_1}}
\end{eqnarray}
and
\begin{eqnarray}
\label{u2del}
{ x_2}&=&1+{ {{ \Delta_3}\,\left ({ x_0}-1\right )}\over{{ \Delta_1}\,{
x_0}}}
\end{eqnarray}

Let us now introduce a new variable $\delta_n$:
\begin{eqnarray}
\delta_{n}\,\,=\,\,{\Delta_{n+2} \over \Delta_{n}}
\end{eqnarray}

Equations (\ref{del4}) and (\ref{u2del}) respectively read:
\begin{eqnarray}
\label{deltan}
\delta_{n+1}\,=\,{\frac { x_{n}\,\left ( x_{n+2}-1\right )}{ x_{n}-1}}
\\
\label{deltan2}
{\frac { \delta_{n+2}+1}{ \delta_{n+1}}}\,\,=\,\,{\frac { 1 +
\delta_n}{ x_{n}\, x_{n+2} \, \delta_n}}
\end{eqnarray}

Eliminating the variable $\delta_n$ in relations (\ref{deltan}) and
(\ref{deltan2}), {\em one recovers recurrence} (\ref{xnIV}) bearing on
a single variable $x_n$:
\begin{eqnarray}
\label{un4}
{{x_{n+3} -1}\over{x_{n+2}\; x_{n+4}-1}}\,=\,{{x_{n+1} -1}\over{x_n \;
x_{n+2}-1}} \cdot x_n \; x_{n+3}
\end{eqnarray}

Eliminating the variable $x_n$ in equations (\ref{deltan}) and
(\ref{deltan2}), {\em one also gets a recurrence bearing on another
single variable}:
\begin{eqnarray}
\label{qn4}
&&\hskip -1cm\left ( \delta_{n}+1\right )\left ( \delta_{n+1}+1\right
)\left ( \delta_{n+3}+1\right )\left ( \delta_{n+4}+1\right )
\delta_{n+2}\,= \\ \nonumber
&& \delta_{n+3}\,\left ( \delta_{n}\, \delta_{n+2}+2\,
\delta_{n}+1\right )\left ( \delta_{n+4}\, \delta_{n+2}+2\,
\delta_{n+2}+1\right )
\end{eqnarray}

One can now study finite order orbits. They read respectively for order
four and order six :

- orbit of order four:
\begin{eqnarray}
 x_{n}\, x_{n+2}-1\,=\,0
\end{eqnarray}
and :
\begin{eqnarray}
1+ x_{n}+x_{n+1}- x_{n}\, x_{n+1}\,=\,0
\end{eqnarray}
that is :
\begin{eqnarray}
 \delta_{n}\, \delta_{n+1}+ \delta_{n}- \delta_{n+1}+1\,=\,0
\end{eqnarray}

- orbit of order six:
\begin{eqnarray}
\label{labelme}
\delta_{n+1}\,\delta_{n+4}-\delta_{n} \,\delta_{n+3} \, = \, 0 \, \, ,
\,\,\,\,\,
\hbox{or}  \,\,\,\,\,  x_{n}\, x_{n+2}- x_{n+1}=0
\end{eqnarray}

\subsubsection{Demonstration of the recurrences for class V}

Let us represent class V, by transposition $t$ exchanging $m_{11}$ and
$m_{23}$.
Then (\ref{delta0ij}) reads

$\Delta_0=[R_{q}]_{2\,3}-[R_{q}]_{1\,1}=m_{23}-m_{11}$, and $P$ given
by (\ref{Pij}) becomes:
\begin{eqnarray}
P=\left [\begin {array}{ccccc}
1&0&{ 0}&{ 0}&\ldots\\
{ 0}&{ 0}&-1&{ 0}&\ldots\\
0&{ 0}&{ 0}&{ 0}&\ldots\\
{ 0}&{ 0}&{ 0}&{ 0}&\ldots\\
\vdots & \vdots & \vdots & \vdots&\ddots
\end {array}\right ]
\end{eqnarray}
Moreover $\Delta_1$ reads:
\begin{eqnarray}
\label{Delta1V}
\Delta_1\,=\,[ \widehat{K}(R_{q}) ]_{2\,3} - [ \widehat{K}(R_{q})
]_{1\,1}
\,=\,[ \widehat{I}(R_{q}) ]_{1\,1} - [ \widehat{I}(R_{q}) ]_{2\,3}
\end{eqnarray}
This transposition perturbing two columns (and two rows), $x_0$ is a
quadratic expression in terms of $\Delta_1$ and reads:
\begin{eqnarray}
\label{x0V}
det(U)\,=\,x_0\,=\,1+T_{0}\,\Delta_{1}+N_0 \,\Delta_{1}^{2}
\end{eqnarray}
with $N_0=\left (m_{31}\,m_{12}-m_{11}\,m_{32}\right )\,$ and
$\,T_0=m_{32}-m_{11}$.

Relation (\ref{IUij}) yields:
\begin{eqnarray}
\label{IUV}
x_0&\cdot&\widehat{I}(U)=\\
&&\left [\begin {array}{ccccc}
\Delta_{1}\, m_{32}+1&0&- \Delta_{1}\, m_{12}&0&\ldots\\
\Delta_{1}\,\left ( \Delta_{1}\, m_{21}\, m_{32}+ m_{21}- \Delta_{1}\,
m_{31}\, m_{22}\right )& x_0& \Delta_{1}\,\left ( \Delta_{1}\, m_{22}\,
m_{11}- m_{22}- \Delta_{1}\, m_{21}\, m_{12}
\right )&0&\ldots\\
\Delta_{1}\, m_{31}&0&1- \Delta_{1}\, m_{11}&0&\ldots\\
- \Delta_{1}\,
\left ( \Delta_{1}\, m_{31}\, m_{42}- \Delta_{1}\, m_{41}\, m_{32}-
m_{41}\right )&0&
\Delta_{1}\,\left ( \Delta_{1}\, m_{42}\, m_{11}- m_{42}- \Delta_{1}\,
m_{41}\, m_{12}\right )& x_0&\ldots\\
\vdots & \vdots & vdots & \vdots & \ddots
\end {array}
\right ]\nonumber
\end{eqnarray}

Matrix $\widehat{K}^2(R_q)$ is obtained from relation (\ref{K2ij}).
Its explicit form is quite involved, and will not be given here.

We will just concentrate on a {\em fixed} number of variables, enabling
to
understand the evolution of $T_0, \, N_0$ and $\Delta_0$,
the action of $\widehat{K}^2$ preserving this set of variables.
\begin{eqnarray}
\label{K2Vr}
m_{11}\, \rightarrow \, \, \,
{{m_{11}-\Delta_1\,N_0}\over{x_0}}-\Delta_2 \, , \, \nonumber \\
m_{12}\, \rightarrow \,\,\, {{m_{12}}\over{x_0}}  \, , \,\nonumber \\
m_{31} \,  \rightarrow  \,\,\, {{m_{31}}\over{x_0}}  \, , \,\nonumber
\\
m_{32} \, \rightarrow \,\,\, m^{<2>}_{32}\,=\,
{{m_{32}+\Delta_1\,N_0}\over{x_0}}
\end{eqnarray}
{}From (\ref{K2Vr}) one gets the following equations:
\begin{eqnarray}
\label{t2V}
T_2&=&{{T_0+2\,\Delta_1\,N_0}\over{x_0}}+\Delta_2
\end{eqnarray}
Finally, coming back to equations (\ref{x0V}) and (\ref{t2V}), one can
eliminate the $T_n$'s and get the following equation:
\begin{eqnarray}
\label{xdV}
{\frac {1- x_{n+2}}{ \Delta_{n+3}}}\,-\,{\frac {1- x_{n}}{ x_{n}\,
\Delta_{n+1}}
}\,+\, \Delta_{n+2}\,-\, \Delta_{n+3}\, N_{n+2}\,-\,{\frac {
\Delta_{n+1}\, N_{n}}{ x_{n}}}\,\,=\,\,0
\end{eqnarray}
On the other hand $N_0$ transforms as follows:
\begin{eqnarray}
\label{n2V}
N_2\,\,=\,\,{{N_0\,(1+\Delta_1\,T_0+\Delta_1^2\,N_0)}\over{x_0^2}}\,+\,\Delta_2\,m^{<2>}_{32}
\,\,=\,\,{{N_0}\over{x_0}}+\Delta_2\,m^{<2>}_{32}
\end{eqnarray}
{}From (\ref{K2Vr}) and relation (\ref{n2V}) one can eliminate the
$m^{<n>}_{32}$'s
and get:
\begin{eqnarray}
\label{ndV}
{\frac { N_{n+4}}{ \Delta_{n+3}\, \Delta_{n+4}}}- N_{n+2}\,\left
(1+{\frac {1}{ \Delta_{n+3}\, \Delta_{n+4}}}+{\frac {1}{ \Delta_{n+2}\,
\Delta_{n+3}}}\right ) x_{n+2}^{-
1}+{\frac { N_{n}}{ x_{n}\, x_{n+2}\, \Delta_{n+3}\,
\Delta_{n+2}}}\,=\,0
\end{eqnarray}

Moreover $\Delta_{n+2}$ satisfies the following relation:
\begin{eqnarray}
\label{delta2V}
{\frac { x_{n}\, \Delta_{n+2}\,+\, \Delta_{n}}{ \Delta_{n+1}}}\,-\,
N_{n}-
P_{n}\,- \, Q_{n}\,+\, \Delta_{n+1}\,
R_{n}\,=\,0
\end{eqnarray}
where:
\begin{eqnarray}
\label{defvar}
P_0\,=\, \det(M_1), \quad Q_0=\det(M_2),\quad R_0=\det(M_3)
\end{eqnarray}
with:
\begin{eqnarray}
\label{M2}
M_1=
\left [\begin {array}{cccc}
m_{22}&m_{23}\\
m_{32}&m_{33}
\end {array}\right ]
, \quad
M_2=
\left [\begin {array}{cccc}
m_{11}&m_{13}\\
m_{21}&m_{23}
\end {array}\right ]
, \quad
M_3=
\left [\begin {array}{cccc}
m_{11}&m_{12}&m_{13}\\
m_{21}&m_{22}&m_{23}\\
m_{31}&m_{32}&m_{33}
\end {array}\right ]
\end{eqnarray}

\begin{eqnarray}
\label{xqrdV}
 Q_{n+2}\,-\,{\frac { Q_{n}}{ x_{n}}}\,+\,{\frac { \Delta_{n+1}\,
 R_{n}}{ x_{n}}}\,\,=\,0
\end{eqnarray}

\begin{eqnarray}
\label{rnxdpV}
 R_{n+2}\,-\,{\frac { R_{n}}{ x_{n}}}\,+\,{\frac { \Delta_{n+2}\,
 N_{n}}{ x_{n}}}\,+\, \Delta_{n+2}\, P_{n+2}\,\,=\,0
\end{eqnarray}

\begin{eqnarray}
\label{xdpnrV}
 P_{n+2}\,-\,{\frac { P_{n}}{ x_{n}}}\,+\,{\frac { \Delta_{n+1}\,
 R_{n}}{ x_{n}}}\,+\,{\frac { N_{n}}{ x_{n}}}- N_{n+2}\,=\,0
\end{eqnarray}
One directly gets  from relations (\ref{xqrdV}) and (\ref{xdpnrV}) a
covariant
expression, enabling to easily eliminate  one variable:
\begin{eqnarray}
\label{ts9}
 Q_{n+2}+N_{n+2}-P_{n+2}\,\,=\,\,{\frac { Q_{n}+N_{n}-P_{n}}{ x_{n}}}
\end{eqnarray}
The elimination of the variables $P_n$, $Q_n$ and $R_n$ in this set of
recurrences ((\ref{delta2V}), (\ref{xqrdV}) and (\ref{rnxdpV}),
(\ref{xdpnrV})) can be performed and yield a third equation between the
$x_n$'s, $\Delta_n$'s and $N_n$'s (the two other equations being
(\ref{xdV}) and (\ref{ndV})).
The elimination of the $N_n$'s yield two equations, relating the
$x_n$'s and the $\Delta_n$'s, of the form:
\begin{eqnarray}
\label{ABCD}
&&A_n+B_n\cdot x_n+C_n\cdot x_n\,x_{n+2}+D_n\cdot
x_n\,x_{n+2}\,x_{n+4}+E_n\cdot x_n\,x_{n+2}\,x_{n+4}\,x_{n+6}\,=\,0
\\
\hbox{and:} \nonumber \\
\label{FGH}
&&F_n+G_n\cdot x_n+H_n\cdot x_n\,x_{n+2}+I_n\cdot
x_n\,x_{n+2}\,x_{n+4}+J_n\cdot x_n\,x_{n+2}\,x_{n+4}\,x_{n+6}\nonumber
\\
&&\,\,+K_n\cdot x_n\,x_{n+2}\,x_{n+4}\,x_{n+6}\,x_{n+8}\,=\,0
\end{eqnarray}
where the $A_n, B_n, \cdots, K_n$ are polynomials in terms of the
$\Delta_n$'s.
One can, in principle, eliminate the $x_n$'s to get an algebraic
relation
between the $\Delta_n$'s only.
The coefficients of these last two
equations are too involved expressions to enable to perform such an
elimination.

Let us just note that, though such a system of recurrences (\ref{ABCD})
and
(\ref{FGH}) is quite involved, one can however get some  finite order
conditions for these recurrences of class V, namely the orbits of order
four :
\begin{eqnarray}
\label{delxfour}
&& \Delta_{n+1}\, \Delta_{n+2}+ \Delta_{n+1}\, \Delta_{n}+
\Delta_{n+2}\,
\Delta_{n+3}+ \Delta_{n+3}\, \Delta_{n}+ \Delta_{n+1}\, \Delta_{n+3}\,
\Delta_{n}\, \Delta_{n+2}\,=\,0\,, \,  \nonumber \\
&&\hbox{or} \, \, \, \, \, \, \, \, \, \, \,  x_{n+2}\,
x_{n}-1\,\,=\,\,0
\end{eqnarray}

\subsubsection{Demonstration of the recurrences for class VI}

Let us now consider a transposition of class VI. It permutes two
entries in
the same column, or row. Let us take for example a transposition
perturbing
only one row: $t$ denotes the
transposition exchanging the entries $m_{11}$ and $m_{12}$ of matrix
$R_q$.
$\Delta_0$ now reads:
$\Delta_0=[R_{q}]_{2\,1}-[R_{q}]_{1\,1}=m_{21}-m_{11}$, and $P$ denotes
the
following matrix:
\begin{eqnarray}
P=\left [\begin {array}{ccccc}
1&-1&{ 0}&{ 0}&\ldots\\
{ 0}&{0}&{ 0}&{ 0}&\ldots\\
0&{ 0}&{ 0}&{ 0}&\ldots\\
{ 0}&{ 0}&{ 0}&{ 0}&\ldots\\
\vdots & \vdots & \vdots & \vdots&\ddots
\end {array}\right ]
\end{eqnarray}
For this transposition $T_0=m_{12}-m_{11}$, and since $t$ perturbes a
row
(and not a column as in the previous demonstration)
matrix $U$ will denote:
$U=\widehat{K}(R_{q}) \cdot R_{q}$ which reads (instead of $U=R_{q}
\cdot \widehat{K}(R_{q})$):
\begin{eqnarray}
\label{UVI}
U\,=\,\,{\cal I}d_{q}-\Delta_1 \;\cdot \Bigl( P \cdot R_{q} \Bigr)
\end{eqnarray}
The action of $\widehat{I}$ on $U$ reads:
\begin{eqnarray}
\label{IUVI}
\widehat{I}(U)\,=\,\,{\cal I}d_{q}+{{\Delta_1 \;\cdot \Bigl( R_{q}
\cdot P \Bigr)}\over{x_0}}
\end{eqnarray}
and equation (\ref{K2ij}) becomes:
\begin{eqnarray}
\label{K2VI}
\widehat{K}^{2}(R_{q})\,=\, t(\widehat{I}(\widehat{K}(R_{q})))\,=\,
t(R_{q} . \widehat{I}(U))
\end{eqnarray}
{}From relations (\ref{UVI}), (\ref{IUVI}) and (\ref{K2VI}) one gets the
explicit form of
$\widehat{K}^2(R_{q})$, namely:
\begin{eqnarray}
\label{K2I}
x_0\,\widehat{K}^2(R_{q})=
\left [\begin {array}{cccc}
\left (m_{12}-\Delta_1\,P_0\right )&
m_{11}&
\left (m_{13}-\Delta_1\,Q_0\right )&\ldots\\
m_{21}&
\left (m_{22}-\Delta_1\,P_0\right )&
\left (m_{23}-\Delta_1\,Q_0\right )&\ldots\\
m_{31}&
m_{32}\,x_0+\Delta_1\,m_{31}\,(m_{12}-m_{22})&
m_{33}\,x_0+\Delta_1\,m_{31}\,(m_{13}-m_{23})&\ldots\\
\vdots & \vdots & \vdots &\ddots
\end {array}\right ] \nonumber \\
\end{eqnarray}
where $P_0$ and $Q_0$ read:
\begin{eqnarray}
\label{N0I}
P_0=m_{11}\,m_{22}-m_{12}\,m_{21}, \quad
Q_0=m_{11}\,m_{23}-m_{13}\,m_{21}
\end{eqnarray}

{}From definition (\ref{x0ij1}), $x_0$, the determinant of matrix $U$,
reads :
\begin{eqnarray}
\label{x0I}
x_0\,=\,\,1+\,\Delta_1\,T_0
\end{eqnarray}
One thus gets the following action of $\widehat{K}^2$ on a {\em finite
set}
of homogeneous variables:
\begin{eqnarray}
\label{K2Ir}
m_{11}\, \rightarrow \,\, {{m_{12}-\Delta_1\,P_0}\over{x_0}}\,\, , \,
\nonumber \\
m_{12}\, \rightarrow \, \, {{m_{11}}\over{x_0}}\,\, , \, \nonumber \\
m_{21}\, \rightarrow \,\,  {{m_{21}}\over{x_0}}\,\, , \, \nonumber \\
m_{22}\, \rightarrow \,\, {{m_{22}-\Delta_1\,P_0}\over{x_0}}
\end{eqnarray}
Relation (\ref{K2Ir}) yields:
\begin{eqnarray}
\label{d2I}
\Delta_2\,\,=\,\,{{-\Delta_0+\Delta_1\,P_0}\over{x_0}}
\end{eqnarray}
Relation (\ref{x0I}) yields with a shift of two (that is replacing
matrix
$R_q$ by $\widehat{K}^2(R_q)$):
\begin{eqnarray}
\label{x2I}
x_2&=&1+\Delta_3\,{{m_{21}-m_{12}+\Delta_1\,P_0}\over{x_0}} \nonumber
\\
&=&1+\Delta_3\,{{m_{21}-m_{11}+m_{11}-m_{12}+\Delta_1\,P_0}\over{x_0}}\nonumber
\\
&=&1+\Delta_2\,\Delta_3+{{\Delta_3\,(x_0-1)}\over{x_0\,\Delta_1}}
\end{eqnarray}
Similarly equation (\ref{N0I}) gives after a shift of two:
\begin{eqnarray}
\label{N2I}
P_2&=&{{1}\over{x_0^2}}\cdot
\Bigl((m_{11}+\Delta_0-\Delta_1\,P_0)\,(m_{22}-\Delta_1\,P_0)-(m_{12}-\Delta_0)\,m_{21}\Bigr)\nonumber
\\
&=&{{1}\over{x_0^2}}\cdot
\Bigl(P_0\,(1-\Delta_1\,(m_{11}-m_{21}))+(m_{21}+m_{22}-\Delta_1\,P_0)\,(\Delta_0\,-\Delta_1\,P_0)\Bigr)\nonumber
\\
&=&{{1}\over{x_0^2}}\cdot
\Bigl(P_0\,x_0-\Delta_2\,x_0\,(m_{21}+m_{22}-\Delta_1\,P_0)\Bigr)\nonumber
\\
&=&{{1}\over{x_0}}\cdot
\Bigl(P_0-\Delta_2\,(m_{21}+m_{22}-\Delta_1\,P_0)\Bigr)
\end{eqnarray}
One immediately gets from equation (\ref{K2Ir}):
\begin{eqnarray}
m_{21}+m_{22} \, \rightarrow \,
{{m_{21}+m_{22}-\Delta_1\,P_0}\over{x_0}}
\end{eqnarray}
and from equation (\ref{N2I}):
\begin{eqnarray}
\label{N4Ix}
P_4&=&{{1}\over{x_2}}\,\Bigl(P_2-\Delta_4\,(({{m_{21}+m_{22}-\Delta_1\,P_0}\over{x_0}})-\Delta_3\,P_2)\Bigr)\nonumber
\\
&=&{{1}\over{x_2}}\,\Bigl(P_2\,(1+{{\Delta_4}\over{\Delta_2}})+{{\Delta_4\,P_0}\over{\Delta_2\,x_0}}-\Delta_4\,(\Delta_2+\Delta_4\,x_2)\Bigr)
\end{eqnarray}
Equation (\ref{d2I}) can be written as follows:
\begin{eqnarray}
\label{N0Ix}
P_0\,\,=\,\,{{x_0\,\Delta_2+\Delta_0}\over{\Delta_1}}
\end{eqnarray}
Relation (\ref{N0Ix}) shifted by two, or four, respectively gives $P_2$
and $P_4$ in
terms of $x_n$'s and $\Delta_n$'s.
In fact the $\Delta_n$'s only appear through their products
$\Delta_n\,\Delta_{n+1}$, we will denote $p_n$ in the following.
Replacing $P_0$, $P_2$ and $P_4$ in relation (\ref{N4Ix}) one gets for
arbitrary $n$:
\begin{eqnarray}
\label{xpI}
 x_{n+4}\, p_{n+5}+ p_{n+4}- p_{n+4}^{2}\left (1+{\frac { p_{n+2}}{
 p_{n+3}\, x_{n+2}}}\right )\left ( p_{n+2}^{-1}+ p_{n+3}^{-1}+1\right
)+{\frac { p_{n+4}^{2} p_{n+2}\,\left ( x_{n}\, p_{n+1}+ p_{n}
\right )}{ p_{n+3}\, p_{n+1}^{2} x_{n}\, x_{n+2}}}=0
\end{eqnarray}
Besides, relation (\ref{x2I}) can also be written in terms of the
$p_n$'s and reads:
\begin{eqnarray}
\label{x2Ip}
{\frac {1- x_{n+2}}{ p_{n+2}}}\,\,=\,\, {\frac {1- x_{n}}{ x_{n}\,
p_{n+1}}
}-1
\end{eqnarray}

The elimination of the $x_n$'s in equations (\ref{xpI}) and
(\ref{x2Ip})
can be performed but it yields a ``huge'' algebraic relation between
the
$p_n$'s.
In principle one should also be able to get another algebraic relation
between the  $x_n$'s, but the calculations are too large.

Let us note that
these algebraic relations between the $p_n$'s, or $x_n$'s, are not (at
first
sight !) recurrences, however for particular conditions on the initial
$p_n$'s or
$x_n$'s one can actually get a (quite involved) recurrence on the
$x_n$'s.

Let us now write the finite order orbits in terms of the variables
$p_n$'s or $x_n$'s for order three and four respectively :

- order three :
\begin{eqnarray}
 p_{n+2}\, p_{n}\, p_{n+1}+ p_{n+2}\, p_{n+1}+ p_{n+2}\, p_{n}+ p_{n}\,
 p_{n+1}\,=\,0
\end{eqnarray}
and :
\begin{eqnarray}
2\, x_{n}\, x_{n+1}\, x_{n+2}- x_{n}\, x_{n+2}- x_{n+1}\, x_{n+2}+1-
x_{n}\, x_{n+1}\,=\,0 \quad
\hbox{or} \quad
x_{n}\, x_{n+1}\, x_{n+2}-1\,=\,0
\end{eqnarray}

- order four :
\begin{eqnarray}
 p_{n+3}\, p_{n+2}\, p_{n}\, p_{n+1}+ p_{n+3}\, p_{n+2}\, p_{n}+
 p_{n+3}\,
p_{n}\, p_{n+1}+ p_{n+3}\, p_{n+2}\, p_{n+1}
+ p_{n+2}\, p_{n}\, p_{n+1}\,=\,0
\end{eqnarray}
\begin{eqnarray}
\hbox{or} \quad \quad  x_{n}\, x_{n+1}\, x_{n+2}\, x_{n+3}-1\,= \,0
\end{eqnarray}

We have given in this last section examples of
conditions corresponding to finite orbits for the
iteration of $K$, for various classes, even when no simple  recurrence
on a single variable exists.
These expressions and these calculations very much depend  on the
specific
birational transformations one considers.
In fact it can be shown that finite order conditions can be obtained in
a
quite general framework not associated to transpositions
of two entries anymore. In this respect the example of finite order
conditions of order
four, given in Appendix A, is quite illuminating.

\section{Class IV revisited}

Among the  classes with exponential growth, class IV is singled out, as
far as two
particular properties are concerned.
On one hand, the mappings of this class have $q^2-3\,$ algebraic
invariants.
More precisely the algebraic varieties, the equations of which
correspond to
these invariants, are actually {\em planes} ({\em depending of course
of the initial
point of} $\C P_{q^2-1}$).
On the other hand, simple recurrences, bearing on the $x_n$'s, emerge.
These recurrences have been shown to have an integrable
subcase~\cite{BoMaRo93d}.

We will now study in detail this class. In particular, we will show
{\em how
to associate with
these mappings in} $\C P_{q^2-1}$, {\em mappings in} $\C P_{2}$, which
are closely related
to the recurrences in the $x_n$'s.
Finally, we will show how the {\em integrability modifies the
factorization
properties}
and provides an example of integrable mapping for arbitrary $q$.

\subsection{Class IV as a  mapping in  two variables}
\label{planes}

{}From equations (\ref{un4}) and (\ref{qn4}) or (\ref{labelme}), one may
have
 the ``prejudice'' that the orbits of transformation
 $K$ in $\C P_{15}$ (or $\C P_{q^2-1}$) should be
 curves~\cite{BoMaRo93d}.
In fact it has been shown in~\cite{BoMaRo93d} that, in some
domain of the parameter space $\C P_{15}$ (or $\C P_{q^2-1}$), these
orbits
look like {\em curves} which may explode in some algebraic surface.

These {\em algebraic surfaces are actually planes}~\cite{BoMaRo93d}
 (depending on the initial matrix one iterates).
This can be shown as follows, coming back to the action of
$\widehat{K}^2$
on a generic matrix $R_q$ (see equation (\ref{K2IV})), which can be
written in the following way:
\begin{eqnarray}
\label{K2p}
\widehat{K}^{2}(R_q)\,=\,{{1}\over {x_0}}\cdot (R_q+\Delta_1\,F+b_1\,P)
\end{eqnarray}
where matrix $P$ still denotes the constant matrix given in (\ref{pIV})
and $b_1$ reading:
\begin{eqnarray}
b_1\,=\, {{x_0\,\Delta_2-\Delta_0}\over{2}}
\end{eqnarray}
and $F$ denotes a $q \times q$ matrix, quadratic in the entries of
matrix $R_q$:
\pagebreak
\begin{eqnarray}
\label{defF}
&&F= \nonumber\\ \nonumber\\
&&\left (
\begin {array}{cc}
m_{21}\, m_{13}-
m_{11}\,m_{23}&((m_{12}+m_{32})(m_{21}-m_{23})+m_{22}(m_{13}-m_{11}+m_{33}-m_{31}))/2\\
0&0\\
m_{21}\, m_{33}- m_{31}\,
m_{23}&((m_{12}+m_{32})(m_{21}-m_{23})+m_{22}(m_{13}-m_{11}+m_{33}-m_{31}))/2\\
m_{21}\, m_{43}- m_{41}\, m_{23}&
m_{42}(m_{21}-m_{23})+m_{22}(m_{43}-m_{41})\\
\vdots & \vdots
\end {array} \nonumber
\right .\\
\noalign {\vskip .5cm}
&&\left .
\hskip 5cm
\begin {array}{ccc}
m_{21}\, m_{13}-
m_{11}\,m_{23}&m_{14}(m_{21}-m_{23})+m_{24}(m_{13}-m_{11})&\ldots\\
0&0&\ldots\\
m_{21}\, m_{33}- m_{31}\,
m_{23}&m_{34}(m_{21}-m_{23})+m_{24}(m_{33}-m_{31})&\ldots \\
m_{21}\, m_{43}- m_{41}\,
m_{23}&m_{44}(m_{21}-m_{23})+m_{24}(m_{43}-m_{41}) &\ldots\\
\vdots & \vdots & \ddots
\end {array}
\right ) \nonumber
\end{eqnarray}
One will recursively show that the successive iterates of
$\widehat{K}^2$ read:
\begin{eqnarray}
\label{K2np}
\widehat{K}^{2\,n}(R_q)\,=\,{{1}\over {x_0\,x_2\ldots
x_{2\,n-2}}}\;\cdot (R_q+a_n\,F+b_n\,P)
\end{eqnarray}
$F$ denotes the same matrix as in relation (\ref{K2p}) : $F$ {\em does
depend on} $R_q$,
{\em but not on the order $n$ of the iteration}.
In other words all the iterates of $\widehat{K}^2$ {\em lie in a plane
which depends on the initial matrix} $R_q$ (or equivalently, on any
other ``even'' iterates of $R_q$).
This plane is led by two vectors, namely a {\em fixed vector} $P$
 and another one $F$, {\em depending on the initial matrix}.

In order to show recursively relation (\ref{K2np}), let us perform the
{\em
right-action} of $\widehat{K}^2$ on equation (\ref{K2np}).
One gets:
\begin{eqnarray}
\label{K2n+2p}
\widehat{K}^{2\,n+2}(R_q)\,\,=\,\,{{1}\over {x_2\,x_4\ldots
x_{2\,n}}}\; \cdot
\Bigl(\widehat{K}^2(R_q)+(a_n)_{\widehat{K}^2}\,(F)_{\widehat{K}^2}+(b_n)_{\widehat{K}^2}\,P
\Bigr)
\end{eqnarray}
Matrix $(\widehat{K}^2)(R_q)$ is given by equation (\ref{K2p}).
Straight calculations show that:
\begin{eqnarray}
(F)_{\widehat{K}^2}\,\,\,=\,\,\,{{F}\over{x_0}}
\end{eqnarray}
The right hand side of (\ref{K2n+2p}) thus reads:
\begin{eqnarray}
\label{final}
\widehat{K}^{2\,n+2}(R_q)\,=\,{{1}\over {x_0\,x_2\ldots x_{2\,n}}}\cdot
\Bigl(R_q+(\Delta_1+(a_n)_{\widehat{K}^2})\,F+(b_1+x_0\,(b_n)_{\widehat{K}^2})\,P
\Bigr)
\end{eqnarray}
which is of the same form as (\ref{K2np}) with the following definition
of  $a_{n+1}$ and $b_{n+1}$:
\begin{eqnarray}
a_{n+1}=\,\Delta_1+(a_n)_{\widehat{K}^2}, \quad
b_{n+1}=\,b_1+(b_n)_{\widehat{K}^2}\cdot x_0
\end{eqnarray}
Defining $a_1=\Delta_1$, the successive $a_n$'s read:
\begin{eqnarray}
a_n=\Delta_1+\Delta_3+\ldots+\Delta_{2\,n-1}
\end{eqnarray}
With this expression of the $a_n$'s, one  notices that $1+T_0\,a_n$  is
directly related to the $x_n$'s:
\begin{eqnarray}
\label{1pt0an}
1+T_0\,a_n\,\,=\,\,x_0\,x_2\,\ldots\,x_{2\,n-2}
\end{eqnarray}
One can also give the expression of the successive $b_n$'s:
\begin{eqnarray}
b_n={{((1+a_n\,T_0)\,\Delta_{2\,n}-\Delta_0)}\over{2}}
\end{eqnarray}
In order to obtain $a_{n+1}$ and $b_{n+1}$ in terms of $a_n$ and $b_n$,
one can consider a generic matrix:
\begin{eqnarray}
A={{1}\over{(1+a\,T_0)}}\cdot (R_q+a\,F+b\,P) \nonumber
\end{eqnarray}
and get $\widehat{K}^2(A)$. Since it is necessarily of the form:
\begin{eqnarray}
\widehat{K}^2(A)={{1}\over{(1+a'\,T_0)}}\cdot (R_q+a'\,F+b'\,P)
\nonumber
\end{eqnarray}
one obtains $a'$ and $b'$ in terms of $a$ and $b$.
In fact, these calculations are quite heavy (see for instance Appendix
B) and it is simpler to use the
recurrences on the $x_n$'s or more precisely the recurrences on the
homogeneous variables $q_n$'s:
\begin{eqnarray}
\label{xq}
x_{n}\, = \, {{q_{n+2}}\over{q_n}}
\end{eqnarray}
{}From recurrence (\ref{xnIV}) bearing on the $x_n$'s, one recovers the
``almost integrable'' recurrence studied in section (8)
of~\cite{BoMaRo93d}:
\begin{eqnarray}
\label{qnIV}
 q_{n+3}\,q_{n+5}\quad
{{q_{n+6}-q_{n+2}}\over{\left ( q_{n+3} - q_{n+5}\right )}}
\,\, =\,\,
 q_{n+1}\,q_{n+3}\quad
{{q_{n+4}-q_{n}}\over{\left ( q_{n+1} - q_{n+3} \right )}}
\end{eqnarray}
which can be partially integrated (see equation (8.18)
in~\cite{BoMaRo93d})
as follows:
\begin{eqnarray}
\label{intalm2}
q_{2\,n+2}+q_{2\,n}+{{\lambda_2}\over {q_{2\,n+1}}}=\rho_2 \\
\label{intalm1}
q_{2\,n+3}+q_{2\,n+1}+{{\lambda_1}\over {q_{2\,n+2}}}=\rho_1
\end{eqnarray}
Let us first give the correspondence between the variables of the
mapping
in $\C P_{q^2-1}$, and the variables of the recurrences (the $q_n$'s,
$\lambda_1$ and $\lambda_2$).

{}From relation (\ref{1pt0an}) one directly gets the $q_n$'s, for $n$
even:
\begin{eqnarray}
\label{q2nan}
q_{2n}\,=\,x_0\,x_2\,\ldots\,x_{2\,n-2}\cdot q_0 \,=\,(1+T_0\,a_n)\cdot
q_0
\end{eqnarray}
One also has $\,\,q_{2n+1}\,=\,x_1\,x_3\,\ldots\,x_{2\,n-1} \cdot
q_1\,\,$, and
recalling the definition of the $x_n$'s:
\begin{eqnarray}
x_n=\det(\widehat{K}^n(R_q))\cdot\det(\widehat{K}^{n+1}(R_q))\nonumber
\end{eqnarray}
one can write the $q_n$'s, for $n$  odd, as:
\begin{eqnarray}
q_{2n+1}\,=\, x_0\,x_2\,\ldots\,x_{2\,n-2} \cdot
{{\det(\widehat{K}^{2\,n}(R_q))}\over {\det(R_q)}}\,\cdot q_1
\,=\,(1+T_0\,a_n)\cdot
{{\det(\widehat{K}^{2\,n}(R_q))}\over {\det(R_q)}}\cdot q_1
\end{eqnarray}
Moreover one has the following  remarkable relation:
\begin{eqnarray}
\label{detanbn}
det(\widehat{K}^{2\,n}(R_q))\,\,=\,\,  {{1}\over
{(1+T_0\,a_n})^{2}}\cdot \Bigl(
det(R_q)+a_n\, {\cal P}_1+b_n\, {\cal P}_2 \Bigr)
\end{eqnarray}
It is not surprising, since the rank of
matrix $P$ is one, that equation (\ref{detanbn}) is {\em linear} in
term of $b_n$.
In contrast its expression in term of $a_n$ is anything but obvious.

The $q_n$'s, for $n$  odd, finally read:
\begin{eqnarray}
\label{q2np1an}
q_{2n+1}\,=\, {{1}\over{(1+T_0\,a_n)}} \cdot \Bigl(1 + a_n \,{{{\cal
P}_1}\over
{\det(R_q)}} + b_n \,{{{\cal P}_2}\over {\det(R_q)}} \Bigr) \cdot q_1
\end{eqnarray}
One now has to get explicit expressions for ${\cal P}_1$ and ${\cal
P}_2$.

Let us first recall $\,T_1=T_0(\widehat{K}(R_q))$, one remarks that one
has:
\begin{eqnarray}
\label{p2}
{{{\cal P}_2}\over {\det(R_q)}}\,=\,-\,T_1
\end{eqnarray}
On the other hand, one notices that recurrences (\ref{intalm2}) and
(\ref{intalm1}) can
also be written, eliminating $\rho_1$ and $\rho_2$, as:
\begin{eqnarray}
\lambda_2\,=\,q_{2\,n}\,q_{2\,n+1}\cdot
{{(x_{2\,n}\,x_{2\,n+2}-1)\,\,x_{2\,n+1}}\over{1-x_{2\,n+1}}}\\
\lambda_1\,=\,q_{2\,n+1}\,q_{2\,n+2}\cdot
{{(x_{2\,n+1}\,x_{2\,n+3}-1)\,\,x_{2\,n+2}}\over{1-x_{2\,n+2}}}
\end{eqnarray}
In order to write these expressions only in terms of  the entries of
the matrix $R_q$, one
has to get rid of these factors $q_{n}\,q_{n+1}$, which correspond to
an artifact
of the homogeneity. In this respect let us remark that
$\,\,T_{n}\,q_{n}=T_{n+2}\,q_{n+2}\,\,$:
$\,\,w_0=T_{n}\,T_{n+1}\,q_{n}\,q_{n+1}\,\,$ {\em is thus a constant}.
Introducing:
\begin{eqnarray}
E_n\,=\,{{x_{n-1}\,x_{n+1}-1}\over{(x_n-1)\,\,T_{n-1}\,T_n}} \cdot
\,x_n
\end{eqnarray}
one obtains $\lambda_2=w_0\,E_{2\,n+1}$ (and similarly
$\lambda_1=w_0\,E_{2\,n+2}$).

Relation $E_{n+2}=E_n$ is always satisfied, and recurrences
(\ref{intalm2}) and (\ref{intalm1}) are finally
equivalent to:
\begin{eqnarray}
\label{lambk}
\lambda_1\,=\,w_0\,E_0, \quad \lambda_2\,=\,w_0\,E_1
\end{eqnarray}
Notice that:
\begin{eqnarray}
\label{E0delt}
E_0\,=\,{{(\Delta_0+\Delta_2)\,x_0}\over {\Delta_1\,T_0^2}}
\end{eqnarray}
or, else, written directly in the entries of matrix $R_q$:
\begin{eqnarray}
\label{e0}
E_0\,=\,{{m_{22}\,(m_{33}-m_{31}+m_{11}-m_{13})}\over{(m_{23}-m_{21})^2}}
\end{eqnarray}
With these expressions, one first notices the following relation
between ${\cal P}_1$ and ${\cal P}_2$:
\begin{eqnarray}
{{{\cal P}_1}\over{\det(R_q)}}=\Bigl(
1+{{\lambda_1}\over{2\,q_0\,q_1}}+{{{\cal
P}_2\,\Delta_0}\over{2\,\det(R_q)}}\Bigr)\cdot
T_0
\end{eqnarray}
One also gets from relations (\ref{lambk}) and (\ref{E0delt}):
\begin{eqnarray}
{{(\Delta_0+\Delta_2)}\over
{\Delta_1\,T_0}}\,=\,{{\lambda_1}\over{q_1\,T_1}}
\end{eqnarray}
Recalling $x_n=q_{n+2}/q_n=1+T_n\,\Delta_{n+1}$, one gets:
\begin{eqnarray}
\Delta_1={{q_2-q_0}\over{q_0\,T_0}},\quad
\Delta_2={{q_3-q_1}\over{q_1\,T_1}}
\end{eqnarray}
yielding an explicit expression for $\,{\cal P}_2=-\;T_1 \cdot
\det(R_q)\,$:
\begin{eqnarray}
{{{\cal
P}_2}\over{\det(R_q)}}\,=\,{{q_0\,q_2\,(q_3-q_1)+\lambda_1\,(q_0-q_2)}\over{q_0\,q_1\,q_2\,\Delta_0}}
\end{eqnarray}
Recalling the two relations between the $q_n$'s, namely recurrences
(\ref{intalm2}) and
(\ref{intalm1}), and substituting, from (\ref{q2nan}) and
(\ref{q2np1an}),
 the $q_n$'s, in terms of the $a_n$'s and $b_n$'s, one gets
$a_{n+1}$ and $b_{n+1}$ in terms  of  $a_n$ and $b_n$ (see for instance
Appendix B).
One now has a representation of transformation $\widehat{K}^2$, {\em as
a
mapping in} $\C P_2$.

Let us also note that, as a consequence of the two simple
matricial relations:
\begin{eqnarray}
t(F)\,\,=\,\,F, \quad t(P)\,=\,-P \nonumber
\end{eqnarray}
transposition $t$, can simply be represented as a
reflection in the $(a_n,b_n)$-plane:
\begin{eqnarray}
t: \,\,\,\,(a,b)\, \,\rightarrow \,\, (a,\Delta_0-b)
\end{eqnarray}
{}From these two representations of $t$ and $\widehat{K}^2$ in the
$(a_n,b_n)$-plane, one gets a representation of
 $\widehat{I}\,t\,\widehat{I}$,  which is actually an involution.

After a change of variables
\footnote{ We would
like to thank M.P. Bellon for many  parallel checks on
most of the calculations detailed in this very section.}:
\begin{eqnarray}
u_n={{q_0\,q_1}\over{q_1\,q_2+q_0\,q_1+\lambda_2}}\cdot
\Bigl(1+T_0\,a_n
\Bigr),
\quad v_n=-{{q_0\,q_1}\over{\lambda_2}}\cdot \Bigl(1+{{{\cal
P}_1}\over{\det(R_q)}}\,a_n+ {{{\cal P}_2}\over{\det(R_q)}}\,b_n\Bigr)
\end{eqnarray}
the involutive transformation $\widehat{I}\,t\,\widehat{I}$ takes the
remarkably simple form ({\em independent of any parameter ! }):
\begin{eqnarray}
\label{tIt}
\widehat{I}\,t\,\widehat{I}: \,\,\,\,(u,v) \,\,\rightarrow
\,\,(u',v')\,
= \,\,\Bigl({{u+v-u\;v}\over{v}},\,\,{{u+v-u\;v}\over{u}}\Bigr)
\end{eqnarray}
and transformation $t$ is represented as the following ({\em two
parameter})
transformation:
\begin{eqnarray}
\label{t}
t: \,\,\,\,(u,v) \,\,\rightarrow \,\, (u,\,\,1+ \epsilon -v + \alpha \;
u)
\end{eqnarray}
where $\epsilon$ and $\alpha$ read:
\begin{eqnarray}
\epsilon\,=\,{{\lambda_1-\lambda_2}\over{\lambda_2}}, \quad
\alpha\,=\,-
\,{{(q_2\,(q_1+q_3)+\lambda_1)\,(q_1\,(q_0+q_2)+\lambda_2)}\over{q_1\,q_2\,\lambda_2}}\,=\,-\,{{\rho_1\,\rho_2}\over{\lambda_2}}
\end{eqnarray}
Note that one has the following relation:
\begin{eqnarray}
{{u_n} \over{v_n}}\, =\,  - \,\,{{\rho_2 \cdot q_{2\,n+1}}
\over{\lambda_2}}
\end{eqnarray}
Let us now recall again that there does exist an integrable
subcase of these mappings associated to class IV.
When $\lambda_1=\lambda_2=\lambda$~\cite{BoMaRo93d}, the $q_n$'s
actually {\em
satisfy two biquadratic equations} $F(x,y)=0$ and $F(y,x)=0$, depending
on
the parity of $n$, with:
\begin{eqnarray}
\label{bic}
F(x,y)\,= \, (xy-\lambda)(x-\rho_1)(y-\rho_2)-\mu
\end{eqnarray}
namely:
\begin{eqnarray}
\label{bicint}
\mu
&=&(\rho_1 - q_{2\,n+1} )\; (\rho_2 - q_{2\,n} ) \; ( {q_{2\,n} \;
q_{2\,n+1} + \lambda} ) \nonumber \\
&=&(\rho_2 - q_{2\,n+2} )\; (\rho_1 - q_{2\,n+1} ) \; ( {q_{2\,n+1} \;
q_{2\,n+2} + \lambda} )
\end{eqnarray}
It is well known that {\em biquadratic equations} are associated with
elliptic curves~\cite{prl2}
\footnote{
One should mention the work of several
authors who have presented a wide class of mappings of the form
$x_{n+1}=f(x_n,x_{n-1})$ which can be seen as discretizations of a
second-order ordinary differential equations related to elliptic
functions and more precisely biquadratic
relations~\cite{QuNi92,QuRoTh88,QuRoTh89,Ra91,Ra91b,Ra92,PaNiCa90}.
}.

With these remarkably simple forms for the two involutions generating
our
group of birational transformations, the integrability condition for
the
birational mappings of $\C P_2$ simply reads $\epsilon =0$ (or
equivalently $E_0=E_1$).
For this integrable $\epsilon =0$ subcase the group generated by
transformations
(\ref{tIt}) and (\ref{t}) yields a foliation of the $(u,v)$-plane in
terms
of curves, which form a {\em linear pencil of elliptic curves}.
This can be seen noticing that,  for $\epsilon =0$, an algebraic
expression ${\cal
I}$ is actually invariant under both transformations
$\widehat{I}\,t\,\widehat{I}$  and $t$:
\begin{eqnarray}
\label{Inveps}
{\cal I}\,\,=\,\,
{{(1-u)\;(1-v)\;(v-\alpha\;u)} \over {u}}
\end{eqnarray}
One should also remark that  $\epsilon =0$ is not the only
integrability
condition for these birational mappings. One can actually verify
straightforwardly that  $\epsilon = -1$ is also an {\em integrability
condition}: it corresponds to a (rational) degeneracy of the mapping.
Condition  $\epsilon = -1$ drastically simplifies
transformation (\ref{t}) which becomes:
\begin{eqnarray}
\label{t123}
t: \,\,\,\,(u,v)\,\, \rightarrow \, \,(u,\, -v + \alpha \; u)
\end{eqnarray}
One immediately gets an algebraic invariant under transformations
(\ref{tIt}) and
(\ref{t}):
\begin{eqnarray}
\label{Inveps2}
{\cal J}\,\,=\,\,\Bigl({{v}\over{u}}-{{\alpha}\over{2}}\Bigr)^2
\end{eqnarray}
This $\epsilon = -1$ case, which corresponds to $\lambda_1=0$, yields a
{\em simple
rational parameterization} of the iteration.

For heuristic reasons let us consider the $\alpha = 0$ case
 (which happens to correspond to a {\em rational
parameterization}, when $\epsilon =0$).
{}From relation (\ref{Inveps}) the variable $u$ can be simply written in
terms of
the variable $v$  and the invariant ${\cal I}$:
\begin{eqnarray}
u\,=\,{{v\,(1-v)}\over{{\cal I}+v\,(1-v)}}
\end{eqnarray}
and the mappings $t$ and $\widehat{I}\,t\,\widehat{I}$ respectively
read:
\begin{eqnarray}
\label{perturb}
&&t: \,\,\,\,({\cal I},v)
\rightarrow \,\,\,
 ({\cal I'},v')\,\, =\,\,
\Bigl({\cal I}\,(1-{{\epsilon}\over{v}})\,(1-{{\epsilon}\over{v-1}})
,\, \,1 - v +\epsilon \Bigr)
\\
\label{perturb2}
&&\widehat{I}\,t\,\widehat{I}: \,\,\,\,({\cal I},v)
\rightarrow \,\,\,
({\cal I'},v')\,\, =\,\,
\Bigl({\cal I} ,\,\, 1-{{{\cal I}}\over{v-1}} \Bigr)
\end{eqnarray}
These very simple representations of the birational transformations of
class IV
enabled us to perform a large number of numerical calculations which
confirm the analysis performed in~\cite{BoMaRo93d}. The iterations of
these transformations often yield orbits which look like
curves~\cite{BoMaRo93d}.
The situation, as far as the visualization of {\em a single orbit} is
concerned, is
similar to the situation encountered in the Henon-Heiles
mapping~\cite{He64}.
Let us note however, that {\em one does not consider perturbations near
a
fixed point anymore, but in the neighborhood of the whole
integrability condition}.

Figure (4) shows a set of
one hundred orbits corresponding to the iteration of $\widehat{K}^2$ in
the
$({\cal I},v)$-plane. The orbits are very regular: figure (4) gives an
illustration of the ``almost integrability'' described
in~\cite{BoMaRo93d}
and also reminiscent of the situation one encounters near
elliptic points in dynamical systems~\cite{Al88,Ar76,Ar80,Mu93}.
At this point it is important to make the following comment: since
they are generated by involutions,  all our
birational transformations are such that transformation $\widehat{K}$
and
 transformation $\widehat{K}^{-1}$ are conjugated ( $\widehat{K} = t
 \cdot
\widehat{I} = t \cdot  \widehat{K}^{-1} \cdot t $). When transformation
$\widehat{K}$ (or more precisely $\widehat{K}^2$) can be reduced to a
mapping on only
two variables this means that one has some area preserving properties
and
one can recover the features of two-dimensional dynamics
(elliptic versus hyperbolic points, Arnold's diffusion
...~\cite{Al88,Ar76,Ar80,Mu93}), and this explains to a great extend
the regularities one
encounters here with class IV, even when the mapping is not
integrable.
However this conjugation properties does not seem sufficient to explain
the
regularities observed for the other (non-integrable) classes for which
the
dynamics cannot be reduced to two-dimension any more: volume-(or
hypervolume) preserving properties are no longer sufficient to explain
our regularities.

\subsection{Factorizations for the Integrable subcase of Class IV}
\label{exegese}
The birational transformations for class IV do not generically
correspond
to integrable mappings~\cite{BoMaRo93d}.
They have been seen to yield an
{\em exponential growth} of the calculations. However, {\em there
exists a subvariety}
of ${\Bbb C} P_{q^2-1}$ on which these  birational transformations {\em
become integrable}, yielding
{\em algebraic elliptic curves} for arbitrary $q$.
Recalling (\ref{e0}), this integrability condition $E_{n+2}=E_{n+1}$
(or equivalently $E_0=E_1$)
can  be given, from (\ref{lambk}), in terms of the
entries of the initial matrix $M_0=R_q$:
\begin{eqnarray}
\label{e0e1}
{{m_{22}\,(m_{33}-m_{31}+m_{11}-m_{13})}\over{(m_{23}-m_{21})^2}}\,=
\,\,
\hbox{same expression with}\, \,\,\,  m_{i,j} \rightarrow K(M_0)_{i,j}
\end{eqnarray}
An example of a one parameter dependent $4 \times 4$  matrix satisfying
this integrability condition is given in Appendix C.

 In this framework a question naturally pops out: how does
the factorizations (\ref{facIV}), (\ref{detIV}) and  (\ref{KIV}) modify
when one
restricts to this integrability condition ? In particular, does the
exponential growth of the calculations becomes a polynomial growth when
restricted to this integrability condition ?

Restricted to this integrable subcase
the factorizations corresponding to the iterations of $K$ for class IV
are such that {\em the first five iterations yield
 the same factorizations} as for the generic (non-integrable) case:
\begin{eqnarray}
\label{factoIVint}
&&\tilde{M}_1=K(\tilde{M}_0),\;\;
\tilde{f}_1=\det(\tilde{M}_0),\;\;
\tilde{f}_2={{\det(\tilde{M}_1)}\over {\tilde{f}_1^{2}}},\;\;
\tilde{M}_2= {{K(\tilde{M}_1)}\over {\tilde{f}_1}},\;\;
\tilde{f}_3={{\det(\tilde{M}_2)}\over {\tilde{f}_1 \cdot
\tilde{f}_2^{2}}},\;\;
\tilde{M}_3={{K(\tilde{M}_2)}\over{\tilde{f}_2}} \nonumber \\
&&\tilde{f}_4={{\det(\tilde{M}_3)}\over{\tilde{f}_1^{3} \cdot
\tilde{f}_2 \cdot \tilde{f}_3^{2}}},\;\;
\tilde{M}_4={{K(\tilde{M}_3)}\over{\tilde{f}_1^{2} \cdot
\tilde{f}_3}},\;\;
\tilde{f}_5={{\det(\tilde{M}_4)} \over {\tilde{f}_1^2 \cdot
\tilde{f}_2^{3} \cdot \tilde{f}_3 \cdot \tilde{f}_4^{2}}},\;\;
\tilde{M}_5={{K(\tilde{M}_4)} \over {\tilde{f}_1 \cdot \tilde{f}_2^{2}
\cdot \tilde{f}_4}}\,\, \cdots
\end{eqnarray}
The difference with the generic factorizations (\ref{facIV}) comes with
the next iterations :
\begin{eqnarray}
\label{factoIVintb}
&&\tilde{f}_6={{\det(\tilde{M}_5)}\over{\tilde{f}_1^3 \cdot
\tilde{f}_2^2 \cdot \tilde{f}_3^{3} \cdot \tilde{f}_4 \cdot
\tilde{f}_5^{2}}},\;\;
\tilde{M}_6={{K(\tilde{M}_5)}\over{\tilde{f}_1 ^2 \cdot \tilde{f}_2
\cdot \tilde{f}_3^{2} \cdot \tilde{f}_5}},\;\;
\tilde{f}_7={{\det(\tilde{M}_6)}\over{\tilde{f}_2^3 \cdot \tilde{f}_3^2
\cdot \tilde{f}_4^{3} \cdot \tilde{f}_5 \cdot
\tilde{f}_6^{2}}},\;\; \nonumber \\
&&\tilde{M}_7={{K(\tilde{M}_6)}\over{\tilde{f}_2 ^2 \cdot \tilde{f}_3
\cdot \tilde{f}_4^{2} \cdot \tilde{f}_6}},\;\;
\tilde{f}_8={{\det(\tilde{M}_7)}\over{\tilde{f}_3^3 \cdot \tilde{f}_4^2
\cdot \tilde{f}_5^{3} \cdot \tilde{f}_6 \cdot \tilde{f}_7^{2}}},\;\;
\tilde{M}_8={{K(\tilde{M}_7)}\over{\tilde{f}_3 ^2 \cdot \tilde{f}_4
\cdot \tilde{f}_5^{2} \cdot \tilde{f}_7}},\;\;
\cdots
\end{eqnarray}
One remarks that,  compared
to factorizations (\ref{facIV}), $\tilde{f}_1$ {\em factorizes one more
time in} $\det(\tilde{M}_5)$ and $K(\tilde{M}_5)$. These additional
factorizations propagate and yield  for arbitrary  $n$:
\begin{eqnarray}
\label{detIVint}
det(\tilde{M}_n)&=& \tilde{f}_{n+1} \cdot \tilde{f}_{n}^{2} \cdot
\tilde{f}_{n-1} \cdot \tilde{f}_{n-2}^{3} \cdot \tilde{f}_{n-3}^2 \cdot
\tilde{f}_{n-4}^{3} \\
\label{KIVint}
K(\tilde{M}_n) &=& \tilde{M}_{n+1} \cdot  \tilde{f}_{n} \cdot
\tilde{f}_{n-2}^{2} \cdot \tilde{f}_{n-3}  \cdot \tilde{f}_{n-4}^{2}
\end{eqnarray}
which yield:
\begin{eqnarray}
\label{KdetIVint}
{{K(\tilde{M}_n)} \over {det(\tilde{M}_n)}}\,\,=\,\, {{\tilde{M}_{n+1}}
\over { \tilde{f}_{n-4}\cdot
\tilde{f}_{n-3} \cdot \tilde{f}_{n-2} \cdot \tilde{f}_{n-1} \;
\tilde{f}_n \cdot \tilde{f}_{n+1}}}\,\, = \, \,\widehat{K}(\tilde{M}_n)
\end{eqnarray}
The correspondence between the $f_n$'s, associated to the generic
(non-integrable) case (see  factorizations (\ref{facIV})),
 and these new factorizing polynomials associated to the integrable
 subcase
(see factorizations (\ref{factoIVint})), denoted $\tilde{f}_n$, read:
\begin{eqnarray}
\label{ffint}
&&f_6=\tilde{f}_6\cdot f_1 ,\;\;\,\,
f_7= \tilde{f}_7\cdot f_2\cdot f_1  ,\;\;\,\,
f_8= \tilde{f}_8\cdot f_3\cdot f_2\cdot f_1^2   ,\;\;
\cdots \nonumber
\end{eqnarray}

{}From equations (\ref{detIVint}), (\ref{KIVint}) and (\ref{KdetIVint}),
one gets linear recurrences
 on the degrees of the polynomials $\det(\tilde{M}_n)$ and
 $\tilde{f}_n$
(respectively the $\tilde{\alpha}_n$'s and $\tilde{\beta}_n$'s):
\begin{eqnarray}
\label{alIVint}
\tilde{\alpha}_{n+2} \,=\, 3\; \tilde{\beta}_{n-2} +
2\,\tilde{\beta}_{n-1}+ 3\; \tilde{\beta}_{n} + \tilde{\beta}_{n+1} +
2\; \tilde{\beta}_{n+2} +\tilde{\beta}_{n+3}
\end{eqnarray}
\begin{eqnarray}
\label{beIVint}
3 \; \tilde{\alpha}_{n+2} \,=\, \tilde{\alpha}_{n+3} + 4\,\Bigl(2\,
\tilde{\beta}_{n-2} +\tilde{\beta}_{n-1}+2\;
\tilde{\beta}_{n}+\tilde{\beta}_{n+2}\Bigr)
\end{eqnarray}
and
\begin{eqnarray}
\label{albeIVint}
  \tilde{\alpha}_{n+3}+ \tilde{\alpha}_{n+2}\, =\,  4\,\Bigl(
  \tilde{\beta}_{n-2} +\tilde{\beta}_{n-1}+ \tilde{\beta}_{n} +
  \tilde{\beta}_{n+1} +  \tilde{\beta}_{n+2} +\tilde{\beta}_{n+3}\Bigr)
\end{eqnarray}
giving the following relations on the generating functions:
\begin{eqnarray}
\label{alIVxint}
x\,\cdot \tilde{\alpha}(x)\,\,= \,\, \tilde{\beta}(x) \cdot
(1+2\,x+x^2+3\,x^3+2\,x^4+3\,x^5)
\end{eqnarray}
\begin{eqnarray}
\label{beIVxint}
(3x -1)\cdot \tilde{\alpha}(x) \,= \, 4\,x \cdot \tilde{\beta}(x) \cdot
(1+2\,x^2+x^3+2\,x^4) - \,4
\end{eqnarray}
and:
\begin{eqnarray}
\label{albeIVxint}
(1+x)\cdot \tilde{\alpha}(x)\, = \, 4\, (1+x+x^2+x^3+x^4+x^5) \cdot
\tilde{\beta}(x)\,+\,4
\end{eqnarray}
The two generating functions $\tilde{\alpha}(x)$ and $\tilde{\beta}(x)$
read:
\begin{eqnarray}
\label{albexIVint}
\tilde{\alpha}(x)={{4\,(1+x-x^2+3\,x^3)}\over{(1+x)(1-x)^3 }}, \quad
\tilde{\beta}(x)={{4\;x}\over{(1+x)(1-x)^3(1+x+x^2)}}
\end{eqnarray}
The linear relations corresponding to the right action of $K$, namely
factorizations (\ref{kfnVI})
 and (\ref{mfnVI}), are {\em still valid} when one restricts to this
 integrable
case. Therefore the generating functions $\tilde{\mu}(x)$ and
$\tilde{\nu}(x)$ still verify relation (\ref{ggI}) and $\tilde{\mu}(x)$
and $\tilde{\nu}(x)$ read:
\begin{eqnarray}
\label{munuxIVint}
\tilde{\mu}(x)={\frac {x\left (2-x+x^{3}+x^{4}-x^{5}\right )}{\left
(1+x\right )
\left (1+x+x^{2}\right )\left (1-x\right )^{3}}}, \quad
\tilde{\nu}(x)={\frac {x\left (1-x+2\,x^{2}\right )}{\left (1-x\right
)^{3}\left (1+x\right )}}
\end{eqnarray}
These relations clearly show that the additional factorizations
occurring
for this integrable subcase remarkably yield factorizations like
 (\ref{KdetIVint}), (\ref{KIVint}), or (\ref{detIVint}) bearing on a
 {\em fixed
number of polynomials} $\tilde{f}_n$ and even more, to a {\em
polynomial growth} of
 the calculations instead of the exponential growth previously
 described
(see section (\ref{class4})). One should however note {\em the
occurrence of a third root
of unity} in the denominator of the generating functions
$\tilde{\beta}(x)$ and
$\tilde{\mu}(x)$ (and not for the generating functions
$\tilde{\alpha}(x)$ and $\tilde{\nu}(x)$).
 Of course the occurrence of this new root of unity, different from
 $\pm
1$, {\em does not rule out the  polynomial growth}.

Finally it is important to note that , in
this integrable subcase, the new polynomials $\tilde{f}_n$ defined
 from the factorization relations (\ref{detIVint}) and (\ref{KIVint})
 {\em
do satisfy recurrences bearing on products of a fixed number of
polynomials} (instead of ``pseudo'' recurrences like (\ref{pseudo})):
\begin{eqnarray}
\label{recucu}
{{\tilde{f}_{n+2}\,\tilde{f}_{n+7}\,\tilde{f}_{n+9} -
\tilde{f}_{n+3}\,\tilde{f}_{n+5}\,\tilde{f}_{n+10}}\over
{\tilde{f}_{n+3}\,\tilde{f}_{n+7}\,\tilde{f}_{n+8} -
\tilde{f}_{n+4}\,\tilde{f}_{n+5}\,\tilde{f}_{n+9}}}\, = \,
{{\tilde{f}_{n+1}\,\tilde{f}_{n+6}\,\tilde{f}_{n+8} -
\tilde{f}_{n+2}\,\tilde{f}_{n+4}\,\tilde{f}_{n+9}}\over{\tilde{f}_{n+2}\,\tilde{f}_{n+6}\,\tilde{f}_{n+7}
- \tilde{f}_{n+3}\,\tilde{f}_{n+4}\,\tilde{f}_{n+8}}}
\end{eqnarray}
Such a recurrence is very similar to recurrences (\ref{recufnqq}) or
(\ref{recufnqq2})
 and can be written introducing  well-suited variables $\tilde{q}_n$'s
 defined by:
\begin{eqnarray}
\label{well}
\tilde{q}_{n+4}\,=\,
{{\tilde{f}_n\,\tilde{f}_{n+5}}\over{\tilde{f}_{n+2}\,\tilde{f}_{n+3}}}
\end{eqnarray}
In terms of these variables  $\tilde{q}_n$, recurrence (\ref{recucu})
becomes :
\begin{eqnarray}
\label{verywell}
{{\tilde{q}_{n+2} - \tilde{q}_{n+1}} \over{ \tilde{q}_{n+1} \cdot
\tilde{q}_{n+3} -  \tilde{q}_{n} \cdot \tilde{q}_{n+2}}} \,=
\,{{\tilde{q}_{n+3} - \tilde{q}_{n+2}} \over{ \tilde{q}_{n+2} \cdot
\tilde{q}_{n+4} -  \tilde{q}_{n+1} \cdot \tilde{q}_{n+3}}}
\end{eqnarray}
which can be integrated as:
\begin{eqnarray}
\label{integ1}
\tilde{q}_{n+2} - \lambda' \cdot \tilde{q}_{n+1} \cdot
\tilde{q}_{n+3}\, =\, \rho'
\end{eqnarray}
and in a second step:
\begin{eqnarray}
\label{integ2}
{{1+ \lambda' \cdot \tilde{q}_{n+1}} \over{\tilde{q}_{n+2}}} \,=\,
{{1+ \lambda ' \cdot \tilde{q}_{n+4}} \over{\tilde{q}_{n+3}}}
\end{eqnarray}
\begin{eqnarray}
\label{integ3}
{{(1+ \lambda' \cdot \tilde{q}_{n+1})(1+ \lambda'  \cdot
\tilde{q}_{n+2})(1+ \lambda' \cdot
\tilde{q}_{n+3})}\over{\tilde{q}_{n+2}}}
 \,=\, {{(1+ \lambda' \cdot \tilde{q}_{n+2})(1+ \lambda' \cdot
 \tilde{q}_{n+3})(1+ \lambda'\cdot \tilde{q}_{n+4})}
 \over{\tilde{q}_{n+3}}}
\,=\, \mu'
\end{eqnarray}
which {\em finally yields a biquadratic relation}:
\begin{eqnarray}
\label{biqintIV}
(1+ \lambda'\cdot  \tilde{q}_{n+1})(1+ \lambda' \cdot \tilde{q}_{n+2})(
\tilde{q}_{n+1} +
\tilde{q}_{n+2} - \rho')\, = \, \mu' \,\, \tilde{q}_{n+1} \,
\tilde{q}_{n+2}
\end{eqnarray}
Recalling the two biquadratic relations (\ref{bicint}) given in section
(\ref{planes}), one notices that, taking into account the homogeneity
of the
$q_n$'s,
one can barter these two biquadratics for a single one (changing
$q_{2\,n}$
into $q_{2\,n}/ \rho_2$ and $q_{2\,n+1}$
into $q_{2\,n+1}/ \rho_1$).
This last biquadratic is apparently different from (\ref{biqintIV}):
one would like to see the relation
between these two biquadratics bearing respectively on the
$\tilde{q}_n$'s introduced here, and the $q_n$'s introduced
in section (\ref{class4}).
The relation between the $q_n$'s and the $\tilde{q}_n$'s reads as
follows:
\begin{eqnarray}
\label{dicoIV}
{{\tilde{q}_{n}} \over {\tilde{q}_{0}}} \,= \, {{\,q_{n} \, \, q_{n+1}}
\over{q_0 \, q_1}}
\end{eqnarray}
After straightforward calculations (corresponding to introduce the
product
$q_{n} \, q_{n+1}$ in recurrences (\ref{intalm2}) and (\ref{intalm1}))
one can show  in the integrable case,
$\lambda_1=\lambda_2=\lambda$, that the biquadratic relation
(\ref{bicint}) yields the biquadratic
relation (\ref{biqintIV}) with the following correspondence:
\begin{eqnarray}
\label{corresp}
\lambda' ={{1} \over{\lambda}} \, \,, \, \,  \,\,\, \rho' =
-{{\lambda^2 +\mu -  \lambda \,\rho_1\,\rho_2} \over{\lambda}}\, \,
,\,\,\,
\,\,\mu'= {{\mu} \over{\lambda^3}}
\end{eqnarray}
or conversely:
\begin{eqnarray}
\label{corresp2}
\lambda ={{1} \over{\lambda'}} \, \,, \,\,\, \,\rho_1\,\rho_2=
{{\lambda'+\lambda'^2\,\rho'+\mu'} \over{\lambda'^2}}\, \,
, \, \, \,\,
\mu = {{\mu'} \over{\lambda'^3}}\,\,\,
\end{eqnarray}

Since variable $\tilde{q}_{n}$ is a homogeneous variable,
 it is tempting to write recurrence (\ref{verywell}) in terms of
 the ratio of two successive
$\tilde{q}_{n}$'s. In fact this ratio {\em happens to coincide exactly
with the
variable} $x_n$ defined by (\ref{xq}):
\begin{eqnarray}
x_{n} \, =\, {{\tilde{q}_{n+1}} \over{\tilde{q}_{n}}}
\end{eqnarray}
The variable $x_n$
can be written either in terms of the $f_n$'s or in terms of the
$\tilde{f}_{n}$'s:
\begin{eqnarray}
x_{n+6} \, = \, {{ f_{n+8}\,f_{n+4}\,f_{n}\,f_{n-4}\,f_{n-8} \cdots}
\over{ f_{n+7} \, f_{n+6}\,
f_{n+2}\,f_{n-2}\,f_{n-6} \cdots}} \,
= \, \,  {{\tilde{f}_{n+8}\,\tilde{f}_{n+4}\,\tilde{f}_{n+3}}
\over{\tilde{f}_{n+7} \, \tilde{f}_{n+6} \, \tilde{f}_{n+2}}}
\end{eqnarray}

The relations between the initial values of variables $q_n$,
$\tilde{q}_n$,
$f_n$,
$\tilde{f}_n$ and of the variables $x_n$'s are given in Appendix D.

With these new variables the integrable recurrence (\ref{verywell})
reads:
\begin{eqnarray}
\label{recint}
{{x_{n+2} -1}\over {x_{n+1}\,x_{n+3} -1}}\,= \,{{x_{n}\,x_{n+2}}
\over {x_{n+1}}} \cdot {{x_{n+1} -1}\over
{x_{n}\,x_{n+2} -1}}
\end{eqnarray}
Let us note that, combining (\ref{recint}) with itself where $n$ has
been
shifted by one, one recovers:
\begin{eqnarray}
\label{recint2}
{{x_{n+3} -1}\over {x_{n+2}\,x_{n+4} -1}}\,= \,x_{n}\,x_{n+3}
\cdot {{x_{n+1} -1}\over
{x_{n}\,x_{n+2} -1}}
\end{eqnarray}
which coincides with relation (\ref{xnIV}).
In fact one can actually show, but this will not be performed here,
that
the $x_n$'s corresponding to class IV do satisfy a {\em whole hierarchy
of
recurrences}
in the same way it has been proved for class I
in~\cite{BoMaRo93a}.

Similarly to~\cite{BoMaRo93a}, one can consider these
recurrences for themselves, without referring to our birational
transformations acting on $q \times q$ matrices anymore. Again one can
see
that some of these recurrences {\em are integrable recurrences} (for
instance
recurrence (\ref{recint})) and some {\em are} (generically) {\em not
integrable}
 (for instance recurrence (\ref{recint2})).
All the analysis performed in~\cite{BoMaRo93a} can be applied to the
hierarchy emerging from (\ref{recint})), in particular the fact that
recurrences like
(\ref{recint})) are equivalent to another recurrence of the hierarchy,
namely:
\begin{eqnarray}
\label{recint3}
{{x_{n+2} -1} \over{x_{n+1}\,x_{n+3} - x_{n+2}}} \,=\, x_n \, x_{n+2}
\cdot {{x_{n+1} -1} \over{x_{n}\,x_{n+2} - x_{n+1}}}
\end{eqnarray}

\section{Conclusion}

The analysis of the factorizations corresponding to the six previously
 defined classes of transpositions has shed some light on the relations
 between different properties and structures related to integrable
 mappings
 such as  the {\em existence of
factorizations} involving a fixed number of polynomials,
{\em the polynomial growth} of the complexity of the iterations of
 these mappings seen as homogeneous transformations,
 the existence of {\em recurrences} bearing on the factorized
 polynomials
$f_n$'s or on other variables such as the $x_n$'s and the {\em
integrability of the mappings}.

A single factorization relation {\em independent of $q$} (relation
(\ref{MGMqq})) has been shown to be satisfied for classes
I, III and V. All these classes {\em satisfy factorizations where
products of a fixed number} of polynomials $f_n$ occur (see
(\ref{detIII}), (\ref{KIII})).
The  polynomial growth of the complexity of the
iterations suites quite well with factorizations where products of a
fixed number of
polynomials occur: this is actually the case for classes I and III
which actually
 verify {\em the same factorizations} on the $f_n$'s.
 In fact, surprisingly, it will be shown in~\cite{BoMaRo93c} (on an
 example
corresponding to the symmetries of a three
dimensional vertex model) that a {\em polynomial growth of the
complexity
can actually
occur even with ``string-like'' factorizations} (like (\ref{KdetI})).
On the other hand, one sees with class V that one can have a
factorization of the
iteration with a {\em fixed number of polynomials and, in the same
time,
 an exponential growth of the complexity} of the iterations.
Among the two classes satisfying exactly the same factorizations, that
is:
I and III, {\em only the first one verifies recurrences on the
polynomials} $f_n$'s.
For class I, for arbitrary $q$, these recurrences are {\em integrable}
recurrences and probably are the birational mappings in $\C
P_{q^2-1}$.
For $q=4$ the iteration of these transformations associated to classes
I, II and III
yield algebraic elliptic curves.
For $q \ge 5$ class III, yields quite regular orbits which often look
like curves, or like a set of curves which seem to lie on higher
dimensional varieties. However, the actual status of these higher
dimensional
varieties is not very clear (abelian varieties ... ).
Class III is thus, for
$q \ge 5$, an interesting class since it would provide {\em an example
of
polynomial growth} (and in fact exactly the same factorization as the
``nice'' class I) {\em with orbits densifying nice algebraic varieties
(abelian surfaces, product of elliptic curves ? ... ), but not curves
anymore}.

A ``string-like'' factorization relation {\em independent of $q$}
(relation
(\ref{facdetI})) is satisfied by the two classes IV,  VI and also for
class
II but only for $q \ge 5$.
These three sets of factorizations with a ``growing''  number of
polynomials
(see relation (\ref{facdetI})) correspond to an exponential growth
 of the complexity of the iterations. However, one remarks that class
 IV
 {\em has actually a recurrence on the variables} $x_n$
 {\em but not of course on the variables} $f_n$'s:
these variables cannot satisfy simple recurrences
 like (\ref{recufnqq}), but {\em they satisfy ``pseudo-recurrences''
 like} (\ref{pseudo}).
{\em This recurrence, on the} $x_n${\em 's, is not generically an
integrable
one}~\cite{BoMaRo93d}.
It however yields orbits which look like {\em curves} for some
domain of the initial conditions~\cite{BoMaRo93d}.
Such a recurrence is an illustration of a transition from integrability
to
 weak chaos, through situations visually similar to the one encountered
 in
the  Henon-Heiles mapping~\cite{He64}, or to the situation encountered
near
elliptic points in the theory of (hamiltonian) dynamical systems of two
variables~\cite{Al88,Ar76,Ar80,Mu93}.
We hope that class II will help understanding the structures of
elliptic
points for three-dimensional mappings (see relation (\ref{affineII})).

Note however that we have been able to prove that the integrable
subcase of class IV is actually valid for
arbitrary $q$: {\em  it thus provides an example of integrable mapping
in
arbitrary dimension even infinite}.

 For all these six classes, one has recurrences on a fixed number of
 variables
(see (\ref{recfinIV})).
The elimination of these variables may yield quite involved algebraic
relations on the remaining variable  $x_n$.
The recurrences associated to classes I and IV are the only one which
are {\em
recurrences} and not involved algebraic relations between the $x_n$'s.
Note that when a recurrence is integrable the
corresponding birational transformations are probably also integrable:
in fact
this appears as the only ``handable'' way to find the integrable
subcase of class IV ...

 To sum up the relations between all these various structures and
 properties
are subtle: the only systematic relation being that
integrability seems to imply the polynomial growth of the iterations.
Let us  however recall that {\em one can have polynomial growth with
orbits
densifying algebraic
surfaces}~\cite{BoMaRo93c}.

The analysis of the factorization of the iteration corresponding to
birational transformations such as the one studied here can be seen as
a
new method to analyze birational mappings and therefore the symmetries
of
lattice models,
 in particular
models in dimension greater than two.
In a forthcoming publication~\cite{BoMaRo93c} we will consider
three-dimensional (and higher dimensional) vertex models and show that
the birational transformations associated with these models yield
algebraic surfaces {\em and} polynomial growth.

\vskip 1cm \noindent {\bf Acknowledgement}: We thank  M. Bellon for
discussions and for many calculations in the analysis of class IV. We
thank C.M.
Viallet for many
 encouragements and discussions, in particular concerning the relation
 of the
behavior of our transformations compared with the Henon-Heiles
mapping.
We thank P. Lochak for discussions on  the relation of the
behavior of our transformations with the situation encountered near
elliptic points, in two dimensional dynamical systems.

\section{Appendix A: a comment on the finite order conditions }

In the framework of finite order conditions it is worth mentioning
the following very general result (it works with the six
classes defined in section (\ref{six}), and can even be generalized to
 arbitrary permutations of a $q\times q$ matrix~\cite{BoMaRo93c}).

For an arbitrary transposition one imposes the two conditions:
\begin{eqnarray}
\label{cond111}
\Delta_1(R_{q})=0
\end{eqnarray}
 {\em and}
\begin{eqnarray}
\label{cond112}
\Delta_1(t(R_{q}))=0
\end{eqnarray}
where $t$ denotes a
transposition of two entries.
The very definition of $\Delta_1$ yields:
\begin{eqnarray}
\label{cond1}
\Delta_1(R_{q})=0 \Rightarrow \widehat{K}(R_{q})=
\widehat{I}(R_{q}) \Rightarrow \widehat{I}(\widehat{K}(R_{q}))= R_{q}
\Rightarrow \widehat{K}^2(R_{q})=t(R_{q})
\end{eqnarray}
Since one assumes $\Delta_1(t(R_{q}))\,=\,0$, one similarly gets:
\begin{eqnarray}
\label{cond2}
\widehat{K}^2(t(R_{q}))\,=\,t(t(R_{q}))\,=\,R_{q}
\end{eqnarray}
Hence for every matrix $R_{q}$ satisfying (\ref{cond111}) and
(\ref{cond112}), transformation $\widehat{K}$ is a transformation of
{\em
order four}, that is:
\begin{eqnarray}
\label{cond3}
\widehat{K}^4(R_{q})\,=\,R_{q}
\end{eqnarray}

Let us note that the (codimension two) variety given by the two
conditions
(\ref{cond111}) and (\ref{cond112}) contains remarkably simple linear
subvarieties.
Let us consider for instance a transposition of class VI, namely $t$
exchanging
$m_{11}$ and $m_{12}$, the following conditions yield (\ref{cond3}):
\begin{eqnarray}
\label{condentries}
m_{32}\,=\,0\,,\,\,\,m_{33}\,=\,0\, ,\,\,\,\ldots,\,\,\,m_{3q}\,=\,0
\end{eqnarray}

These considerations can be generalized for very general permutations
of
the entries.
 For instance, let us consider the case where the transposition $t$ is
replaced by a 3-cycle $C$. One can also find remarkably
simple conditions
 on the entries of the matrix such that $\widehat{K}$ is of finite
 order.
Considering the 3-cycle C:
\begin{eqnarray}
m_{11} \,\,\rightarrow \,\,  m_{13}\, ,  \nonumber \\
m_{13} \,\, \rightarrow \,\, m_{12} \, ,\nonumber \\
m_{12} \, \,\rightarrow \,\, m_{11}
\end{eqnarray}
and the following vanishing conditions on the entries:
\begin{eqnarray}
\label{condentries2}
m_{42}\,=\,0\,,\,\,\,m_{43}\,=\,0\,,\,\, \ldots, \,\,m_{4q}\,=\,0
\end{eqnarray}
Restricted to conditions (\ref{condentries2}) transformation
$\widehat{K}$ is such
that $\widehat{K}^2=C$, hence, $\widehat{K}^6={\cal I}d$, when one
restricts to the (linear)
subvariety (\ref{condentries2}).

\section{Appendix B}

{}From the parameterization of the $(a_n,b_n)$-plane (see relation
(\ref{planes})), one can actually  get the
mapping of $\C P_2$ corresponding to $\widehat{K}^2$:
\begin{eqnarray}
a_{n+1}&=&\,-a_n+k_4+{{k_5\,b_n+k_6}\over{k_1\,a_n+k_2\,b_n+k_3}}\nonumber
\\
b_{n+1}&=&\,\Delta_0-b_n+(k_5\,b_n+k_6)\cdot
\Bigl(k_7+{{k_8}\over{1+T_0\,a_n}}+{{k_9\,(k_5\,b_n+k_6)}\over{(1+T_0\,a_n)\,(k_1\,a_n+k_2\,b_n+k_3)}}\Bigr)
\end{eqnarray}
where the $k_{\alpha}$'s denote {\em rational} expressions depending
only on the initial
matrix $R_q$.
However the $k_{\alpha}$'s are not independent, {\em they do satisfy
additional
relations}:
\begin{eqnarray}
\label{sixpara}
k_6\,=\,{{k_5\,(k_3\,T_0-k_1)}\over{k_2\,T_0}}, \quad
k_7\,=\,{{2}\over{k_5}}, \quad
k_8\,=\,-{{2+k_4\,T_0}\over{k_5}}, \quad
k_9\,=\,-{{T_0}\over{k_5}}
\end{eqnarray}
Therefore $\widehat{K}^2$ is represented as a birational transformation
in
$\C P_2$ depending on {\em six parameters}.
These parameters are of course functions of the $q^2$ homogeneous
entries of
the initial matrix $R_q$. Among the six parameters one can actually
drop
out four parameters corresponding to (independent) dilatations and
translations
of variables $a_n$ and $b_n$, yielding to only two remaining parameters
like in section (\ref{planes}).

\section{Appendix C}

 Let us  give
 an example of one parameter dependent $4 \times 4$  matrix satisfying
 the
integrability condition (\ref{e0e1}):

\begin{eqnarray}
\label{int u}
{{M_0} \over{3450543}} \, \,  = \, \,
 \left [\begin {array}{cccc}
-85&
2636150642/3450543-u &
-78 &-35\\
97& 50& 100 & 56\\
49&
u-55& 62&-59\\
45 & -8 & 62 & 92
\end {array}\right ] \nonumber \\
\end{eqnarray}

\section{Appendix D}

Let us give the expression of the first  $x_n$'s in terms of the
first  $f_n$'s:

\begin{eqnarray}
x_{0} \, = \, {{f_{2}} \over{ f_{1}}} \, ,\,\,\,
x_{1} \, = \, {{f_{3}} \over{ f_{2} \, f_{1}}} \, \, ,\, \,\,x_{2} \, =
\, {{f_{4}} \over{ f_{3} \, f_{2}}} \,\, ,\,\,\,  x_{3} \, = \,
{{f_{5}\, f_{1}} \over{ f_{4} \, f_{3}}}\,\, ,\,\,\,x_{4} \, = \,
{{f_{6}\, f_{2}} \over{ f_{5} \, f_{4}}}
\end{eqnarray}

The expression of the first $x_n$'s in terms of the first
$\tilde{f}_n$ read:

\begin{eqnarray}
x_{0} \, = \, {{\tilde{f}_{2}} \over {\tilde{f}_{1}}}\,\, ,\,\,\,
x_{1} \, = \, {{\tilde{f}_{3}} \over {\tilde{f}_{2} \,
\tilde{f}_{1}}}\,\, ,\,\,\,
x_{2} \, = \, {{\tilde{f}_{4}} \over {\tilde{f}_{3} \,
\tilde{f}_{2}}}\,\, ,\,\,\,
x_{3} \, = \, {{\tilde{f}_{5}\,\tilde{f}_{1}} \over {\tilde{f}_{4} \,
\tilde{f}_{3}}}\,\, ,\,\,\,\,
x_{4} \, = \, {{\tilde{f}_{6}\,\tilde{f}_{2}} \over {\tilde{f}_{5} \,
\tilde{f}_{4}}}
\end{eqnarray}

The first homogeneous variables  $q_n$ are given in terms of  the
$f_n$'s as:

\begin{eqnarray}
q_2 &=& \, {{f_{2}} \over{ f_{1}}} \cdot q_0 , \, \, \,\, q_3 \, = \,
{{f_{3}\,
f_{1}} \over{ f_{2}}} \cdot q_1 , \, \, \,\,q_4 \, = \, {{f_{4}} \over{
f_{3} \, f_{1}}} \cdot q_0 \,\,, \, \,\,\,
 q_5 \, = \, {{f_{5}} \over{ f_{4}\, f_{2}}} \cdot q_1 , \, \, \,\,q_6
 \, = \,
{{f_{6}\, f_{2}} \over{ f_{5} \, f_{3}\, f_{1}}} \cdot q_0 , \,
\,\,\,\\ \nonumber
 q_7 &=& \, {{f_{7}\,f_{3}} \over{ f_{6}\, f_{4}\, f_{2}\, f_{1}}}
 \cdot
q_1 , \, \, \,\,q_8 \, = \,
{{f_{8}\, f_{4}} \over{ f_{7} \, f_{5}\, f_{3}\, f_{1}}} \cdot q_0  ,
\, \,\,\,
 q_9 \, = \, {{f_{9}\,f_{5}} \over{ f_{8}\, f_{6}\, f_{4}\, f_{2}}}
 \cdot q_1  , \,\,\,
  q_{10} \, = \,
{{f_{10}\, f_{6} \, f_{2}} \over{ f_{9} \, f_{7}\, f_{5}\, f_{3}\,
f_{1}}}
\cdot q_0 , \, \,\\ \nonumber
 q_{11} &=& \, {{f_{11}\,f_{7}\,f_{3}} \over{ f_{10}\, f_{8}\, f_{6}\,
f_{4}\, f_{2}}} \cdot q_1  , \, \,\,\,
 q_{12} \, = \,
{{f_{12}\, f_{8} \, f_{4}} \over{ f_{11} \, f_{9}\, f_{7}\, f_{5}\,
f_{3}\, f_{2}}}
\cdot q_0
\end{eqnarray}

\end{document}